\documentclass[12pt]{article}
\usepackage{amsmath,amssymb,amsthm}
\usepackage{graphicx}
\usepackage{hyperref}
\usepackage{float} 
\usepackage[most]{tcolorbox}
\usepackage{booktabs}
\usepackage{tabularx}
\usepackage{caption}
\usepackage{placeins}

\newcommand{\pass}{\textcolor{green!50!black}{\checkmark}}
\newcommand{\fail}{\textcolor{red!70!black}{$\times$}}

\title{Causal Event Networks: Entangled Tensor Graphs, Connectome Quotient Topology, and Subjective Experience as Geometric Dual}
\author{
  Vahid R. Ramezani\thanks{Email: \texttt{vahid@engramsciences.com}} 
}
\date{November 2026}

\theoremstyle{plain}

\newtheorem{proposition}{Proposition}

\theoremstyle{definition}

\theoremstyle{remark}
\newtheorem{remark}{Remark}

\newtheorem{corollary}{Corollary}

\begin{document}
\sloppy
\maketitle

\begin{abstract}
We introduce a chromatically constrained tensor-product construction that generates sparse recurrent computational graphs whose coloring bounds the number of sequential execution phases: vertices are computational units, edges are causal message-passing channels, and the graph can grow arbitrarily large while its chromatic causal depth stays fixed. Coarsened to the scale of the Budapest Reference Connectome, the community quotient of a simple local construction reproduces the connectome's broad degree profile, passes a nine-criterion topological test and comes close to the connectome on most of a further battery of complexity, spectral, controllability and clique measures, while the fine parent graph shows a scale separation: multiscale organization is intrinsic to the fine graph, and short-path integration appears only under coarsening. Placing short-path integration directly in the fine substrate, with chromatically admissible bridges and targeted sparsification, yields quotients far less modular than the connectome across a wide range of coarsening schemes. Relocating edges into the interiors of quotient nodes instead produces a quotient that matches the connectome's degree profile, edge count and fiber-bundle distribution, passes the full nine-criterion test and comes close on most complexity and spectral measures; its fine substrate fails only the two near-random path criteria. Used as the deep-feature pathway of a U-Net encoder-decoder, the graph-derived recurrent module improves image inpainting over the convolutional baseline. We close with a conjectural, structural account of subjective experience in holographic terms: the graph acts as an evolution operator that fixes the admissible causal order and the geometries the substrate can support, and experience occupies the position of the bulk dual of a particular activity state.

\end{abstract}

\section{Introduction}
\label{sec:intro-triangle}

This paper grows out of the author's 2014 essay~\cite{Ramezani2014CausalEvent}. There we described cognition as a network of causal events and argued that a conscious moment can contain only a limited number of sequential computations. The present paper is self-contained, but the essay supplies much of its philosophical motivation, as well as the central mathematical ideas we develop further here.

What kind of computational graph could support the highly distributed and concurrent operations of the brain, while remaining consistent with the broad architecture of the human connectome? We approach this question by constructing a sparse graph whose chromatic number places a bound on the number of sequential execution phases. We then examine whether a coarse-grained representation of this graph can reproduce the main topological features of the connectome.

Consider first the growth of the mammalian cortex. Over the course of evolution, the cortical sheet expanded enormously, while its thickness remained comparatively bounded. The radial unit hypothesis provides a developmental explanation for this difference \cite{Rakic1988,Rakic2000}. Changes in the proliferation of founder cells in the ventricular zone increase the number of radial columnar units without substantially increasing the number of neurons within each unit. Symmetric cell divisions expand the progenitor population and, consequently, the cortical surface. Asymmetric divisions contribute neurons within each unit and increase cortical thickness. The result is a pattern of growth in which local neural organization is repeated over an expanding surface. Much of the increase in cortical complexity comes from adding units, not from increasing the depth of each unit.

A second observation concerns time. Mammalian brains differ enormously in size, number of neurons and complexity of connections. Yet the characteristic timescales of neural processing and distinguishable perception remain broadly comparable, typically ranging from tens to hundreds of milliseconds \cite{BuzsakiLogothetisSinger2013,Thorpe1996Speed,FabreThorpe1998Monkeys,Poppel1997Hierarchical}. Why should this be so? If increasingly complex brains required proportionally more sequential stages of computation to produce a percept, we would expect perceptual times to increase accordingly. The observed similarity of these timescales suggests that much of the additional complexity is accommodated in another way.

We propose that this way is concurrency. As neural systems become more complex, the number of computations that can take place concurrently increases far more than the number of successive computational stages. This is not parallel processing in the conventional computer-science sense. We cannot identify a collection of independent processors in the brain, each carrying out a well-defined task. Neural computation takes place through overlapping pathways in a recurrent and dynamically coupled network. The question is whether such a network can grow in complexity while the number of required sequential phases remains bounded. We investigate the graph-theoretic consequences of this possibility and, subsequently, what they might tell us about the organization of subjective experience.

Consider a collection of computational tasks distributed among a set of units. Some tasks can take place simultaneously, while others must be separated in time. We can represent this constraint by a graph whose vertices are task--unit pairs and whose edges join pairs that cannot execute in the same phase.
An edge specifies that two tasks must be temporally separated. It does not, by itself, determine which task must come first. The graph is therefore undirected. A particular execution schedule assigns an order to the tasks and thereby orients the active constraints from earlier to later events. How many sequential phases are needed to execute the entire task set? If we require only that adjacent vertices occupy different phases, the answer is the chromatic number of the graph, $\chi(G)$. A proper coloring assigns a phase to every vertex, and the smallest number of colors gives the minimum number of phases.

We take the preservation of this minimum phase requirement as a form of {\it cognitive action principle}. Colors represent causal phase classes, and we call their minimum number the \emph{chromatic causal depth}.

We next ask whether a computational graph can grow in size and complexity without a corresponding increase in chromatic causal depth. One way to accomplish this is through graph tensor products. Starting from a small tripartite seed, we construct increasingly large graphs by repeated tensor composition. Graph products have been studied extensively in combinatorics \cite{HammackImrichKlavzar} and have also been used in the construction of complex network models \cite{LeskovecKronecker2010}. A useful property of the tensor product is that

\begin{equation}
\label{eq:chrom}
\chi(G \otimes H) \;\le\; \min\{\chi(G),\chi(H)\}.
\end{equation}

Thus, tensor composition can multiply the number of vertices without increasing the minimum number of execution phases.

The construction has a simple interpretation. Each vertex is expanded into a collection of new vertices, and each edge generates a corresponding pattern of connections between the expanded collections. Repeating this operation produces a larger graph in which the local structure is reproduced at successive scales. The number of computational units can increase enormously while the chromatic depth remains bounded.

The inequality \eqref{eq:chrom} follows directly from the definition of a proper coloring. A coloring of either factor induces a coloring of the product through the corresponding projection. A more difficult question is whether equality always holds. Hedetniemi conjectured in 1966 that it does \cite{Hedetniemi1966}. More than fifty years later, Shitov disproved the conjecture by constructing graphs whose tensor product has a smaller chromatic number than either factor \cite{shitov2019counterexamples}. Earlier results had established that such a reduction is impossible when the smaller chromatic number is at most four \cite{ElZaharSauer1985}. The known counterexamples occur above this threshold \cite{Wrochna2022}.

For our construction, equality follows from a simpler argument. The seed graph $S$ contains a triangle. If $G$ and $H$ each contain a triangle, with vertices $(g_1,g_2,g_3)$ and $(h_1,h_2,h_3)$, then the three vertices $(g_1,h_1)$, $(g_2,h_2)$ and $(g_3,h_3)$ form a triangle in $G\otimes H$. Hence the tensor product also requires at least three colors. Writing $M_k$ for the graph at tensor scale $k$, we have $\chi(M_k)\geq 3$. But \eqref{eq:chrom} gives $\chi(M_k)\leq 3$. Consequently,

\[
\chi(M_k)=3
\]

at every tensor scale. The graph can grow without bound while the chromatic causal depth remains exactly three. We will introduce further operations that alter its structure while preserving three-colorability.

There is another useful aspect of tensor composition. In quantum mechanics, tensor products provide a way of describing composite systems in terms of their constituent spaces. Graph tensor products provide a related compositional structure, although the objects being composed are graphs, not quantum states. The adjacency matrix of the product satisfies

\[
A(G\otimes H)=A(G)\otimes A(H).
\]

The resulting graph therefore retains an exact factorization into its original components. In a $k$-fold product of the seed, each vertex is represented by a $k$-tuple. The entries identify positions at successive scales: the first determines a coarse block, and the remaining entries specify progressively finer positions within it.

What happens if we modify this graph by adding or deleting edges? In general, the resulting adjacency matrix can no longer be expressed as the Kronecker product of the original factors. The modifications destroy the exact tensor decomposition and introduce relations among scales that were previously separable. We call this \emph{graph entanglement}.

The term is used in a structural sense. Quantum entanglement ordinarily concerns states in a tensor-product Hilbert space. Here we are concerned with the factorization of the graph itself. The connection is to operator entanglement \cite{Zanardi2001}. Relative to the chosen tensor decomposition, a pure Kronecker-product adjacency matrix has operator Schmidt rank one. When the modified matrix admits no Kronecker factorization at all, the rank becomes greater than one. Graph entanglement is therefore a property of the computational structure on which the dynamics takes place, not a claim about the quantum state of the system.

We impose one important restriction on these modifications. The chromatic bound must not increase. Removing an edge cannot increase the chromatic number, and an added edge is admissible if its endpoints belong to different classes of the inherited three-coloring. We can therefore introduce new connections, remove existing ones and break the tensor-product symmetry while retaining a computational graph that requires no more than three sequential phases.

These operations also have a possible biological interpretation. Pruning, local closure and long-range bridging resemble some of the structural changes that take place in developing neural networks. The resemblance is only at the level of graph structure. We do not attempt to reconstruct axonal development or synaptic learning. Our purpose is to investigate what kinds of large-scale organization can be produced under a fixed constraint on sequential computation.

We can now return to the causal interpretation. Neural computation is not a single sequence of events. Many processes occur concurrently, while others influence subsequent processes through the propagation of signals. A particular history of neural activity can therefore be represented as a partially ordered set of events. Two events are causally ordered when information propagates from one to the other. Events that have no causal relation in the chosen description may remain unordered, even if they occur at approximately the same time or participate in a common pattern of activity.

We distinguish this latter possibility as {\it coactivation}. Units coactivate when they settle jointly into one coordinated state, through rapid mutual exchange in which no transmitter and receiver can be distinguished; occurring at the same time is not enough. In our framework, causal signaling is represented by the active connections between computational units, while coactivation need not correspond to an edge. The undirected graph specifies the admissible structure of causal interactions; a particular execution history determines their temporal ordering. In this sense, causal computation involves one cluster or center of activity ``listening'' while another ``speaks.'' The resulting order is between groups, not merely between individual units. Because a coactivating group settles jointly, an input that reaches any one of its members shapes the whole group, so a single edge between two groups is enough to order them as wholes; it does not take a line between every pair of their units.

\subsection{Why three phases}

The preceding argument motivates a bound on chromatic depth, but does not determine what this bound should be. We choose three. Other values are worth investigating. With more colors, it may still be possible to group the phases into a small number of functional macro-phases at a higher scale. Our choice of three follows from two independent lines of thought.

The first is phenomenological. A central problem in the phenomenology of time is how we experience a temporally extended object, such as a melody or a sentence, as a unified whole and not merely as a sequence of isolated moments. Husserl's account \cite{Husserl1991} describes what he later called the ``living present'' in terms of three distinguishable but inseparable processes: retention, primal impression, and protention. These preserve what has just occurred, register what is occurring, and anticipate what is about to occur.

A growing body of research suggests that subjective experience does not reside in isolated neurons or brain regions. It emerges from coordinated dynamics distributed across large-scale neural networks, constrained by the structure of the connectome \cite{BullmoreSporns2009,DehaeneChangeux2011}. It is therefore fruitful to ask whether the temporal organization described by phenomenology can be projected onto a computational model whose underlying graph reproduces the major topological properties of the connectome. Such a model would provide a common framework for studying temporal integration, recurrent computation, and brain-like network organization.

The second line of thought is functional and comes from recent mechanistic studies of language models. Gurnee et al.\ provide a particularly clear parallel. Using the Jacobian lens, they identify a privileged workspace-like set of representations in language models, the J-space, whose contents are reportable, subject to directed modulation, used in internal reasoning, flexibly available to downstream computations, and selective \cite{Gurnee2026GlobalWorkspace}. Their work is particularly valuable because it demonstrates that global-workspace-like organization can be studied mechanistically within a modern artificial system, without relying only on a high-level cognitive metaphor. Their results divide transformer depth into three functional regimes: early sensory-like layers, intermediate workspace-like layers, and late motor-like layers. In a transformer, broadcast takes place along the depth and sequence axes, without explicit recurrent loops. This suggests that a sufficiently deep non-recurrent stack may simulate, through composition, some aspects of scale simultaneity, a notion we will discuss in the following sections.

Global-workspace accounts, in their standard formulation, concern long-range cortical connectivity. A distributed set of regions, with extensive long-range projections concentrated in prefrontal, parietal, and cingulate cortex, makes selected content available to many specialized processes \cite{Baars1988,DehaeneChangeux2011}. This organization is largely at the scale of white-matter pathways between anatomical regions, which is precisely the scale represented by a parcel-level structural connectome. The structural signatures associated with workspace organization include connector hubs that bridge otherwise segregated modules, a rich-club core of densely interconnected high-degree regions, and short paths that permit rapid global reach. These are all measurable properties of such a graph, and all appear in the tests reported in this paper.

\subsection{A Holographic Analogy}
\label{sec:holographic-analogy}

Our construction contains three interlocking structures. The first is a correlational layer, where same-phase units may exhibit coordinated activity without sharing a graph edge or exchanging a message. The second is causal signaling, carried by cross-color edges and counted in color transitions. The third is a geometry, which the pure tensor product does not yet possess, and which becomes nontrivial only after the symmetry-breaking edits differentiate it (Sections~\ref{subsec:symmetry-metric}--\ref{sec:emergent-geometry}).

Frameworks that take causal order as primitive, such as causal-set theory \cite{BombelliLeeMeyerSorkin1987}, naturally accommodate the causal layer. Correlations between causally unrelated events enter these frameworks through quantum fields defined on the order, not through the order itself. Holography \cite{Maldacena1998,GubserKlebanovPolyakov1998,Witten1998} provides a developed framework in which all three structures belong to a single description. A boundary theory fixes the admissible causal structure. Entanglement is a relational structure of the boundary state and is not itself a signaling channel. Geometry emerges in the bulk \cite{RyuTakayanagi2006,VanRaamsdonk2010}, and bulk causal propagation is defined relative to this emergent geometry. Our analogy is based on this structural correspondence and no other. In both settings, correlation, geometry, and signaling are three aspects of a single underlying relational object, not independent ingredients.

We introduce holography not to suggest that the brain realizes an AdS/CFT duality, but to clarify the kind of emergence that our construction may require. The general pattern is the familiar duality between space and quantity, where a space is recovered from an algebra of distinctions, and its points are whatever that algebra can jointly distinguish. Holography is an instance of this pattern in which the recovered space also carries locality and causal propagation, properties required of a cognitive geometry. If a dual description exists here, the emergent geometry would not reconstruct physical spatial distance. It would instead describe the richer geometry of cognitive content, in which perceptual proximity, object membership, semantic relatedness, and attentional grouping define different notions of nearness over the same underlying events. The comparison is therefore methodological. Holography provides a concrete example in which a non-spatial relational structure supports an emergent geometry and causal organization that are not explicit in the fundamental description. We develop the analogy and tabulate its exact extent and limits in Section~\ref{sec:emergent-geometry}, after introducing the construction that supports it.

Our contributions are as follows.

\begin{enumerate}

\item A chromatically constrained tensor-product construction (growth, pruning, local repair, triangle closure) that produces sparse recurrent graphs of unbounded size while keeping chromatic causal depth fixed at three phases.

\item A demonstration that the community quotient of this local construction reproduces the degree profile of the Budapest Reference Connectome and passes a nine-criterion topological test, together with complexity, spectral, and controllability comparisons. Only node count, edge count, approximate modularity, and low-degree periphery nodes are used as targets. These properties are explicit design targets. Parameters were adjusted against them, and they therefore indirectly constrain the construction of the parent graph.

\item A parent--quotient scale separation in which multiscale organization is intrinsic to the fine graph, while short-path integration emerges under coarsening. We also develop an extension using chromatically admissible bridges and multiscale-targeted sparsification that restores integration at the fine scale. In our experiments, no tested strict coarsening of this integrated parent recovers the modular organization of the reference connectome. This obstruction motivates an inverse formulation in which the coarse-graining map is held fixed and the parent substrate is rebuilt under it. Edges are relocated into quotient-node interiors while every remaining cross-supernode relation is represented. The resulting quotient matches the reference degree profile, edge count, and fiber-bundle distribution and passes the full nine-criterion test.

\item A recurrent architecture derived from the graph, with same-phase coactivation modeled by attention, evaluated as the deep-feature pathway of an encoder--decoder on image inpainting.

\item A conjectural theory of subjective experience grounded in the causal event structure of the construction. It describes cognitive objects as realized causal structure, acausally constituted coactivations with no extent in causal time, a symmetry-breaking origin for the metric, and experience as the geometric side of a structural duality whose other side is the causal web.

\end{enumerate}

\subsection{Organization of the paper}

The remainder of the paper is organized as follows.

Section~\ref{sec:related-work} places our construction in the context of existing work on generative connectome models, graph products, and bounded-depth concurrent computation.

Section~\ref{sec:tensor-growth} introduces the growth rule: tensor expansion of a tripartite seed, together with chromatically constrained edge additions and deletions that break exact product decomposability while preserving the three-coloring.

Sections~\ref{sec:local_repair_quotient} and~\ref{sec:full-parent-and-budapest} develop the construction in two stages. The distinction between these stages is central to the argument of the paper.

Section~\ref{sec:local_repair_quotient} introduces a ``local'' pipeline consisting of pruning, repair, admissible triangle closure, and coarsening. We test the resulting graph at both scales. Its Budapest-scale quotient passes the standard connectome diagnostics, but the fine parent graph fails two integration criteria: near-random path length and near-random global efficiency. This identifies what coarsening supplies.

Section~\ref{sec:full-parent-and-budapest} introduces chromatically admissible cross-community bridges and multiscale-targeted sparsification to place short-path integration directly in the parent graph. However, strict coarsening of this integrated parent does not recover the modular organization of the reference connectome across the partitioning schemes we tested. This obstruction motivates an inverse formulation. We hold the coarse-graining map fixed and rebuild the parent substrate under it. Edges are relocated into quotient-node interiors instead of being suppressed between supernodes, and every remaining cross-supernode relation is represented in the quotient. The within-supernode relations are explicitly realized as edges in the parent graph. We compare the resulting \(1{,}015\)-node quotient with the Budapest Reference Connectome using the nine-criterion network-topology test and additional complexity, spectral, combinatorial, and controllability measures.

Section~\ref{sec:dynamical-realization} considers the dynamical interpretation. The three color classes become three recurrent computational phases, with sparse cross-phase message-passing inherited from the graph and same-phase coactivation represented separately. We evaluate a graph-derived recurrent module as the deep-feature pathway of a U-Net \cite{Ronneberger2015U-Net} encoder--decoder on CelebA-HQ image inpainting. This section also develops the event grain and the routing constraints imposed by the graph.

Section~\ref{sec:ads-cft-correspondence} develops the emergent geometry. We prove that a pure tensor power of any seed is metrically degenerate by symmetry, and that this degeneracy can be removed only by symmetry-breaking edits after which the graph is no longer a tensor power. We distinguish same-phase collective correlation from cross-phase causal propagation, examine the extent and limits of the holographic analogy, and develop the proposed account of cognitive objects and cognitive time.

Section~\ref{sec:conclusion} concludes the paper with a discussion of the interpretation and limitations of the results and directions for future work.

The appendices contain detailed construction and validation material. Appendix~\ref{app:global_parent_graph} documents the parent graphs and their tests, including the full \(9/9\) test of the original parent. The remaining appendices describe the coarsening obstruction and partition sweep; the inverse relocation procedure, bundle matching, degree matching, and strict quotient derivation; parameter accounting and further diagnostics of the final quotient; the image-inpainting architecture and training procedure; the network-topology test and null models; and a statistical-mechanical precedent for the hierarchical description of dynamical variables. We retain the original parent graph because it is the starting substrate for the relocation construction and provides the before-relocation comparison.

\section{Related Work}
\label{sec:related-work}

\paragraph{Generative and statistical models of connectome topology.}

Our construction is related to stochastic Kronecker graphical models \cite{LeskovecKronecker2010}. These models generate large realistic networks by forming a factorizable Kronecker probability template \(P_k=P_1^{\otimes k}\) and then sampling the final graph from this template. At the level of the realized graph, the sampling is effectively a probabilistic pruning. Starting from the Kronecker-defined edge support, only a selected subset of edges is retained, without a separate mechanism for enforcing a bound on chromatic execution depth.

Our use of tensor products is different. The tensor product provides the growth scaffold, but we subsequently modify the graph using chromatically constrained edits that break its product form. In the local construction, pruning, repair, and triangle-closing additions take place inside the growth loop, so that local additions can become part of later tensor products. In the global construction, pruning dominates the tensor-growth phase and produces a sparse tensor-derived skeleton. Larger additive operations, including bridge insertion and later refinement, act on this skeleton without fitting an empirical edge set. The final graph is therefore not merely a sampled or thinned Kronecker template. It is a tensor-derived graph whose product ancestry is altered through constrained deletion, repair, closure, and integration operations while preserving the chromatic bound. There is also a structural distinction. A pure tensor power has a large automorphism group under which almost all vertices are metrically indistinguishable. This degeneracy persists as long as the graph remains a tensor power (Section~\ref{subsec:symmetry-metric}). In our construction, the edits that break the product form individuate positions.

Furthermore, the Kronecker graph models of Leskovec et al.\ \cite{LeskovecKronecker2010} can generate realistic network statistics, but do not identify the sparse computational graph that may be ``hidden'' beneath an observed connectome. We already have empirical connectome graphs. If the computational architecture could be explicitly recovered from these graphs, we could simply use them directly as neural graphical models. Our problem is different. We seek to construct a fine-scale computational graph whose vertices and edges define a bounded-depth recurrent substrate and whose clustered quotient has connectome-like topology. In this interpretation, large-scale connectome edges arise as axon-bundle-like aggregates of many finer computational graph edges, while connectome nodes correspond to clusters of fine computational units.

A substantial literature has attempted to explain connectome topology through compact generative or statistical rules. Early mechanistic models emphasized spatial embedding, wiring cost, and developmental growth. Kaiser and Hilgetag showed that spatially constrained growth can generate clustered and interconnected networks without conventional preferential attachment \cite{kaiser2004spatial}. Their work suggested that the physical embedding of a neural system can strongly constrain the topology of the resulting graph.

Later models described brain-network formation in terms of a trade-off between the cost and topological value of a connection. V\'ertes et al.\ introduced an economical clustering model of human functional networks in which connection probability depends on both physical distance and the number of neighbors shared by two regions \cite{vertes2012simple}. Penalizing distance alone did not reproduce the observed combination of clustering, modularity, global efficiency, and degree distribution. A substantially better agreement was obtained when spatial cost was combined with a preference for connections between regions that share neighbors.

Betzel et al.\ systematically extended this approach to human structural connectomes \cite{betzel2016generative}. Their models add edges according to rules of the general form

\begin{equation}
P_{ij} \propto D_{ij}^{\eta} K_{ij}^{\gamma},
\label{eq:cost-value-model}
\end{equation}
where $D_{ij}$ is the physical distance between regions $i$ and $j$, and $K_{ij}$ is a dynamically updated measure of topological value. The choices examined for $K_{ij}$ included degree-based, clustering-based, and homophily-based quantities, with the last measured using common neighbors or the matching index. Models combining geometric cost with homophilic attachment, particularly those using the matching index, reproduced several properties of empirical connectomes, including properties not explicitly included in the model optimization.

Akarca et al.\ used related cost--value models to study neurodevelopmental and inter-individual variation in structural brain organization \cite{akarca2021generative}. They found that varying a small number of parameters controlling spatial cost and topological value produces substantial diversity in network topology. Oldham et al.\ extended this framework by comparing spatial, developmental, physiological, and topological constraints on human brain connectivity \cite{oldham2022modeling}. Their best-fitting models assigned connection value according to the similarity of regional gene expression instead of topology. All models, however, had difficulty reproducing the spatial embedding of connectome topology. These studies support the view that connectome formation reflects an interaction between economical wiring and the functional, topological, or physiological value of particular connections.

A complementary approach treats an observed connectome as a realization from a fitted probability distribution over graphs. Exponential random graph models describe the probability of an entire network using interacting local statistics, including edges, stars, triangles, and shared-neighbor configurations. Simpson et al.\ applied this framework to whole-brain functional networks and showed that combinations of local motifs can be fitted to reproduce aspects of global brain-network organization \cite{simpson2011exponential}. These models are statistically principled and can generate synthetic networks from an estimated ensemble. However, their fitted parameters do not necessarily identify a unique biological growth mechanism.

Stochastic block models assign nodes to latent classes and estimate the probability or weight distribution of connections between these classes. Faskowitz et al.\ applied weighted stochastic block models to human structural connectomes across the lifespan \cite{faskowitz2018weighted}. Unlike standard modularity-based models, stochastic block models are not restricted to assortative communities with dense internal connectivity. They can describe modular, core--periphery, disassortative, and mixed mesoscale organizations. These methods provide flexible statistical descriptions of connectome structure, but fit mesoscale organization directly instead of deriving it from a microscopic compositional process.

Klimm et al.\ compared human diffusion-imaging connectomes with thirteen synthetic network models incorporating spatial embedding, temporal growth, random connectivity, preferential attachment, and other structural constraints \cite{klimm2014resolving}. Their results showed that models designed to reproduce one network property generally failed to reproduce the full collection of connectome characteristics. This observation is particularly relevant to our work. Reproducing clustering, modularity, path length, degree heterogeneity, or hierarchy separately is substantially easier than reproducing their characteristic combination in a single sparse graph.

Our construction belongs to the mechanistic generative tradition, but differs from cost--value models and fitted statistical graph ensembles. We do not estimate edge probabilities directly from an empirical connectome. Nor do we begin with a fixed set of anatomically embedded regions and add edges individually. We begin with a small tripartite seed and expand it compositionally through graph tensor products. The resulting scaffold is modified by community-aware pruning, cross-community bridge insertion, and sparsification while preserving a global chromatic constraint. Our model is therefore an abstract combinatorial mechanism for generating connectome-like complexity under bounded chromatic causal depth, not a literal simulation of axonal development.

The models reviewed above describe the connectome primarily as a spatially embedded graph, the result of developmental wiring, or a realization from a statistical graph ensemble \cite{betzel2016generative,akarca2021generative,oldham2022modeling,simpson2011exponential,faskowitz2018weighted}. Although they reproduce important structural properties, they generally do not derive connectome topology from a fundamental property of the computation that the network must perform. The connectome, however, is not simply an anatomical wiring diagram. It is the physical substrate of a distributed biological computation, where many neural operations take place concurrently and others must be causally ordered. It is therefore natural to ask whether some aspects of connectome topology can be generated from the computational constraints themselves.

There is another important distinction. Our comparison target is not the empirical connectome graph itself. We do not use anatomical coordinates, tractography weights, subject-specific adjacency matrices, or a fitted probability model over edges. Our construction therefore cannot be interpreted as a reconstruction of a particular connectome. Instead, we compare the broader organization of the graph: its network-topological profile, algebraic and combinatorial invariants, and, in subsequent dynamical analyses, its capacity to support efficient state propagation and control-like behavior. Our objective is not to reproduce the exact edge set of the brain. We ask whether a combinatorial rule that increases complexity under bounded chromatic execution depth can generate the same class of structural complexity exhibited by empirical connectomes.

\paragraph{Inference-based accounts of brain function.}

A different tradition describes the brain primarily in terms of what it computes. Predictive coding treats cortical processing as hierarchical inference, where predictions and prediction errors are exchanged between levels \cite{RaoBallard1999}. Hierarchical Bayesian models extend this picture to the visual cortex as a whole \cite{LeeMumford2003}, while the broader Bayesian-brain program treats neural computation as inference under uncertainty \cite{KnillPouget2004}. These accounts are often expressed as message passing on a graphical model, which is close to the causal layer of our construction. However, the graph itself is usually assumed to be given rather than generated.

The combination of directed and symmetric dependence used here also has a counterpart in graphical models. A chain graph has directed edges between blocks and undirected edges within blocks. Its distribution can be generated as the equilibrium of a dynamic model with feedback \cite{LauritzenRichardson2002}, corresponding naturally to fast local settling within a population.

Our construction differs in three respects. First, the blocks are not specified in advance. Same-phase dependence is wrapped by causal paths, since the possible locations of coactivation are determined by the causal environment shared by the units. This organization is already present in the substrate before clustering, which only exposes it. Second, within-block dependence is represented by an explicit operator that transforms the state carried forward, not by an implicit factor of a joint distribution. Third, because the substrate is generated through tensor composition, the same wrapping of dependence by causal paths recurs at every scale (Section~\ref{subsec:full-parent-operators}).

The free-energy principle goes beyond these inference accounts by grounding them in statistical physics. A system that persists in a changing environment, separated from it by a Markov blanket of sensory and active states, can be described as minimizing variational free energy and therefore as performing approximate inference \cite{Friston2010FreeEnergy,Friston2013Life}. Both approaches draw a boundary between a system and what lies outside it, but the boundaries are different.

A Markov blanket is a statistical boundary of a persisting system, defined by conditional independence between its internal and external states. In our construction, the boundary is drawn anew for each event complex. The complex is carved out by causal closure and is self-contained in the sense that its internal construction is not separately visible. Everything outside it, whether in the world or elsewhere in the nervous system, enters only as input that can modulate the transformations performed by the complex, without becoming part of it.

The explanatory objective is also different. Inference-based accounts are formulated using probabilistic quantities such as prediction error, free energy, and entropy. Here the problem is geometric. Experience directly presents a geometry, with things near and far, beside and behind (Section~\ref{subsec:cognitive-geometry}). We ask how such a geometry can emerge from the causal--correlational organization of events.

The two approaches are complementary, not competing. Inference may describe what the dynamics computes. Our construction concerns the structure within which that computation is organized and, according to our conjecture, experienced. This structure is a graph generated under a bound on causal depth, with an explicit distinction between directed and same-phase dependence and an emergence of geometry through symmetry breaking and coarse-graining.

\section{Tensor-Product Growth with Chromatically Constrained Connectivity Modification}
\label{sec:tensor-growth}

Let \(G_k=(V_k,E_k)\) denote the graph at scale \(k\), and let
\(S=(V_S,E_S)\) be a fixed seed graph. The basic growth operation is the
tensor product
\[
G_k \longmapsto G_k\otimes S,
\]
with
\[
V(G_k\otimes S)=V_k\times V_S
\]
and
\[
E(G_k\otimes S)
=
\left\{
\big((u,a),(v,b)\big):
(u,v)\in E_k,\ (a,b)\in E_S
\right\}.
\]
Thus, at each scale, every vertex of the previous graph is replaced by a
copy of the seed's vertex set, and each edge of the previous graph gives
rise to connections between the corresponding copies, following the
adjacency pattern of the seed: two new vertices \((u,a)\) and \((v,b)\) are
adjacent exactly when \(u\) and \(v\) are adjacent in \(G_k\) and \(a\) and
\(b\) are adjacent in \(S\).

The edited growth rule is
\[
G_{k+1}
=
(G_k\otimes S)+A_k-D_k ,
\]
where \(A_k\) is a set of added edges and \(D_k\) is a set of deleted
edges. More explicitly, if \(G=(V,E)\), then
\[
G+A := (V,E\cup A),
\qquad
G-D := (V,E\setminus D),
\]
so that
\[
(G_k\otimes S)+A_k-D_k
=
\left(
V_k\times V_S,\,
\big(E(G_k\otimes S)\cup A_k\big)\setminus D_k
\right).
\]

The essential constraint is that the graph remains properly
\(3\)-colored. Let
\[
c_k:V_k\to\{0,1,2\}
\]
be the carried coloring at scale \(k\). After each tensor expansion, a
vertex inherits the color of its first coordinate:
\[
c_{k+1}((u,a)):=c_k(u).
\]
For a colored vertex set \(V\), define the chromatically admissible edge set
\[
\mathcal E_\chi(V)
=
\left\{
\{x,y\}\subset V:
c(x)\neq c(y)
\right\}.
\]
Every added edge is required to belong to \(\mathcal E_\chi(V)\).
Deletions require no chromatic restriction, because deleting an edge cannot
invalidate a proper coloring. Connectivity and non-bipartiteness are not
guaranteed by these rules, since deletions can disconnect a graph or remove
all of its odd cycles; they are maintained separately, as formalized in the
following proposition.

\begin{proposition}[Chromatic invariance under admissible edge modification]
\label{prop:invariance}
Let \(S\) be a finite connected non-bipartite graph with a proper
\(3\)-coloring \(c_S\), and set \(G_0=S\) with carried coloring
\(c_0=c_S\). Suppose recursively that
\[
G_{k+1}=(G_k\otimes S)+A_k-D_k,
\]
with carried coloring \(c_{k+1}((u,a))=c_k(u)\), where:
\begin{enumerate}
\item[(A1)] every edge in \(A_k\) joins different classes of
\(c_{k+1}\);
\item[(A2)] after the modifications, \(G_{k+1}\) is connected and
non-bipartite.
\end{enumerate}
Then, for every \(k\), the carried coloring \(c_k\) is a proper
\(3\)-coloring of \(G_k\), and \(G_k\) is connected with
\[
\chi(G_k)=3.
\]
\end{proposition}

\begin{proof}
Proceed by induction on the stronger statement that \(c_k\) is a proper
\(3\)-coloring of \(G_k\). The base case holds because \(c_0=c_S\) is
proper on \(G_0=S\), and \(S\) is connected and non-bipartite, so
\(\chi(G_0)=3\). Assuming \(c_k\)
proper, every tensor edge of \(G_k\otimes S\) projects in the first
coordinate to an edge of \(G_k\), so \(c_{k+1}\) is proper on
\(G_k\otimes S\); this is the classical bound~\eqref{eq:chrom}
\cite{Hedetniemi1966} realized by an explicit coloring. Edges of \(A_k\)
are admissible for \(c_{k+1}\) by (A1), and deletions cannot invalidate a proper
coloring, so \(c_{k+1}\) is a proper \(3\)-coloring of \(G_{k+1}\) and
\(\chi(G_{k+1})\leq3\). By (A2), \(G_{k+1}\) is non-bipartite and
therefore \(\chi(G_{k+1})\geq3\), while connectivity is also supplied
by (A2).
\end{proof}

The same argument covers admissible additions and arbitrary deletions made after
the last tensor step, such as the triangle closure, bridge insertion, and
sparsification used in the constructions below: each preserves the carried
coloring, so only connectivity and non-bipartiteness need to be checked.

\section{Local Construction}
\label{sec:local_repair_quotient}

We first introduce a simple ``local'' construction.  It uses a
tripartite seed expanded repeatedly by graph tensor product, pruned,
locally repaired, and then enlarged by chromatically admissible triangle
closure. Unlike the integrated construction developed later, this construction does
not include cross-community bridge insertion or multiscale-targeted
sparsification.  Before describing it, we
record a property of the pure tensor powers that will matter throughout,
and then explain why approaches based on constant diameter are not
suited to generating concurrency.

\begin{proposition}
\label{prop:diameter-flat}
Let $S$ be connected and non-bipartite. For $u,v \in V(S)$ let $\ell(u,v)$
be the least $N$ such that $u$--$v$ walks of every length $m \ge N$ exist
(finite, since $S$ is connected and contains an odd cycle), and set
$\Lambda(S) = \max_{u,v} \ell(u,v)$. Then $S^{\otimes k}$ is connected and
non-bipartite for every $k \ge 1$, and
\[
\operatorname{diam}(S)
\;\le\;
\operatorname{diam}\bigl(S^{\otimes k}\bigr)
\;\le\;
\operatorname{diam}\bigl(S^{\otimes (k+1)}\bigr)
\;\le\;
\Lambda(S)
\qquad \text{for all } k \ge 1 .
\]
In particular, if $\operatorname{diam}(S)=\Lambda(S)$ then
$\operatorname{diam}(S^{\otimes k})=\operatorname{diam}(S)$ for every
$k\ge1$. The seed used in this paper, listed with its coloring in Appendix~\ref{app:construction-parameters}, satisfies
$\operatorname{diam}(S)=\Lambda(S)=4$, so every tensor power has
diameter exactly $4$.
\end{proposition}

\begin{proof}

Connectivity of the loopless tensor product of connected graphs holds when
a factor is non-bipartite \cite{Weichsel1962}. Non-bipartiteness persists
because an odd cycle in \(S\) lifts diagonally to an odd cycle in every
tensor power; in particular, for the seed used here, the triangle itself
lifts to a triangle at every tensor scale by the argument of Section~\ref{sec:intro-triangle}. 
A walk in $S^{\otimes k}$ is exactly a tuple of equal-length walks in the factors,
so
\[
d\bigl((u_i),(v_i)\bigr)
=\min\{\, m : \text{for every } i \text{ there is a }
u_i\text{--}v_i \text{ walk of length } m \,\}.
\]
Every coordinate pair admits walks of every length $m \ge \Lambda(S)$,
giving the upper bound. For monotonicity, extend a diameter-realizing
pair of $S^{\otimes k}$ by an arbitrary $(k{+}1)$-st coordinate: the set
of lengths feasible for all $k{+}1$ coordinates is a subset of the set
feasible for the first $k$, so its minimum is at least as large. Taking
$k=1$ in the same argument gives the lower bound
$\operatorname{diam}(S)\le\operatorname{diam}(S^{\otimes k})$. For the
seed used in this paper, direct computation
gives $\ell(0,4)=4$ (there is no $0$--$4$ walk of length $3$, while walks of every length $m\ge4$
exist) and $\ell(u,v)\le4$ for all other pairs, so $\Lambda(S)=4$;
and $d(0,4)=4$ in $S$, so $\operatorname{diam}(S)=4$.
\end{proof}
\subsection{Constant Diameter versus Bounded Chromatic Depth}
\label{subsec:diameter-chromatic-tradeoff}

Leskovec et al.~\cite{LeskovecKronecker2010} observe that the tensor
product of connected graphs need not remain connected, and therefore restrict
their deterministic construction to initiator graphs having a self-loop on
every vertex. Under this assumption they obtain a particularly strong result:
if the initiator has diameter \(D\), then every Kronecker power has the same
diameter \(D\). The self-loops therefore serve as a mechanism for avoiding
connectivity difficulties under repeated tensor products and for obtaining
constant diameter.

This mechanism is not compatible with the chromatic interpretation adopted
here. In the ordinary definition of proper vertex coloring, a graph containing
a self-loop has no proper coloring, since the incident vertex would have to
receive a color different from itself. More importantly, even if the
self-loops are removed only after the tensor construction, their presence
during tensor multiplication can cause chromatic complexity to grow rapidly.

For example, suppose that the loopless support of an initiator \(K\)
contains a triangle with vertices \(a,b,c\), while each of these vertices
also carries a self-loop. After \(k\) tensor powers, consider the \(3^k\)
vertices
\[
(x_1,\ldots,x_k),
\qquad
x_i\in\{a,b,c\}.
\]
For any two distinct tuples, at every coordinate either the entries are equal,
in which case adjacency is supplied by the self-loop, or they are distinct
vertices of the triangle, in which case they are joined by an ordinary edge.
Every pair of distinct tuples is therefore adjacent in the tensor product.
Consequently, after removing the self-loops, the resulting loopless graph
\(\widehat{K^{\otimes k}}\) still contains a clique of size \(3^k\), so that
\[
\chi\!\left(\widehat{K^{\otimes k}}\right)\geq 3^k.
\]
Thus the device that gives the Kronecker construction its
constant-diameter property is unsuitable when chromatic number is interpreted
as bounded execution depth.

Self-loops are not, however, necessary merely to obtain connectivity of a
tensor product. A classical result for simple graphs states that if \(G\) and
\(H\) are connected graphs with at least two vertices, then
\[
G\otimes H
\quad\text{is connected if and only if}\quad
G\ \text{or}\ H\ \text{is non-bipartite}
\]
\cite{Weichsel1962}. Thus a connected non-bipartite seed can generate
connected loopless tensor powers. In particular, the seed used here contains
a triangle and is therefore non-bipartite. For the unmodified tensor
recursion, this provides a loopless route to connectivity without introducing
the self-loops used in the classical Kronecker construction.

At the same time, the loopless tensor product satisfies
\[
\chi(G\otimes H)
\leq
\min\{\chi(G),\chi(H)\}.
\]
Since our seed is properly three-colored and contains a triangle, the pure
tensor powers inherit a three-coloring while retaining the lower bound
\(\chi\geq3\). Hence, before subsequent modifications,
\[
\chi(G_k)=3.
\]
The triangle therefore plays two useful roles: it makes the seed
non-bipartite, supporting connectivity of the loopless tensor product, and it
provides the lower chromatic bound that  fixes the chromatic causal depth at three.

Our construction, however, does not use the tensor-product connectivity theorem as an invariant of the complete algorithm. At each generation step, the tensor-expanded graph is pruned. These edge deletions take the graph outside the assumptions under which connectivity of the pure tensor product would be sufficient. We therefore check connectivity explicitly after pruning and restore it, when necessary, through a repair stage that respects the chromatic constraint. In the runs reported here, pruning disconnects the graph at every scale, producing \(31\), \(136\), \(653\), and \(6{,}826\) components at the four expansions. The repair stage reconnects the graph in each case.

The two approaches make different structural choices. The self-looped Kronecker construction obtains a strong constant-diameter result through reflexive tensor growth. The loopless powers of our seed also have constant diameter, as shown in Proposition~\ref{prop:diameter-flat}, but this follows from the walk-length bound \(\Lambda(S)\), not reflexivity. Our construction remains loopless and preserves bounded chromatic depth. Once pruning and repair are introduced, connectivity becomes a constraint that must be explicitly verified. Short paths and stronger global integration are introduced separately through chromatically admissible modifications. Thus, constant diameter, connectivity, and bounded chromatic causal depth are distinct structural requirements. They do not follow from a single tensor-product mechanism.

In principle, we could impose hard structural constraints on a loop-free stochastic Kronecker construction and statistically fit the remaining parameters to a target adjacency structure. In our setting, however, the computational target graph is not directly observable. Even if such a target were available, we argue below that the relevant complexity should be intrinsic to the construction and not supplied primarily by statistical fitting. We can instead ask whether a small set of symmetry-breaking operations is sufficient to produce the desired network and topological properties of the quotient graph after clustering. We consider it worthwhile to search for suitable parameters without a statistical fitting backbone. Our objective is not to construct a general-purpose generative model capable of producing a broad class of networks. For our purpose, a class of constructions that produces connectome-like properties is sufficient.

\subsection{Algorithm}

We now show that a simple two-part algorithm with a small number of parameters can produce a graph matching numerous properties of the connectome reported in the literature.

In the next section, we extend this algorithm to what we call a ``global'' construction. We emphasize that the simple algorithm already succeeds at the connectome task. The refinement concerns the underlying computational graph, for which no observable target exists. In the extension, we attempt to transfer as many properties of the connectome as possible to this computational graph.

We begin with a small tripartite seed and expand it repeatedly through tensor products. After each expansion, we prune the graph, perform local repair, and add triangles subject to the chromatic constraint. Finally, we coarsen the resulting parent graph to a \(1{,}015\)-node quotient.

\begin{tcolorbox}[
breakable,
colback=white,
colframe=black,
boxrule=0.5pt,
arc=0pt,
left=6pt,
right=6pt,
top=6pt,
bottom=6pt,
title={Algorithm 1A: Parent construction by local repair}
]
\small

\textbf{Input.}
A tripartite seed graph \(S=(V_S,E_S)\) with proper 3-coloring
\(c_S:V_S\to\{0,1,2\}\); tensor depth \(K\); minimum degree
\(d_{\min}\); pruning fraction \(p\); triangle-addition fraction \(a\);
and a fixed random seed for the pseudorandom choices in repair and closure.

\medskip
\textbf{Definitions.}
At scale \(k\), let \(n_k=|V(G_k)|\) and define the geometric-mean edge budget
\[
b_k
=
\operatorname{round}\!\left(
n_k\sqrt{\frac{n_k-1}{2}}
\right).
\]
An admissible edge is an edge \(\{x,y\}\) satisfying \(c(x)\ne c(y)\).
Only admissible edges are added; deletions are unrestricted. The triangle support
of an edge \(\{u,v\}\) is
\[
|N(u)\cap N(v)|.
\]

\medskip
\textbf{Initialize.}
\[
G_0=S,
\qquad
c=c_S.
\]

\medskip
For \(k=0,1,\ldots,K-1\), perform the following steps.

\begin{enumerate}

\item \textbf{Tensor expansion.}
Set
\[
\widetilde G_{k+1}=G_k\otimes S.
\]
The vertices of \(\widetilde G_{k+1}\) are pairs \((u,a)\), where
\(u\in V(G_k)\) and \(a\in V_S\). An edge joins \((u,a)\) and \((v,b)\)
exactly when
\[
\{u,v\}\in E(G_k)
\qquad\text{and}\qquad
\{a,b\}\in E_S.
\]
Project the coloring according to
\[
c\big((u,a)\big)=c(u).
\]
Thus every tensor edge remains cross-color, and the proper 3-coloring is
preserved.

\item \textbf{Triangle-support pruning.}
Score every edge of \(\widetilde G_{k+1}\) by its triangle support.
Retain the
\[
\operatorname{round}(p\,b_{k+1})
\]
highest-support edges and delete the remaining edges.

\item \textbf{Local chromatic repair.}
For each vertex whose degree is below \(d_{\min}\), add admissible cross-color
edges through the vertex's two-hop neighborhood whenever possible,
preferentially selecting edges that close local triangles. If the local
candidate pool is exhausted, use any available admissible endpoint. If the graph
is disconnected, add admissible bridging edges until connectivity is restored.

\item \textbf{Chromatic triangle closure.}
Visit vertices in a pseudorandom order and collect the open wedges
\(u-v-w\) centered at each visited vertex \(v\), sampling at most \(60\)
of its neighbors; admit the closing edge \(\{u,w\}\) as a candidate only
when
\[
c(u)\ne c(w).
\]
Stop collecting once there are more than five times as many candidates as
edges still needed. Rank the candidates by the number of collected wedges
that support them and add them until the total edge count reaches
\[
\operatorname{round}(a\,b_{k+1}).
\]
\item \textbf{Non-bipartiteness safeguard.}
Before pruning, record one odd cycle of \(\widetilde G_{k+1}\). If the
edited graph has become bipartite, restore the missing edges of that cycle;
they are tensor edges and therefore admissible. In the runs reported here this
was never needed.
\end{enumerate}

Set \(G_{k+1}\) equal to the resulting graph. After \(K\) tensor steps,
\(G_K\) is the parent graph.

\medskip
\textbf{Output.}
The locally repaired parent graph \(G_K\), together with its proper
3-coloring \(c\).

\end{tcolorbox}

\begin{tcolorbox}[
breakable,
colback=white,
colframe=black,
boxrule=0.5pt,
arc=0pt,
left=6pt,
right=6pt,
top=6pt,
bottom=6pt,
title={Algorithm 1B: Coarsening to the quotient}
]
\small

\textbf{Input.}
The parent graph \(G_K\) produced by Algorithm~1A; Louvain resolution
\(\gamma\); target quotient size \(N\); and a fixed random seed for
randomized tie-breaking.

\medskip
\textbf{Definitions.}
For two communities \(A\) and \(B\), let \(w(A,B)\) denote the number of
edges of \(G_K\) with one endpoint in \(A\) and the other in \(B\). Define
their intercommunity edge density by
\[
\rho(A,B)
=
\frac{w(A,B)}{|A|\,|B|}.
\]

\medskip
\begin{enumerate}

\item \textbf{High-resolution partition.}
Run Louvain community detection on \(G_K\) at resolution \(\gamma\), chosen
so that the number of detected communities exceeds the target \(N\).

\item \textbf{Density-based merging.}
While the number of communities is greater than \(N\), choose a smallest
community \(A\). Among the communities adjacent to \(A\), select a community
\(B\) maximizing
\[
\rho(A,B)
=
\frac{w(A,B)}{|A|\,|B|},
\]
and merge \(A\) into \(B\). Continue until exactly \(N\) supernodes remain.

\item \textbf{Quotient construction.}
Construct the quotient graph \(H_{\rm LR}\) with one vertex for each
surviving supernode. Two distinct supernodes are adjacent in \(H_{\rm LR}\)
if and only if at least one edge of \(G_K\) has endpoints in the two
corresponding communities.

\end{enumerate}

\medskip
\textbf{Output.}
The local quotient \(H_{\rm LR}\).

\end{tcolorbox}

For the concrete run reported here, the seed has \(|V_S|=7\) vertices and
\(|E_S|=9\) edges, listed with its coloring in
Appendix~\ref{app:construction-parameters}. We use tensor depth \(K=4\), minimum degree
\(d_{\min}=3\), pruning fraction \(p=0.10\), triangle-addition fraction
\(a=0.18\), Louvain resolution \(\gamma=40\), target quotient size
\(N=1{,}015\), and random seed \(42\).  Degree repair takes precedence
over the nominal budget: when pruning and repair already exceed
\(a\,b_k\), the closure step adds nothing.  This occurs at the first two
expansions.  At \(n=49\), minimum degree \(3\) alone requires at least
\(74\) edges against a target of \(43\); at \(n=343\), repair after
pruning already exceeds the target of \(807\).  The realized counts there
are \(102\) and \(837\) edges.  From \(n=2{,}401\) onward the budget
binds, and the final parent has exactly \(0.18\,b_4=277{,}319\) edges.
With these values, the local construction produces a quotient
\(H_{\rm LR}\) with
\[
n=1{,}015,\qquad |E|=67{,}094,\qquad \rho=0.1304,\qquad Q=0.524.
\]

This is the whole local construction.  It has no empirical adjacency matrix,
no anatomical coordinates, no tractography weights, and no fitted edge
probabilities.  It is a local growth-and-repair mechanism followed by
quotienting.  Only node count, edge count, modularity, and low-degree
nodes were used as targets.
\subsection{Basic outcome}

The local quotient has
\[
n=1{,}015,\qquad |E|=67{,}094,\qquad \rho=0.1304,
\]
with mean degree \(132.2\), degree standard deviation \(68\), minimum degree
\(3\), maximum degree \(502\), and modularity
\[
Q=0.524.
\]
The Kolmogorov--Smirnov distance from the Budapest degree distribution is
\[
KS=0.111.
\]
Thus the graph is slightly sparser than the Budapest Reference Connectome
snapshot used below, but it lies at the same node scale and in the same broad
density regime.

\begin{figure}[t]
\centering
\includegraphics[width=\textwidth]{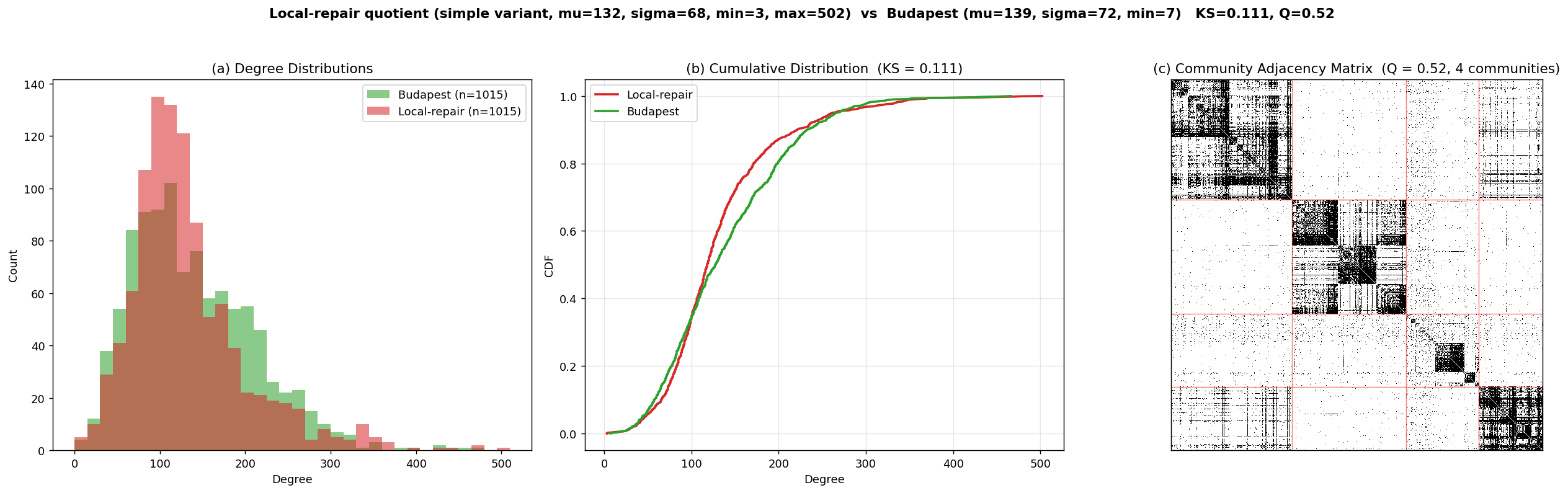}
\caption{Degree structure and modular organization of the local
quotient, compared with the Budapest Reference Connectome at
the \(n=1{,}015\) node scale.  (a) Degree distributions shown as overlaid
filled histograms, with counts on the vertical axis and bin width \(15\):
Budapest in green and the local quotient in red.  The local
quotient reproduces Budapest's broad right-skewed degree profile, with
standard deviation \(68\) versus approximately \(72\) for Budapest.
(b) Empirical cumulative degree distributions, with the local quotient
in red and Budapest in green, and Kolmogorov--Smirnov distance
\(KS=0.111\).  (c) Community-ordered adjacency matrix of the local
quotient, with nodes ordered by hierarchical Louvain communities and hubs
sorted toward block corners.  The graph shows four large modular blocks with
modularity \(Q=0.52\), close to Budapest's \(Q\approx0.56\).  Node ordering is
a display choice only and does not alter the graph.}
\label{fig:local-repair-degree-modularity}
\end{figure}

Figure~\ref{fig:local-repair-degree-modularity} visualizes the two coarse
properties most relevant to the local construction: degree profile and
modular organization.  The point is not that the local graph is identical
to Budapest.  Rather, the simple local mechanism already recovers the
broad right-skewed degree profile and a clear modular block structure at the
same node scale, before the integrated construction developed in the next section.

\paragraph{Edge-budget normalization.}
The local construction was initially parameterized from the side of graph
complexity.  For a graph on \(n\) vertices, the sparsest connected graph has
\(O(n)\) edges, while the complete graph has \(n(n-1)/2\) edges.  Many graph
complexity measures are small near both extremes: trees and paths are too
regular, while complete graphs are also highly regular.  The natural
intermediate scale is therefore the geometric mean between the sparse
connected scale and the complete-graph scale,
\[
E_{\rm gm}
=
n\sqrt{\frac{n-1}{2}},
\]
up to the immaterial distinction between using \(n\) and \(n-1\) for the
sparse connected reference.  This is the edge scale associated with maximal
graph complexity in several graph-complexity measures
\cite{KimWilhelm2008ComplexGraph}.

In the local construction we did not use the full complexity scale.  We used
a fraction of it,
\[
E_{\rm LR}=\alpha E_{\rm gm},
\qquad
\alpha=0.18 .
\]
At the final parent size \(n=16{,}807\), this gives
\[
E_{\rm LR}=277{,}319.
\]
Thus the local graph lies well below the maximal-complexity scale.  This was
empirically necessary: pushing the budget closer to \(E_{\rm gm}\) produced
graphs that were too dense, too modular, and lost the low-degree periphery
seen in the Budapest graph.

The global construction uses the same sparse regime but expresses it from
the opposite direction.  Instead of taking a fraction down from the
maximal-complexity scale, it starts from the connectivity scale, of order
\(n\log n\), and works upward by a controlled multiple.  At \(n=16{,}807\),
the local budget \(277{,}319\) is already only \(1.18\,n\log_2 n\), or about
\(3.39\) times the classical connectivity threshold \(n\ln n/2\).

\subsection{Quotient topology test}

After coarsening, the local quotient passes the same nine
scored topology criteria used below for the final quotient: clustering
enrichment, near-random path length, small-worldness,
modularity enrichment, near-random global efficiency, local efficiency
enrichment, connector hubs, rich-club organization, and multi-resolution
VI-stable modularity.

\begin{table}[htbp]
\centering
\small
\setlength{\tabcolsep}{4pt}
\renewcommand{\arraystretch}{1.05}
\caption{Nine-criterion topology test for the local quotient
(\(n=1{,}015\), \(|E|=67{,}094\)) against the Budapest Reference Connectome
(\(n=1{,}015\), \(|E|=70{,}654\)).
Values in parentheses are multiples of the mean over degree-preserving null
graphs; for example, \(2.46\times\) means \(2.46\) times the null mean. Both graphs
pass all nine scored criteria.}
\label{tab:local-repair-nine-criteria}
\begin{tabularx}{\textwidth}{p{0.26\textwidth} X X c}
\toprule
Criterion & Local quotient & Budapest & Verdict \\
\midrule
Clustering enriched &
\(C=0.515\)\,(\(2.46\times\)) &
\(C=0.670\)\,(\(3.04\times\)) &
\pass{} \\
Path length near random &
\(\mathrm{APL}=1.97\)\,(\(1.05\times\)) &
\(\mathrm{APL}=2.22\)\,(\(1.19\times\)) &
\pass{} \\
Small-worldness (degree-pres.) &
\(\sigma_{\mathrm{dp}}=2.30\) &
\(\sigma_{\mathrm{dp}}=2.22\) &
\pass{} \\
Modularity enriched &
\(Q=0.524\)\,(\(8.24\times\)) &
\(Q=0.557\)\,(\(9.08\times\)) &
\pass{} \\
Global efficiency near random &
\(E_{\mathrm{glob}}=0.548\)\,(\(0.97\times\)) &
\(E_{\mathrm{glob}}=0.510\)\,(\(0.90\times\)) &
\pass{} \\
Local efficiency enriched &
\(E_{\mathrm{loc}}=0.737\)\,(\(1.24\times\)) &
\(E_{\mathrm{loc}}=0.834\)\,(\(1.38\times\)) &
\pass{} \\
Connector hubs &
\(4\) hubs &
\(4\) hubs &
\pass{} \\
Rich-club organization &
\(294\) sig.\ thresholds; run \(177\) &
\(212\) sig.\ thresholds; run \(212\) &
\pass{} \\
Multi-resolution VI stability &
stable partition &
stable partition &
\pass{} \\
\midrule
\multicolumn{4}{l}{\textbf{Score: \(9/9\) for both graphs.}} \\
\bottomrule
\end{tabularx}
\end{table}
For the local quotient, the main criteria are tested against \(100\)
degree-preserving null graphs and the rich club against \(1{,}000\). For the
rich club we report two numbers, both from the same \(1{,}000\) nulls: how
many degree thresholds are significant (\(p<0.05\)), and the longest run of
consecutive significant thresholds. Only thresholds \(k\) at which at least
\(20\) nodes have degree above \(k\) are counted. The connector-hub entry is
the number of hubs identified by their participation coefficient, and a
value such as \(2.46\times\) is a multiple of the mean over the
degree-preserving nulls.

Budapest lies deeper within the enrichment criteria for clustering
(\(3.04\times\) versus \(2.46\times\)), modularity
(\(9.08\times\) versus \(8.24\times\)), and local efficiency
(\(1.38\times\) versus \(1.24\times\)). The local quotient is comparable or
stronger on small-worldness (\(\sigma_{\mathrm{dp}}=2.30\) versus \(2.22\))
and shows comparable rich-club organization, with \(294\) versus \(212\)
significant thresholds and longest runs of \(177\) and \(212\), respectively.

We therefore treat the quotient test as a Budapest-scale coarse-graph result,
not as proof that all nine properties are intrinsic to the fine computational
substrate. The parent-graph test in Appendix~\ref{app:global_parent_graph} is the
stronger substrate-level check. The difference between the parent score of
\(7/9\) and the quotient score of \(9/9\) shows that coarsening supplies the
missing integration in the local construction.

\subsection{Independent graph battery}

The more important test is the independent graph battery, summarized in
Tables~\ref{tab:local-repair-combinatorial}--\ref{tab:local-repair-cliques}.
These quantities
were not used to construct the local quotient.  They include algebraic and
combinatorial measures, graph-complexity measures, spectral quantities,
network-controllability statistics \cite{Gu2015Controllability}, and clique-census measures \cite{SizemoreCliques2018}.

\begin{table}[t]
\centering
\small
\setlength{\tabcolsep}{4pt}
\begin{tabular}{lcc}
\toprule
Measure & Local & Budapest \\
\midrule
\(\lambda_{\max}(A)/\langle d\rangle\) & \(1.380\) & \(1.285\) \\
\(\log\tau(G)/n\) & \(4.739\) & \(4.774\) \\
Fiedler value \(\lambda_2\) & \(2.964\) & \(2.654\) \\
Hoffman bound & \(3.96\) & \(7.91\) \\
Spectral (Fiedler) bisection cut/\(m\) & \(0.4472\) & \(0.0123\) \\ Random-walk eigengap & \(0.0789\) & \(0.0197\) \\
Maximum matching, normalized & \(0.999\) & \(0.999\) \\
Fractional matching & \(1.000\) & \(1.000\) \\
Fractional vertex cover & \(0.500\) & \(0.500\) \\
Vertex cover, 2-approx. normalized & \(0.974\) & \(0.982\) \\
Minimum spanning forest & \(1014\) & \(1014\) \\
\bottomrule
\end{tabular}
\caption{Algebraic and combinatorial comparison between the local
quotient and the Budapest Reference Connectome.  Several quantities are
close, including spanning-tree entropy and the Fiedler value.  Matching,
fractional matching, fractional vertex cover, and minimum spanning forest
size are nearly saturated at this density and are reported for completeness
rather than interpreted as strong independent evidence.}
\label{tab:local-repair-combinatorial}
\end{table}
The local quotient is close to Budapest on spanning-tree entropy and
Fiedler value.  The strongest mismatches are the spectral (Fiedler)
bisection cut and the random-walk eigengap; the eigengap also remains a mismatch for the final quotient (Table~\ref{tab:complexity-spectral}).
The local quotient does separate cleanly along its own communities (a
community-aligned bisection cuts only \(6.27\%\) of its edges), but its
Fiedler direction does not follow them (\(\mathrm{ARI}=0.059\) between
the two splits), whereas Budapest's Fiedler bisection coincides exactly
with its community split.  The low Hoffman bound,
\(1+\lambda_{\max}/|\lambda_{\min}|\), reflects a comparatively large
negative eigenvalue, a multipartite-like spectral signature that may be
inherited from the three-colored parent. It is only a lower bound on the
chromatic number: with a clique of size \(43\), the quotient itself is far
from three-colorable.

\begin{table}[t]
\centering
\small
\setlength{\tabcolsep}{4pt}
\begin{tabular}{lcc}
\toprule
Measure & Local & Budapest \\
\midrule
Off-diagonal complexity & \(5.446\) & \(5.406\) \\
Degree entropy & \(5.283\) & \(5.405\) \\
Degree-pair entropy & \(9.733\) & \(9.899\) \\
Kirchhoff index & \(11{,}228\) & \(10{,}849\) \\
\(\log_{10}\) Estrada index & \(79.25\) & \(77.72\) \\
Graph energy & \(5{,}887\) & \(5{,}778\) \\
Graph energy per node & \(5.80\) & \(5.69\) \\
\bottomrule
\end{tabular}
\caption{Graph-complexity and spectral comparison.  The local
quotient is close to Budapest on entropy-like measures, the Kirchhoff index,
and graph energy.  Because the local quotient is slightly sparser than
Budapest (\(67{,}094\) versus \(70{,}654\) edges), the Kirchhoff-index and
graph-energy comparisons should be read as qualitative rather than
density-matched claims.}
\label{tab:local-repair-complexity}
\end{table}

This table shows why the local variant is informative.  The local quotient is not uniformly worse than the final quotient. On several bulk spectral and resistance-based quantities, especially graph energy and
Kirchhoff index, the local quotient is close to Budapest.  The simple
rule therefore already captures much of the large-scale graph regime.

\begin{table}[t]
\centering
\small
\setlength{\tabcolsep}{4pt}
\begin{tabular}{lcc}
\toprule
Measure & Local & Budapest \\
\midrule
Average controllability, mean & \(1.094\) & \(1.100\) \\
Average controllability, CV & \(0.137\) & \(0.101\) \\
Modal controllability, mean & \(0.9961\) & \(0.9957\) \\
\(\mathrm{corr}(d,\mathrm{avgctrl})\) & \(+0.818\) & \(+0.875\) \\
\(\mathrm{corr}(d,\mathrm{modalctrl})\) & \(-1.000\) & \(-1.000\) \\
\bottomrule
\end{tabular}
\caption{Network-controllability comparison.  The local quotient
reproduces the qualitative controllability profile of Budapest: high-degree
nodes have high average controllability, while low-degree or peripheral
nodes have high modal controllability.}
\label{tab:local-repair-controllability}
\end{table}

The local quotient reproduces the main controllability pattern.  The
degree--average-controllability correlation lies in the same positive
regime as Budapest (\(+0.818\) versus \(+0.875\)); this correlation is not
algebraically forced and is therefore informative.  The
degree--modal-controllability correlation, by contrast, is \(-1.000\) in
both graphs because it is an identity rather than a finding: for an
unweighted graph under the normalization \(W=A/(1+\lambda_{\max})\),
modal controllability equals \(1-s^2d_i\) with \(s=1/(1+\lambda_{\max})\),
an exact affine function of degree, so it carries no comparative
information.

\begin{table}[H]
\centering
\small
\setlength{\tabcolsep}{4pt}
\renewcommand{\arraystretch}{1.12}
\begin{tabular}{lcc}
\toprule
Measure & Local & Budapest \\
\midrule

Clique number \(\omega\) &
\(43\) &
\(64\) \\

Total maximal cliques &
\(5.03\,\mathrm{M}\) &
\(15.4\,\mathrm{M}\) \\

Maximal-clique histogram peak &
\(k=30\) &
\(k=40\) \\

Participation Gini &
\(0.815\) &
\(0.812\) \\

Participation-degree correlation &
\(0.742\) &
\(0.779\) \\

High-participation nodes &
\(62\) \((6.1\%)\) &
\(64\) \((6.3\%)\) \\

\bottomrule
\end{tabular}
\caption{Higher-order clique comparison between the local quotient
and the Budapest Reference Connectome.  The local quotient reaches
\(\omega=43\) versus Budapest's \(64\), and its maximal-clique-size
distribution peaks at \(k=30\), close to Budapest's \(k=40\).  The
participation Gini (\(0.815\) versus \(0.812\)), the participation-degree
correlation (\(0.742\) versus \(0.779\)), and the fraction of
high-participation nodes (\(6.1\%\) versus \(6.3\%\)) are all close to
Budapest.}
\label{tab:local-repair-cliques}
\end{table}

The clique census identifies a limitation of the local variant.  The local
quotient reaches \(\omega=43\), below the Budapest value \(\omega=64\).
Its maximal-clique-size distribution nevertheless peaks at \(k=30\), compared
with Budapest's \(k=40\), while participation Gini, participation--degree
correlation, and the fraction of high-participation nodes are all close to
the reference values.

\subsection{Adding short paths without losing multiscale structure}

When we coarsen the local construction to the scale of the connectome and examine the resulting clustered graph (the quotient), it passes the main network and topological tests characteristic of the connectome: strong local clustering, modular community structure, connector hubs and a rich club, and short, small-world paths. In this sense, the local algorithm already produces a connectome-like graph.

Can these network and topological properties also be intrinsic to the underlying large graph, present in the fine parent itself and not only in its clustered representation? We examine whether the parent has the same structural characteristics: greater local clustering than its randomized counterparts, modular community structure, enriched local efficiency, connector hubs, a rich club of interconnected high-degree nodes, and a stable nested organization. The parent passes almost all of these tests, failing only two: characteristic path length and global efficiency. These failures are closely related. Both concern global integration and reflect the same deficiency: the parent lacks short paths. Its typical inter-node distances are longer than those expected of a comparable small-world graph, and its global efficiency, which measures the average closeness of nodes through these distances, is lower than that of its randomized counterparts.

One might expect that adding a few long-range edges would be sufficient to shorten the paths. The difficulty is that these edges may also disturb the multiscale organization that we wish to preserve. Hierarchical organization and short-path integration can impose competing structural requirements. Shortcuts that improve global integration may cut across the organization established at finer scales. We therefore need to shorten paths while preserving the multiscale structure already present in the parent.

The quotient provides an important clue. We know that short paths appear when the parent is clustered to the scale of the connectome. This suggests that the integration missing from the parent is supplied by grouping nodes into communities, and that the missing connections are between these communities. To obtain short paths in the parent itself, we should therefore introduce connections between its communities.

We do this in a way that respects the existing scale structure. We measure this structure using box covering, which groups nodes into boxes at different scales, from fine to coarse. A new edge is permitted only if it satisfies two conditions. First, its endpoints must belong to different communities, so that it connects distinct parts of the graph. Second, the endpoints must belong to the same box at every tested pre-bridge box-covering scale. The edge must also connect different color classes to preserve the chromatic constraint. In this way, we can introduce connections between communities while keeping the new edges local with respect to all the tested multiscale coverings. Such edges can shorten paths without crossing the boundaries defined by these coverings.

Adding edges, however, can make the graph too dense and weaken its sparse organization. We therefore introduce another pruning step, removing edges that contribute least to the local structure. We measure the contribution of an edge by the number of triangles in which it participates. An edge belonging to many triangles helps maintain a locally connected neighborhood. An edge belonging to no triangle provides less support for local clustering and may be removed with comparatively little structural cost.

We remove edges in increasing order of their triangle support. Edges that are bridges at the beginning of each removal round are excluded. We also skip any removal that would reduce the degree of an endpoint below the prescribed minimum. Since an edge may become a bridge during a removal round, these conditions alone do not guarantee connectivity. We therefore verify connectivity after sparsification. The parent graph reported here is connected.

The resulting parent graph is generated from the seed using the integrated pipeline of Section~\ref{subsec:full-parent-operators}, not by modifying the local parent. It retains the multiscale organization of the local construction while also exhibiting short paths in its own fine-scale structure. These short paths, which previously appeared only after coarsening, are now intrinsic to the parent graph.

\FloatBarrier

\section{Global Parent Graph and Budapest-Scale Quotient}
\label{sec:full-parent-and-budapest}
\subsection{Global Parent-Graph Operators}
\label{subsec:full-parent-operators}

We distinguish two problems of different difficulty. The first is to produce a coarse graph in the connectome regime, with the correct order, density, and modular organization. This problem is essentially solved by a compact algorithm consisting of expansion, pruning, and clustering. The second is to produce the corresponding structural properties directly in an \emph{underlying} fine graph, derived from tensor products with \(\chi=3\) maintained throughout, instead of obtaining these properties only after coarsening. The two requirements interact. An operation that improves the fine graph may substantially change the quotient, while an operation intended to improve the quotient must remain compatible with the chromatic and multiscale constraints of the parent. We address the parent-scale problem by introducing integration directly into the fine substrate. However, as we show below, strict coarsening of this integrated parent does not recover the modular organization of the reference connectome. This obstruction motivates the inverse construction developed later in this section.

The integrated construction addresses the two parent-scale failures of the local construction. These ``failures'' are defined relative to the connectome-inspired test and should not be interpreted as necessary properties of the hypothesized computational substrate. We do not know whether the sparse concurrent computational graph that we hypothesize operates beneath the connectome must itself exhibit near-random path length and global efficiency. The question is whether these properties can also be realized directly in a sparse graph of bounded chromatic causal depth while preserving the multiscale constraints.

The local parent already has intrinsic clustering, modularity, local efficiency, connector hubs, rich-club organization, and multi-resolution modular stability. However, it lacks intrinsic fine-scale integration. Its paths are too long and its global efficiency is too low relative to tripartite degree-preserving nulls. We therefore modify the parent graph so that integration becomes a property of the computational substrate itself, instead of appearing only after coarsening.

The design involves three competing requirements. Tensor growth and pruning produce a self-similar chromatic skeleton. Cross-community bridge edges introduce short paths and increase global efficiency. Multiscale-targeted sparsification then reduces the graph toward a box-covering target, so that the additional connections do not destroy its hierarchical organization. Our objective at the parent scale is to balance short paths, weak self-similar structure, and multi-resolution modular stability while preserving the chromatic constraint.

We now describe the operators used in the construction. The operator \(\mathcal T_S\) performs tensor expansion:

\[
\mathcal T_S(G)=G\otimes S.
\]

The operator \(\mathcal P_{\eta,B}\) performs community-aware pruning. It keeps the graph within an edge budget \(B\), reserving a fraction \(\eta\) of this budget for inter-community edges. The operator \(\mathcal R_\chi\) performs chromatically admissible repair. It adds cross-color edges, when necessary, to restore the minimum-degree floor and reconnect disconnected components.

The repeated tensor-growth step is therefore

\[
G_{k+1}
=
\mathcal R_\chi
\mathcal P_{\eta,B}
\mathcal T_S(G_k).
\]

At each scale we expand the graph by tensor product, prune it, and then repair it. We repeat this procedure until the desired tensor scale is reached.

After the final tensor-pruned and repaired graph is obtained, we apply three additional operations. First, \(\mathcal A_{\triangle}\) adds admissible triangle-closing edges. Given an open wedge \(u-v-w\), the edge \((u,w)\) may be added only if its endpoints belong to different color classes. Second, \(\mathcal B_{\mathrm{box}}\) adds admissible same-box, cross-community bridge edges, using a multiscale box covering to select candidates. Third, \(\mathcal S_{d_B}\) performs multiscale-targeted sparsification, removing edges with low triangle scores until the target box-covering condition is reached.

We can therefore write the final construction as

\[
G_{\mathrm{final}}
=
\mathcal S_{d_B}
\mathcal B_{\mathrm{box}}
\mathcal A_{\triangle}
(G_K),
\]

where \(G_K\) is obtained recursively from

\[
G_{k+1}
=
\mathcal R_\chi
\mathcal P_{\eta,B}
\mathcal T_S(G_k),
\qquad
G_0=S.
\]

The complete construction consists of the following operations:

\begin{enumerate}
\item tensor growth;
\item community-aware pruning;
\item chromatic repair;
\item triangle closure;
\item same-box bridges;
\item multiscale-targeted sparsification.
\end{enumerate}

Every added edge is chromatically admissible, and every deleted edge is chromatically safe. The graph can therefore increase in complexity while preserving the \(3\)-coloring constraint.

\paragraph{Visual structure of the parent graph.}

\begin{figure}[t]
\centering
\includegraphics[width=\textwidth]{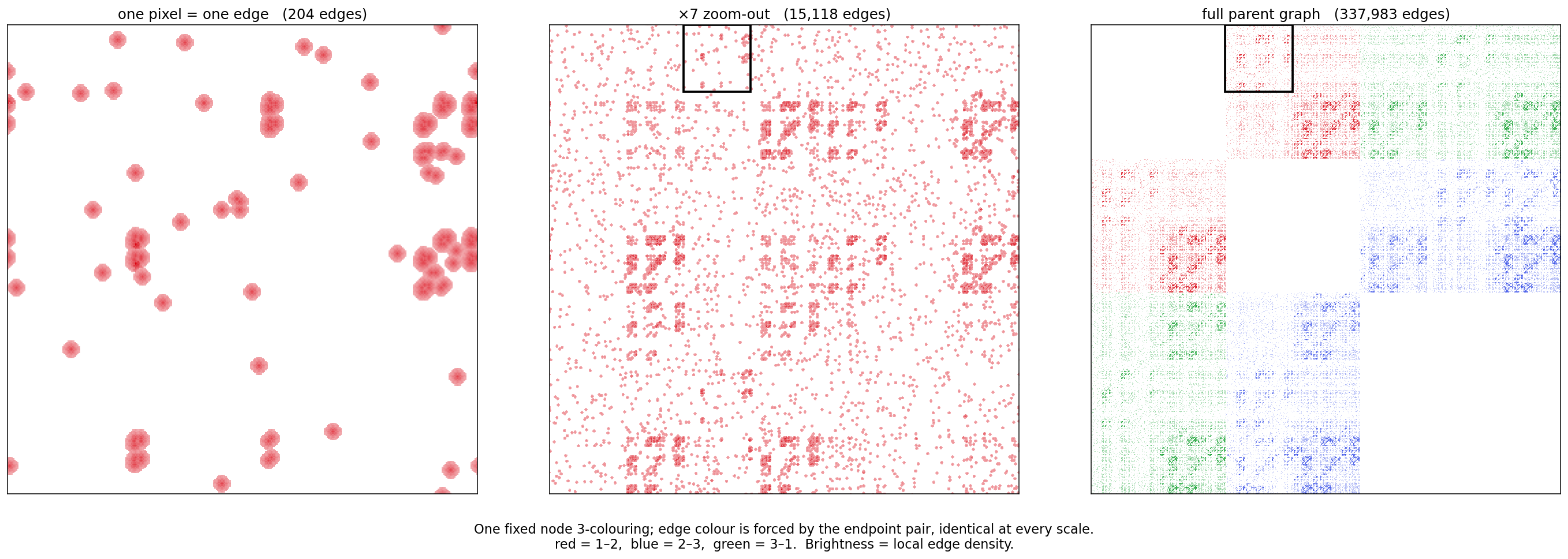}
\caption{Tensor-order zoom-out of the \(16{,}807\)-node parent graph before
quotienting. Left: native local window, where one pixel corresponds to one
edge. Middle: sevenfold zoom-out with resolution preserved, showing the
same edge texture over a larger tensor-ordered window. Right: full parent
graph with \(337{,}983\) edges. Edge colors indicate the three chromatic edge
classes: \(1\!-\!2\) red, \(2\!-\!3\) green, and \(1\!-\!3\) blue; brightness
indicates local edge density.}
\label{fig:parent-zoomout}
\end{figure}

Figure~\ref{fig:parent-zoomout} shows the resulting \(16{,}807\)-node parent graph before quotienting. We do not use this figure as a quantitative test. It illustrates the qualitative structure produced by tensor growth and the chromatically constrained modification operations. At its native scale, the graph is sparse and locally clustered. When we zoom out in tensor order, a similar edge texture remains visible over a larger window. The full adjacency view shows broad block organization across the parent graph. Thus, even before comparison with the Budapest connectome, the construction produces a sparse local edge structure together with large-scale organization.

\paragraph{Scale simultaneity.}

The construction produces a graph with organization at several scales simultaneously. This does not imply strict self-similarity. The box-covering exponent reported in Appendix~\ref{app:global_parent_graph} measures how the number of covering boxes changes with box size. It is a metric property of the graph and does not imply that the graph has identical structure at different resolutions.

We must interpret the box-covering measurement carefully. Over the four scales permitted by the graph diameter, \(l_B\in\{2,4,6,8\}\), the number of boxes decreases from \(2615\) to \(8\). A power law fits this decrease with \(R^2=0.94\) and slope \(d_B=4.03\). However, an exponential fits the same four points better, with \(R^2=0.997\), and the confidence interval for \(d_B\) is \([0.88,\,7.18]\). Exponential decay of the box count is characteristic of a non-fractal network, corresponding to an infinite fractal dimension \cite{Song2005}. We therefore conclude that the parent graph is not fractal. We report \(d_B\) as a descriptive quantity, not as evidence of self-similarity.

The absence of fractal structure does not imply an absence of organization across scales. The scale structure we find is of a different kind, consisting of modular nesting and organization at multiple resolutions. A Louvain sweep from \(\gamma=0.5\) to \(\gamma=3.0\) resolves the parent into \(4\) communities at the coarsest scale and \(49\) at the finest. Modularity decreases smoothly from \(Q=0.61\) to \(Q=0.28\). The partitions largely nest rather than merely subdivide, with a mean nesting score of \(0.756\). We also find a scale that is reproducible across seeds. At \(\gamma=1.2\), the variation of information reaches a local minimum of \(1.57\) nats, with \(Q=0.461\). This is a relative minimum over the sweep, not a small absolute value. For partitions of \(16{,}807\) vertices, the attainable range extends to \(\ln n\approx 9.73\) nats. The parent therefore exhibits organization at several scales simultaneously. However, as we show below, strict coarsening does not generally preserve the modular regime of the reference connectome.

Figure~\ref{fig:parent-zoomout} illustrates this distinction. At the finest scale, the graph appears as a sparse collection of edges with local clustering. As we zoom out, broader blocks and regions of different edge density become visible. These are not simply enlarged copies of the local structure. Graph entanglement is essential to the formation of this organization (Section~\ref{subsec:symmetry-metric}). By \emph{graph entanglement} we mean, as in the Introduction, the non-factorizability of the edited tensor-derived graph. As long as the graph remains a tensor power, it retains the metric degeneracy inherited from the tensor construction. Pruning, repair, closure, and integration can break the product form, remove this degeneracy, and produce richer higher-order organization.

The entanglement approach gives us a more expressive generating graph grammar. Different operators act at different characteristic scales. Triangle closure acts locally. Same-box bridges use agreement of box labels across several covering scales. Multiscale-targeted sparsification is controlled by a global box-covering condition. Consequently, zooming out does not merely reveal a rescaled version of the fine graph with some details missing. It reveals organization introduced at coarser scales by the construction itself. The scales are related, but they are not identical. Each is sparse, modular, and locally clustered, while the particular modules, bridges, and regions of different density at one scale are not simply magnified versions of those at another.

We call this property \emph{scale simultaneity}. Computation in our model takes place on the fine graph. The fine vertices are computational units, the fine edges carry messages, and recurrent updates operate on this microscopic substrate. The coarse levels are not separate computational systems placed above the fine graph. They are patterns that become visible when we view or pool the same graph at larger scales. A group of fine vertices may form a local module. Groups of these modules may form larger communities, and still larger structures may appear through further coarse-graining. These higher-level patterns are emergent descriptions of the fine substrate, not independent substrates that replace it.

There is a loose structural analogy with economic organization. Households, firms, cities, regions, and states are not repeated copies of one another. Different patterns emerge at each scale, although all are ultimately constructed from distributed local interactions. The analogy has an important limitation. Economic cycles and social coordination take place over many heterogeneous timescales, while a concurrent computational system requires a much tighter temporal organization.

In our construction, this temporal organization follows from the invariant chromatic bound. As the graph increases in size and complexity, its chromatic causal depth remains fixed at three phases per recurrent sweep. The emergent patterns become dynamically connected through sequential causal computations and resonant feedback. Consider a painting. Its details and larger patterns exist simultaneously across space, although everything ultimately exists in the details. Similarly, at the large-scale network level, structural complexity can increase without increasing the chromatic causal depth of an individual recurrent sweep. We propose that near-present perception is not merely an evolutionary adaptation, but a fundamental relation between complexity and time.

Thus scale simultaneity has both a spatial and a temporal meaning.  Spatially,
new coarse patterns can emerge under pooling or zoom-out while remaining
embedded in the fine graph.  Temporally, those patterns remain tied to a
bounded chromatic causal depth per recurrent sweep.  This is what connects the construction to the
{\it binding problem}: many distributed processes may contribute to a unified
state, but their interaction must occur within a short effective temporal
window. The chromatic constraint provides an abstract version of such a
window, keeping large-scale concurrent organization compatible with bounded
chromatic causal depth.

We are proposing that much of what appears as higher-scale computation in
the brain underlying subjective experience may not result from the sequential funneling of lower-scale outputs into deep higher-level processors.  Rather, it may consist of effective collective dynamics obtained by variations of the same nested cognitive \emph{motifs}  at progressively coarser resolutions.  Because the relevant multiscale organization is embedded in the network's modular ``hardware'', these coarse descriptions are not arbitrary: they identify stable and functionally meaningful collective variables that coexist with, rather than succeed, the underlying fine-scale computation.

Within a color class, in the parent graph, no edges exist, and so no message
is passed.  Units of the same color may nevertheless fire together, and there
are at least three reasons they might.  They may be synchronized by a shared
oscillation, as brain waves are commonly held to provide.  They may receive
input arriving at the same time.  Or they may coactivate: firing together in
a way that cannot be ordered in the sense we mean here, which is one system
{\it broadcasting} and another {\it listening}.  Any of these may well be
causal at the level of physics.  But causal at the cognitive level means
something narrower (a transmitter and a receiver), and between two units
that merely fire together there is neither, and so no causal order to
recover.  We use the term broadcasting because in our model a color class,
when active, transmits messages to many units of the other classes at once.

We implement this distinction computationally through sparse cross-phase
routing and attention-based same-phase coactivation, developed in
Sections~\ref{subsec:feeder-supernode-comparison}--\ref{subsec:coactivation-attention}.

The corresponding parent-graph recipe is controlled by eleven grouped
parameters, listed below; two further fixed conventions, the Louvain
resolution used during pruning and the box-covering scale set used for the
sparsification estimate, are recorded with the reported-run values in
Appendix~\ref{app:construction-parameters}.

{\centering\small
\begin{tabularx}{\textwidth}{lX}
\toprule
Parameter & Role \\
\midrule
\(K\)
& Number of tensor-product expansion steps. \\
\(B(n)\)
& Edge budget used during scale-wise pruning. \\
\(\eta\)
& Fraction of the edge budget reserved for between-community edges. \\
\(d_{\min}^{\mathrm{repair}}\)
& Minimum degree enforced during chromatic/connectivity repair. \\
\(\alpha_{\triangle}\)
& Amount of triangle-closing edge addition. \\
\(\mathcal L_B\)
& Set of box-covering scales used to identify bridge candidates. \\
\(\alpha_{\mathrm{bridge}}\)
& Maximum budget for same-box bridge edges. \\
\(d_B^{\ast}\)
& Target descriptive box-covering slope used during multiscale-targeted
sparsification. \\
\(\Delta m\)
& Batch size for edge removals during sparsification. \\
\(m_{\min}\)
& Minimum edge count allowed during sparsification. \\
\(d_{\min}^{\mathrm{sparse}}\)
& Minimum degree preserved during sparsification. \\
\bottomrule
\end{tabularx}\par}

These parameters define the integrated \(16{,}807\)-node parent graph.
The subsequent coarsening analysis and inverse relocation construction have
separate parameter accounting.

\paragraph{Brittleness after tensor expansion.}
Exact tensor decomposability is brittle after the tensor expansion has
already been formed.  In the tensor, or direct, product, vertices carry
product coordinates and adjacency is defined coordinatewise.  A change to
one factor therefore induces a coordinated family of changes throughout
the product, rather than an isolated local edit.  Consequently, a generic
edge addition or deletion made directly in the expanded graph will not be
representable as an edit of the factors and will typically break the
inherited tensor factorization.  This is the sense in which the product
structure is brittle: factorization survives only under highly
coordinated, product-aligned modifications.

This does not mean that every pruning destroys every possible
factorization.  Structured blockwise pruning may preserve the original
product structure, or in special cases induce a smaller or different
factorization.  The point is narrower: generic chromatically admissible
repairs produce a tensor-derived graph that retains a product ancestry but
is no longer itself an exact tensor product in the inherited coordinates.
This is consistent with robust tensor-graph factorization, where one
starts from a tensor graph with a small fraction of its edges deleted and
recovers a nearby exact product rather than assuming that the damaged graph
remains exactly decomposable \cite{BrakensiekDaviesTensorGraphs}.  For finite connected non-bipartite graphs, prime factorization with respect to the direct product
is unique up to isomorphism and permutation of factors when loops are
allowed, and can be computed in polynomial time
\cite{ImrichFactorization,HammackImrichKlavzar}; among loopless graphs it
need not be unique.  Thus the construction separates bounded chromatic execution depth from exact tensor decomposability.

\subsection{Community Quotients and the Inverse Problem}

We begin with the ordinary community quotient of the integrated parent graph.
Let \(G=(V,E)\) be the parent graph, and let
\[
\pi_{\rm comm}:V\rightarrow\{1,\ldots,N\}
\]
be a partition obtained by community clustering. For two distinct communities
\(a\) and \(b\), define
\[
W_{ab}
=
\left|
\left\{
\{u,v\}\in E:
\pi_{\rm comm}(u)=a,\;
\pi_{\rm comm}(v)=b
\right\}
\right|.
\]
The community quotient contains the edge \(\{a,b\}\) if and only if
\(W_{ab}>0\), with \(W_{ab}\) equal to the number of parent edges supporting
that coarse relation. Parent edges whose endpoints lie in the same community
remain present in the parent but disappear under quotienting because their
endpoints collapse to the same supernode.

This is a strict community quotient: once the parent graph and its community
partition are fixed, every cross-community relation must appear in the
quotient. The quotient cannot be improved by selectively suppressing
undesirable cross-community edges while leaving the parent unchanged.

Applied to the integrated parent, this strict requirement exposes a limitation
of the forward construction. The integrated parent has the desired
parent-scale properties, including short-path integration, but its community
quotient is substantially less modular than the Budapest Reference Connectome.
More generally, across the partitioning schemes examined in Appendix~\ref{app:inverse-relocation} (Table~\ref{tab:strict-coarsening-sweep}), no tested
strict coarsening of this fixed parent reaches the Budapest modular regime.
Community detection concentrates densely connected vertices within
communities, but those internal edges disappear when the communities are
collapsed. The quotient is therefore determined by the remaining
between-community relations, which are precisely the relations community
detection tends to minimize.

We therefore abandon the stronger requirement that the \emph{unchanged} fine
graph must itself produce the connectome-like coarse graph under strict
collapse.  Importantly, however, we do not abandon the community organization
of the original integrated parent.  The \(1{,}015\)-supernode partition is
obtained by clustering that fine causal-dependency graph and is then held
fixed throughout the reported inverse construction.

This distinction is central.  In the interpretation developed here, the
parent graph is a causal-dependency substrate: an edge indicates that its
endpoints require a causal ordering, whose direction is resolved by the
recurrent phase schedule.  Community detection is therefore applied directly
to the topology of the causal substrate.  The resulting supernodes define
mesoscale populations derived from that causal organization rather than
arbitrary bins introduced only for coarse comparison.

Let
\[
\pi:V\longrightarrow\{1,\ldots,1015\}
\]
denote this fixed community-derived map.  The reported inverse problem is then
not to solve for a new partition, but to reorganize the fine edge set while
preserving \(\pi\):
\[
(G_{\rm int},\pi)
\;\longrightarrow\;
(\widetilde G,\pi)
\;\xrightarrow{\text{strict collapse}}\;
H_{\rm final}.
\]
Thus every fine vertex remains assigned to the same supernode throughout the
relocation procedure.  The mesoscale causal clustering is preserved; what
changes is the pattern of fine causal-dependency edges within and between
those fixed populations.

The key operation is therefore causal-edge relocation under a fixed
partition.  If an edge currently joins two distinct supernodes, moving its
support into a supernode interior replaces an inter-supernode dependency by
an intra-supernode dependency.  Conversely, adding an admissible fine edge between
two supernodes increases the fine causal support for that mesoscale relation.
Every added edge remains chromatically admissible, so the proper three-coloring
and bounded chromatic causal depth are preserved.

The quotient remains strict throughout.  Every parent edge whose endpoints
lie in different supernodes contributes to the corresponding coarse
relation, while an edge disappears under coarse-graining only when both
endpoints lie in the same supernode.  The inverse construction therefore does
not edit coarse relations independently: it reorganizes the fine causal
substrate while preserving its original causal clustering, and the coarse
graph is then recomputed from that substrate.

This inverse construction still achieves the central objective of the paper:
a sparse computational graph of bounded chromatic causal depth can underlie a
connectome-like coarse organization while retaining almost all of the
parent-scale properties in our test. The relocated parent fails only the two
near-random path criteria (Table~\ref{tab:parent-relocation-comparison}). These failures are relative to the
degree-preserving null thresholds used in the test and should not be read as
showing that computation in the parent proceeds along intrinsically long
paths. Its paths remain short in absolute terms, although no longer close
enough to those of its randomized counterparts to satisfy those two criteria.

The inverse construction is carried out in three stages (Appendix~\ref{app:inverse-relocation}). In the first stage,
cross-supernode relations are reorganized so that the distribution of bundle
weights approaches that of Budapest. For a mapped quotient edge
\(\{a,b\}\), its bundle weight is
\[
W_{ab}
=
\left|
\left\{
\{u,v\}\in E:
\pi(u)=a,\;
\pi(v)=b
\right\}
\right|,
\]
the number of parent edges collapsed onto that quotient edge. Cross-supernode edges that would otherwise create unwanted coarse relations are removed, and the parent's edge budget is then restored by adding edges inside supernodes, where they disappear under coarse-graining (Appendix~\ref{app:bundle-redistribution}). In the second stage, a mesoscale support refinement removes the \(6{,}500\) lowest-weight coarse relations joining different groups inside the four largest blocks, together with the \(6{,}503\) parent edges that support them, and adds \(6{,}500\) new relations within those groups, each backed by one new parent edge; the bundle-weight distribution is therefore essentially unchanged, and the parent ends three edges short (Appendix~\ref{app:mesoscale-support-refinement}).

In the third stage, starting from this refined construction, a small
number of additional chromatically admissible relocations are used to improve the
quotient degree profile. The aim is to reduce the discrepancy between the
mapped quotient and the Budapest degree distribution while preserving the
bundle structure established in the first stage. The parent graph remains the actual substrate throughout: every change to the quotient, including the second-stage swap, is made by editing parent edges, and the quotient is always recounted from the parent.

After relocation, the quotient is recomputed directly from the parent:
\[
W_{ab}
=
\sum_{\{u,v\}\in E}
\mathbf 1\!\left[
\{\pi(u),\pi(v)\}=\{a,b\},
\ a\neq b
\right].
\]
Thus every cross-supernode parent edge contributes to exactly one quotient
relation, and every quotient weight is exactly the number of parent edges
supporting that relation.

In the final construction, the \(337{,}980\) parent edges divide into
\(147{,}217\) cross-supernode edges and \(190{,}763\) intra-supernode edges.
The saved weighted quotient is reproduced exactly by the strict collapse of
this parent under the fixed map \(\pi\). The resulting quotient contains
\(1{,}015\) nodes and \(71{,}098\) edges.

We therefore use two distinct coarse-graining constructions in what follows.
The first is the forward \emph{community quotient}, obtained by clustering a
fixed parent and strictly collapsing the resulting communities. The second is
the \emph{inverse mapped quotient}, in which the map is held fixed, the parent
is relocated under it, and the quotient is then obtained by the same strict
collapse rule. The first reveals the obstruction; the second provides the
inverse solution.

In this sense, the construction also provides a model of the connectome's
weighted axon-bundle structure. A quotient edge is not a primitive connection:
its weight \(W_{ab}\) is the multiplicity of fine computational edges joining
the corresponding supernodes. The parent graph may therefore be viewed as a
fine-scale multigraph-like substrate lying beneath the coarse connectome, with
many microscopic computational connections collapsing onto a single weighted
mesoscale edge. The first stage matches the distribution of these bundle weights to the
reference connectome's fiber counts, so the inverse construction reproduces
its coarse axon-bundle organization at the level of that distribution by
design. This is a structural correspondence rather than an anatomical
one: no identification is claimed between individual parent vertices or edges
and specific neurons, axons, or brain regions. 

Finally, community clustering remained essential to the inverse solution.
The forward community quotient provides the mesoscale organization from which
the inverse problem is formulated: it identifies the community structure that
must be respected and reveals which relations survive at the coarse scale.
That community organization therefore supplies the scaffold for constructing
the coarse-graining map and for determining where edge relocation can occur.
The inverse construction is thus built from, rather than independent of, the
community quotient.

\subsection{Numerical Comparison with the Budapest Connectome}
\label{subsec:numerical-budapest}

We compare the final \(1{,}015\)-node inverse mapped quotient with the
Budapest Reference Connectome
\cite{Szalkai2015Budapest,Szalkai2017Budapest,VargaGrolmusz2021Braingraph}
at the same node scale. As the empirical reference, we use the
\(1{,}015\)-node all\_20k Budapest snapshot with
\[
n=1{,}015,
\qquad
m=70{,}654.
\]
The final quotient has
\[
n=1{,}015,
\qquad
m=71{,}098,
\]
giving a density of \(0.1382\), against \(0.1373\) for Budapest.
Thus the graphs are identical in node count and closely matched in edge
count and density.

This comparison is distributional rather than anatomical. No quotient node
is identified with a particular Budapest region, and no individual edge is
fitted to a particular tractographic connection. The comparison asks instead
whether the strict quotient derived from the relocated computational substrate
occupies the same broad graph-theoretic regime as the empirical connectome.

\paragraph{Degree profile.}
The degree distributions are closely aligned, as the final stage of the inverse construction was designed to achieve. The final quotient has mean
degree \(140.1\), compared with \(139.2\) for Budapest, and median degree
\(130\), compared with \(127\). The minimum and maximum degrees are
\(13\) and \(420\) for the quotient and \(7\) and \(466\) for Budapest.
The standard deviations are approximately \(69.5\) and \(71.5\),
respectively.

More importantly, the full empirical degree distributions are close:
the Kolmogorov--Smirnov distance is
\[
KS=0.0187.
\]
Thus the match is not confined to the mean degree. The histogram and
empirical cumulative distribution nearly overlap across the bulk of the
degree range, with the principal residual difference occurring in the
extreme low- and high-degree tails.

\begin{figure}[t]
\centering
\includegraphics[width=\textwidth]{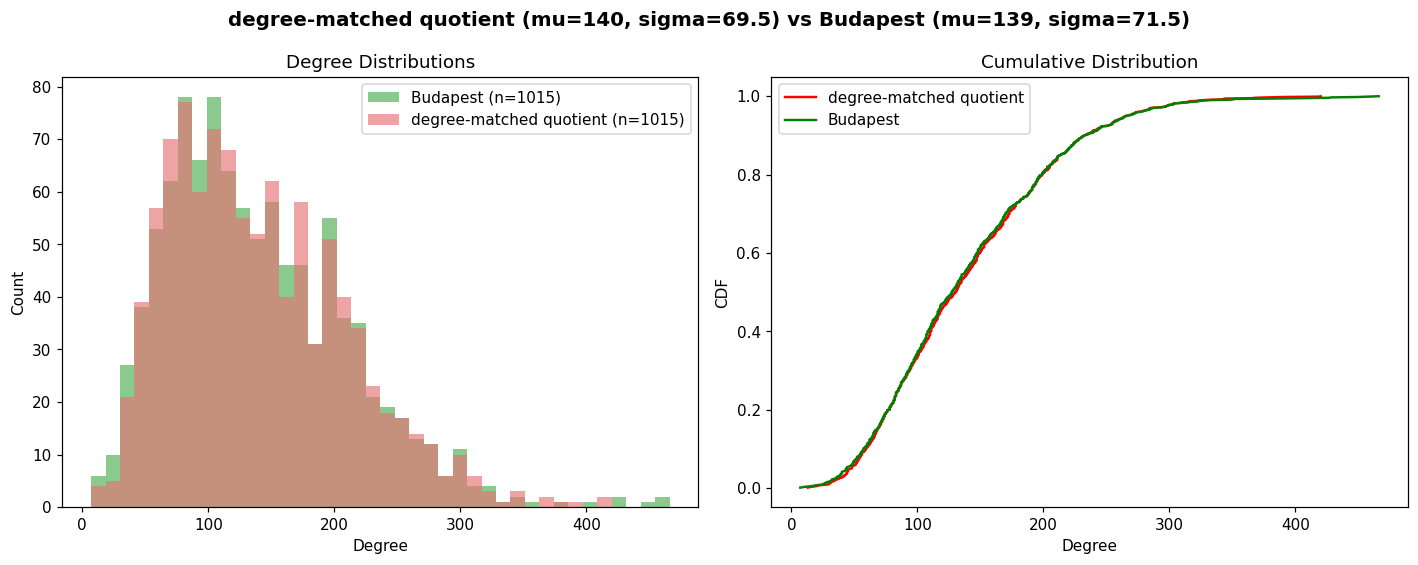}
\caption{
Degree-profile comparison between the final inverse quotient
and the Budapest Reference Connectome. Left: overlaid degree histograms.
Right: empirical cumulative degree distributions. Both graphs contain
\(1{,}015\) nodes; the quotient has mean degree \(140.1\) and standard
deviation approximately \(69.5\), compared with \(139.2\) and \(71.5\)
for Budapest. The empirical degree distributions are closely aligned
(\(KS=0.0187\)).
}
\label{fig:degmatch-degree-budapest}
\end{figure}

\paragraph{Community-ordered adjacency structure.}
The community-ordered adjacency matrices show substantial modular
organization in both graphs, but they also make an important residual
difference visible. The quotient contains strong block-diagonal structure
together with distributed off-diagonal connectivity, whereas the Budapest
matrix separates more sharply into large blocks.

The four principal communities of the quotient have sizes
\(389\), \(265\), \(239\), and \(122\). Budapest's corresponding large
communities have sizes \(307\), \(306\), \(201\), and \(201\), producing
two nearly balanced pairs. This paired organization is consistent with the
bilateral structure of the anatomical connectome and has no counterpart in
the present construction.

\begin{figure}[t]
\centering
\includegraphics[width=\textwidth]{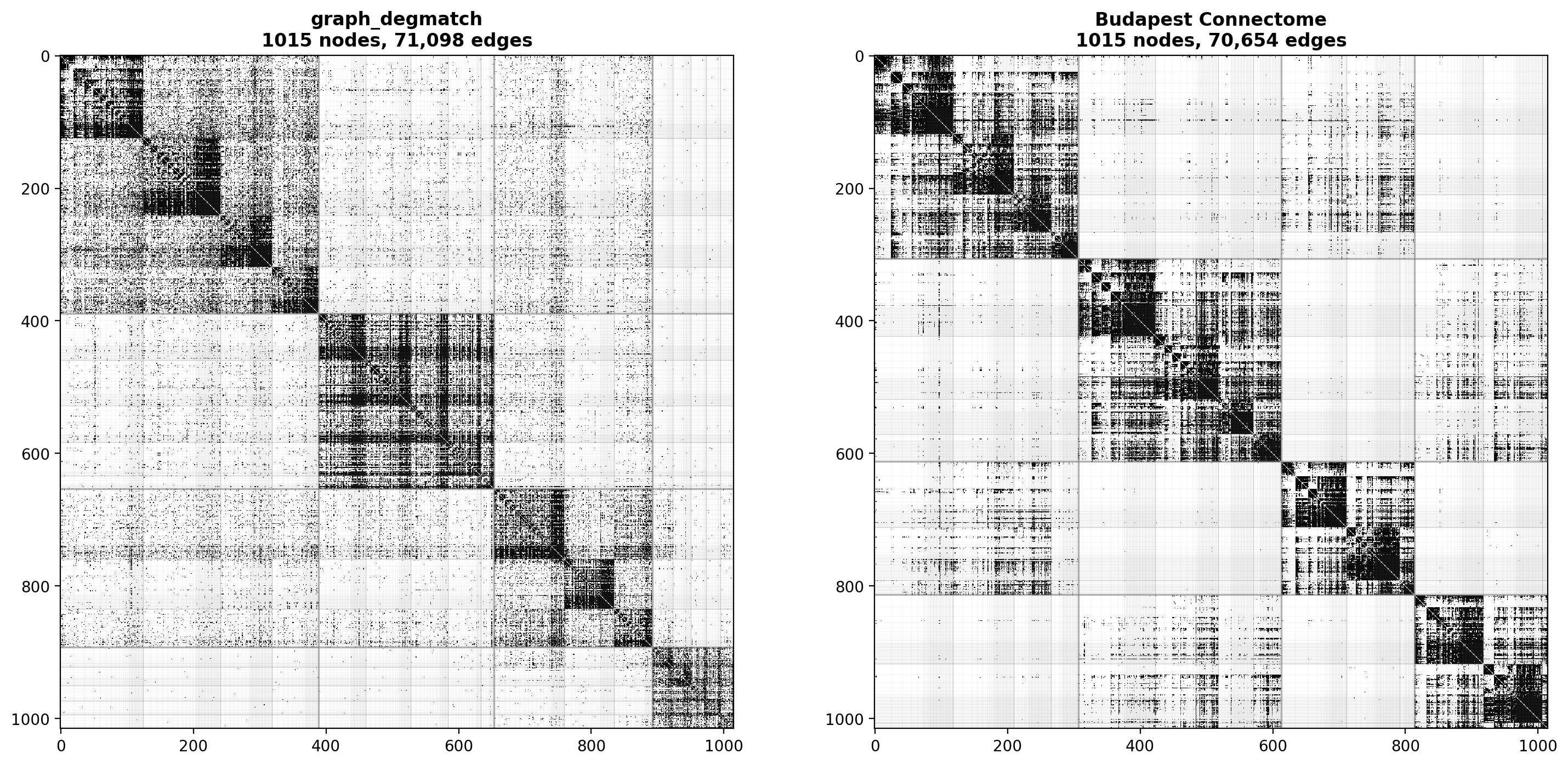}
\caption{
Community-ordered adjacency matrices for the final inverse quotient
(left) and the Budapest Reference Connectome (right), each with
\(1{,}015\) nodes. Black pixels denote edges. Both graphs exhibit pronounced
block-diagonal organization, but Budapest shows a sharper paired large-scale
structure. The construction contains no anatomical left--right axis,
hemispheric labels, spatial embedding, or corpus-callosum-like bottleneck,
and no nodewise correspondence between the two orderings is implied.
}
\label{fig:degmatch-adjacency-budapest}
\end{figure}

The static differences are nevertheless modest on several basic quantities.
The quotient has clustering coefficient \(0.4724\), transitivity \(0.4234\),
average path length \(1.9811\), diameter \(4\), and \(90\%\) effective
diameter \(2.162\). The corresponding Budapest values are \(0.6696\),
\(0.5578\), \(2.2190\), \(5\), and \(2.731\). Modularity remains lower in
the quotient,
\[
Q_{\mathrm{quotient}}=0.4908,
\qquad
Q_{\mathrm{Budapest}}=0.5574,
\]
but is now in the same broad regime while being obtained from a complete
strict collapse of the underlying relocated substrate.

\paragraph{Fiber-bundle weights.}
The weighted quotient also provides a direct comparison with the
tractography-derived bundle weights of the Budapest graph. For a quotient
edge \(\{a,b\}\), the weight
\[
W_{ab}
\]
is exactly the number of parent-graph edges whose endpoints lie in
supernodes \(a\) and \(b\). It therefore has a natural interpretation as
the multiplicity of microscopic computational connections supporting one
mesoscale relation.

As the first stage of the inverse construction was designed to achieve,
the distribution of these bundle weights closely tracks the Budapest
fiber-count distribution. The quotient has mean bundle weight \(2.071\),
compared with \(2.033\) for Budapest; median \(2.0\), compared with
approximately \(1.67\); maximum \(155\), compared with approximately
\(154.9\); and Gini coefficient \(0.306\), compared with \(0.305\).
There are \(41\) quotient edges with bundle weight greater than \(18\),
compared with \(40\) in Budapest.

Thus the construction matches not only the unweighted mesoscale degree
organization but also the distribution of multiplicities underlying its
coarse edges. The claim remains structural rather than anatomical: the
parent is a fine computational substrate whose multiple microscopic edges
collapse onto weighted mesoscale relations, not a reconstruction of
individual biological axons.

\subsection{Network-Topology Test Against the Budapest Connectome}
\label{subsec:network-topology-test}

We evaluated the final inverse quotient using the same nine-criterion
network-topology test applied throughout the paper and to the Budapest
Reference Connectome
\cite{Szalkai2015Budapest,Szalkai2017Budapest,VargaGrolmusz2021Braingraph}.
The quotient was evaluated against \(1{,}000\)
degree-preserving null graphs. The nine scored criteria test clustering
enrichment, near-random path length, small-worldness, modularity enrichment,
near-random global efficiency, local-efficiency enrichment, connector hubs,
rich-club organization, and multi-resolution VI-stable modularity. The
criteria and thresholds are defined in
Appendix~\ref{app:network-topology}.

\begin{table}[t]
\centering
\small
\setlength{\tabcolsep}{5pt}
\begin{tabular}{p{0.40\linewidth}p{0.23\linewidth}p{0.23\linewidth}}
\toprule
Criterion & Final quotient & Budapest \\
\midrule
Clustering enriched
& \pass{} (\(2.198\times\))
& \pass{} (\(3.04\times\)) \\

Path length near random
& \pass{} (\(1.063\times\))
& \pass{} (\(1.19\times\)) \\

Small-worldness, degree-preserving
& \pass{} (\(\sigma_{\mathrm{dp}}=1.918\))
& \pass{} (\(\sigma_{\mathrm{dp}}=2.223\)) \\

Modularity enriched
& \pass{} (\(8.00\times\))
& \pass{} (\(9.08\times\)) \\

Global efficiency near random
& \pass{} (\(0.966\times\))
& \pass{} (\(0.90\times\)) \\

Local efficiency enriched
& \pass{} (\(1.216\times\))
& \pass{} (\(1.38\times\)) \\

Connector hubs
& \pass{} (\(7\) hubs)
& \pass{} (\(4\) hubs) \\

Rich-club organization
& \pass{} (run \(=272\))
& \pass{} (run \(=212\)) \\

Multi-resolution VI stability
& \pass{} (\(3\) stable scales)
& \pass{} (\(2\) stable scales) \\
\midrule
Total
& \(\mathbf{9/9}\)
& \(\mathbf{9/9}\) \\
\bottomrule
\end{tabular}
\caption{
Nine-criterion network-topology test for the final inverse quotient and the
Budapest Reference Connectome. The final quotient is evaluated against
\(1{,}000\) degree-preserving null graphs. Parenthetical values
are ratios to null means except for the small-world statistic, the rich-club
run, and the hub and stable-scale counts. For the null-tested enrichment
measures, the quotient reaches the
\(p=0.001\) empirical floor associated with \(1{,}000\) null realizations.
Both graphs pass all nine scored criteria.
}
\label{tab:nine-criteria-passfail}
\end{table}

The corresponding raw statistics are given in Table~\ref{tab:nine-criteria-values}.

\begin{table}[t]
\centering
\small
\setlength{\tabcolsep}{6pt}
\begin{tabular}{lcc}
\toprule
Measure & Final quotient & Budapest \\
\midrule
Edges & \(71{,}098\) & \(70{,}654\) \\
Density & \(0.1382\) & \(0.1373\) \\
Mean degree & \(140.1\) & \(139.2\) \\
Median degree & \(130\) & \(127\) \\
Minimum degree & \(13\) & \(7\) \\
Maximum degree & \(420\) & \(466\) \\
Clustering coefficient & \(0.4724\) & \(0.6696\) \\
Transitivity & \(0.4234\) & \(0.5578\) \\
Average path length & \(1.9811\) & \(2.2190\) \\
Diameter & \(4\) & \(5\) \\
Effective diameter (90\%) & \(2.162\) & \(2.731\) \\
Global efficiency & \(0.5492\) & \(0.510\) \\
Local efficiency & \(0.7304\) & \(0.834\) \\
Modularity \(Q\) & \(0.4908\) & \(0.5574\) \\
Assortativity & \(+0.0826\) & \(+0.0215\) \\
\bottomrule
\end{tabular}
\caption{
Raw network statistics for the final inverse quotient and the Budapest
Reference Connectome. The graphs are closely matched in node count, edge
count, density, degree profile, and path scale, while Budapest retains
stronger clustering and modular separation.
}
\label{tab:nine-criteria-values}
\end{table}

The test therefore supports a criterion-level rather than an identity
claim. The quotient satisfies the same nine qualitative network-topological
criteria as Budapest, but Budapest remains more strongly clustered and more
modular. The quotient is also more assortative,
\(r=0.0826\) versus \(0.0215\), although substantially less so than the
relocated parent from which it is derived.

Descriptive box-covering scaling is not included in the \(9/9\) score.
It fails for the final quotient, but it also fails for Budapest; the
descriptive scaling result is therefore not treated as a criterion of
quotient-level agreement.

A further six-test static comparison (Appendix~\ref{app:inverse-quotient-diagnostics}, Figure~\ref{fig:graph-tests}) gives a mean normalized error of
\(0.3033\) across degree distribution, effective diameter, hop plot, scree,
network value, and triangle participation. The degree test is particularly
strong: \(KS=0.0187\), the closest degree-distribution match among the
constructed graphs considered in this work. The six-test battery is not
uniformly optimized by the present quotient, however, and we therefore report
it as an additional comparison rather than as the defining result. The
distinguishing property of the final quotient is that it combines the degree
match with complete cross-supernode representation, matched fiber-bundle
weights, and the full \(9/9\) topology test.

\subsection{Algebraic, Spectral, and Combinatorial Comparison with Budapest}
\label{subsec:advanced-combinatorial}

The topology test checks the connectome-associated network organization. As a
separate comparison, we evaluate deterministic algebraic, spectral, and
combinatorial quantities. These measures are not scored against null
ensembles and are not treated as additional pass--fail criteria.

Several quantities agree closely despite not being part of the topology
test (Table~\ref{tab:complexity-spectral}). The normalized adjacency spectral radius is \(1.288\) versus
\(1.285\), the logarithmic spanning-tree count per node is \(4.791\)
versus \(4.774\), and the normalized Kirchhoff index is \(0.00985\)
versus \(0.01053\). These agreements indicate similar global spectral,
combinatorial, and resistance-distance scales.

The clearest spectral discrepancy is associated with bilateral organization.
Budapest has leading adjacency eigenvalues
\[
\lambda_1=178.9,
\qquad
\lambda_2=171.4,
\]
whose near-degeneracy is consistent with two nearly balanced hemispheric
halves. Its four principal communities have sizes \(307,306,201,201\),
which pair naturally into halves of \(508\) and \(507\). The quotient's
four large communities, \(389,265,239,122\), have no analogous pairing.
The construction contains no left--right axis or bilateral parent coordinate,
so this structure is not expected to emerge from the present relocation
procedure.

\subsection{Graph-Complexity and Controllability Measures}
\label{subsec:complexity-spectral}

We also compare the final quotient and Budapest using deterministic
graph-complexity, spectral, and controllability measures
(Table~\ref{tab:complexity-spectral}). These quantities
are descriptive rather than fitted objectives.

\begin{table}[t]
\centering
\small
\setlength{\tabcolsep}{4pt}
\begin{tabular}{lccc}
\toprule
Measure & Final quotient & Budapest & Interpretation \\
\midrule
\(\log_{10}\) Estrada index \cite{Estrada2000}
& \(78.34\) & \(77.72\) & close \\

Graph energy per node \cite{Gutman1978}
& \(7.009\) & \(5.693\) & higher \\

Kirchhoff index/\(n^2\) \cite{KleinRandic1993}
& \(0.00985\) & \(0.01053\) & close \\

Off-diagonal complexity \cite{Claussen2007}
& \(5.318\) & \(5.406\) & close \\

Degree-distribution entropy
& \(5.410\) & \(5.405\) & close \\

\(\lambda_{\max}(A)/\langle d\rangle\)
& \(1.288\) & \(1.285\) & close \\

Hoffman bound \cite{Hoffman1970}
& \(6.612\) & \(7.907\) & same regime \\

Random-walk eigengap
& \(0.0938\) & \(0.0197\) & different \\

\(\log\tau(G)/n\)  \cite{Kirchhoff1847}
& \(4.791\) & \(4.774\) & close \\

Average controllability, mean
& \(1.0932\) & \(1.1004\) & close \\
\bottomrule
\end{tabular}
\caption{
Complexity, spectral, and controllability comparison between the final
inverse quotient and the Budapest Reference Connectome. Several global
invariants are closely matched, particularly normalized spectral radius,
normalized Kirchhoff index, spanning-tree entropy, and average
controllability. Graph energy and the random-walk eigengap retain appreciable
differences.
}
\label{tab:complexity-spectral}
\end{table}

The agreement is therefore selective rather than universal. The quotient
tracks Budapest closely on normalized spectral radius, resistance-distance
scale, spanning-tree entropy, Estrada scale, average controllability,
off-diagonal complexity (\(5.318\) versus \(5.406\)), and
degree-distribution entropy (\(5.410\) versus \(5.405\) nats).
These quantities describe different aspects of global graph organization and
were not used as direct relocation targets, except that degree-distribution
entropy follows from the degree matching of the third stage.

Other measures remain distinct. In particular, graph energy per node is
\(7.009\) versus \(5.693\).
The random-walk eigengap also remains substantially larger. These differences
reinforce the restricted interpretation of the result: the inverse quotient
reproduces a broad connectome-like network regime, but it is not a spectral
or anatomical replica of Budapest.

The strongest unresolved structural feature is bilateral organization.
Budapest's paired community sizes and near-degenerate leading eigenvalues
reflect an anatomical organization that is absent from the construction.
This cannot be generated merely by changing the coarse-graining map or
relocating edges while leaving the parent without a bilateral axis; it would
require an additional bilateral organization in the underlying substrate.

The final quotient should therefore be understood as the strict coarse
representation of a sparse, chromatically bounded computational substrate
whose edge count, degree law, bundle-weight distribution, topology test,
and several global invariants closely reproduce those of the Budapest
Reference Connectome, while leaving specifically anatomical bilateral
structure and several higher-order spectral features outside the present
construction.

\section{Dynamical Realization on a Recurrent Neural Network}
\label{sec:dynamical-realization}

The preceding sections evaluate the construction as a static graph.  In
this section we place dynamics on it.  The purpose is narrow: to show
that the constructed graph can support useful recurrent computation when
used as a trainable substrate.  The model is a toy.  It does not
represent biological circuits, biological learning, or the dynamics of
real neural populations, and nothing in the interpretation developed
later depends on its particular architecture or training procedure.
What it establishes is that a graph with the properties tested above is
not computationally inert.

A biological connectome is not a graph whose nodes and edges are
elementary objects.  Each macroscopic region represents a population of
neurons, and each macroscopic connection represents a bundle or aggregate
of many lower-level axonal relations.  The global graph should be
interpreted in the same way.  A supernode depicts a coarse-grained
population hiding an underlying computational graph.  A supernode edge is
likewise not a single axon-like connection, but an aggregate relation
induced by sets of lower-level edges.  In fact, by designing a large
sparse graph first and clustering it, we hoped to mirror this biological
property.

This distinction is essential for the neural model.  The Budapest
comparison is performed at the \(1{,}015\)-node scale, but the
computational substrate underlying that comparison remains the
\(16{,}807\)-node fine graph.  The supernode assignment is fixed after
clustering the original integrated causal-dependency graph.  The
subsequent inverse construction preserves this mesoscale causal
partition while reorganizing fine causal-dependency edges within and
between the fixed populations.  The fine edge budget is restored rather
than conserved edge-for-edge: the substrate carries \(337{,}983\) edges
before the inverse construction and \(337{,}980\) after.

Relocating an edge into a supernode interior converts an
inter-supernode dependency into an intra-supernode dependency, whereas
adding a fine edge between two supernodes increases the fine causal
support for that mesoscale relation.  Thus the construction preserves
the causal clustering while altering the fine-scale causal connectivity
that realizes it.  Here ``preserves'' means that the community-derived vertex-to-supernode
assignment is held fixed; it does not mean that reclustering the modified
parent would necessarily recover the same partition.
The final substrate is obtained through bundle-matching relocation,
mesoscale support refinement, and a final degree-matching stage, all
while the \(1{,}015\)-supernode assignment remains fixed.  The strict
quotient is then recomputed from this modified fine substrate.  If
\(A\) and \(B\) are fixed supernodes with member sets \(V_A\) and \(V_B\),
the weight of their coarse relation is
\[
W_{AB}
=
\#\bigl\{
\{u,v\}\in E_{\mathrm{fine}}:
u\in V_A,\ v\in V_B
\bigr\}.
\]
A coarse edge is therefore not a primitive connection: it summarizes
the multiplicity of fine causal-dependency edges joining two populations.
Fine edges whose endpoints lie in the same supernode remain part of the
computational substrate but disappear naturally under the strict
coarse-graining.

This preserves the central multiscale interpretation of the construction.
The artificial connectome hides an underlying concurrent computational
structure: supernodes are coarse populations, quotient edges are bundles
of fine relations, and the actual computation takes place on the
fine-scale graph.  Different fine wirings could in principle realize
similar coarse relations, so the quotient by itself does not uniquely
specify a microscopic substrate.  In the reported construction, however,
no separate post hoc lifting operation is used for the neural model.
The final fine substrate itself is the runtime realization, and its
strict \(1{,}015\)-node quotient is its macroscopic connectome-scale
description.

The recurrent neural network therefore operates directly on this final
substrate,
\[
16{,}807\text{ nodes},
\qquad
337{,}980\text{ edges},
\qquad
1{,}015\text{ fixed supernodes},
\]
with its inherited proper three-coloring.  The fixed supernodes remain
available as the mesoscale grouping used by the same-phase attention
mechanism below; there is no separate reconstruction of a parent graph
from the quotient before training.

\subsection{Three Phases as a Recurrent Computation}
\label{subsec:three-phase-dynamics}

The final fine graph retains the chromatic structure of the construction.
Its vertices are organized into three color classes, interpreted as
three causal phase classes.  An undirected fine edge means that causal
ordering is required between its endpoints while the direction is
unresolved at the structural level; the concrete recurrent schedule
resolves that direction during execution.  Nodes in the same color class
can therefore participate in the same causal phase, while graph edges
join vertices whose updates must be ordered across phases. Because units that coactivate settle jointly, a single edge between two coactivating groups orders the groups as wholes, not only the two units it joins.

In the neural implementation these three color classes become three
recurrent populations,
\[
a_1,\qquad a_2,\qquad a_3 .
\]
The first population receives information collected from the visual
encoder, sparse graph operators propagate activity among the populations,
and the third population supplies information back to the decoder.  The
implemented forward operators are
\[
1\to2,
\qquad
1\to3,
\qquad
2\to3,
\]
with recurrent feedback
\[
3\to2,
\qquad
3\to1.
\]
The module executes five recurrent passes.  Feedback from later
populations therefore affects earlier populations on subsequent passes,
allowing causal dependencies to extend across recurrent sweeps.

This is the dynamical meaning of the chromatic constraint.  The graph may
be large and highly structured, but each admissible sweep requires only
the causal phase classes imposed by its coloring.  We call the associated
quantity the \emph{chromatic causal depth},
\[
D_\chi(G)=\chi(G).
\]
For the present graph,
\[
D_\chi(G)=3.
\]
This is the minimum number of causal phase classes required in each admissible sweep. It does not place a bound on the total temporal, recurrent, or dependency depth of the computation.

The recurrent interpretation therefore leads to two distinct levels of description. The coloring constrains the global coordination schedule, while recurrent iterations allow information to pass through dependencies extending beyond a single sweep. Computation within a phase may also be more complex than what a single graph edge represents. At the structural level, the fine graph determines where causal ordering is required. The neural dynamics determine how activity is transformed and propagated through this structure.

We use the graph model as the deep computational pathway of a \(256\times256\) CelebA-HQ image-inpainting encoder--decoder. Features from six encoder scales are collected into the first graph population. Sparse recurrent computation then takes place on the full \(16{,}807\)-node substrate, and a multiscale readout maps the resulting graph state back to the decoder. The deepest decoder inputs are obtained through learned projections over the entire third population. The conventional encoder skips at shallower scales are retained and modulated by graph outputs. We give the complete architectural and optimization details in Appendix~\ref{app:inpainting_model}.

\subsection{Adding Same-Phase Coactivation}
\label{subsec:feeder-supernode-comparison}

The fine causal graph provides sparse routing between color classes, but contains no same-color causal edges. We treat same-phase coactivation as distinct from causal propagation, and therefore require a separate mechanism for its realization. In the graph model reported here, we use supernode-level linear attention within non-empty same-color groups of each fixed supernode. This allows activity belonging to one causal phase to mix within a mesoscale population without introducing same-color edges into the causal graph.

We did not expect a large effect from this mechanism, for structural reasons. Attention operates only within one color class of one supernode, and these groups are generally small. The median group contains \(3\) nodes in each of the first two populations and \(4\) in the third, with means of \(4.9\), \(4.8\), and \(7.1\). Among the \(3{,}001\) non-empty groups, \(84.3\%\) contain eight nodes or fewer. The largest groups in the three populations, however, contain \(94\), \(102\), and \(137\) nodes. Of the \(3{,}045\) color--supernode slots, \(44\) contain no vertices and therefore have no attention block. Altogether, the attention blocks introduce only \(12{,}864\) parameters, corresponding to \(0.017\%\) of the full model.

The complete graph-plus-attention model has \(75{,}965{,}171\) trainable parameters. Of these, \(579{,}248\) are associated with the sparse graph and \(12{,}864\) with the attention blocks. The matched U-Net baseline has \(87{,}103{,}619\) parameters. Thus, the baseline is the larger model, and the comparison does not favor the graph architecture through parameter count.

Our primary experiment compares the graph model with attention against the matched U-Net baseline. We train each model from scratch using three independent random seeds. Both use the same image data, held-out split, mask-generation procedure, \(50\)-epoch training schedule, checkpoint-selection rule, and benchmark evaluation. We also examine a graph variant without attention. Only one run was performed for this variant, so we use it as a mechanism check, not as a separately replicated experimental arm.

All runs complete the full \(50\) epochs with batch size \(2\) and a cosine learning-rate schedule that decays to zero. We do not use early stopping. After training, we select checkpoints according to a rule fixed before the final runs. Validation loss is rounded to four decimal places. After epoch \(40\), we identify the first sequence of three consecutive epochs having the same rounded validation loss and select the earliest checkpoint in that sequence. We use \(1{,}000\) of the \(2{,}000\) held-out images for checkpoint selection and reserve the remaining \(1{,}000\) for final reporting.

For the three graph-plus-attention runs, the rule selects epochs \(45\), \(44\), and \(45\). For the three U-Net runs, it selects epochs \(47\), \(46\), and \(46\). The single no-attention run selects epoch \(47\).

The primary benchmark uses the broader mask-target distribution

\[
U(0.02,0.70).
\]

We compute L1, PSNR, SSIM, and LPIPS separately in six achieved mask-coverage buckets and then average the six values. Thus, the reported quantities are means over coverage buckets, not direct means over individual images.

\begin{table}[H]
\centering
\small
\setlength{\tabcolsep}{7pt}
\renewcommand{\arraystretch}{1.12}
\begin{tabular}{@{}lcc@{}}
\toprule
Metric
& U-Net
& Graph + attention\\
\midrule
L1 (\%) \(\downarrow\)
& \(1.5618\pm0.0105\)
& \(1.4533\pm0.0024\)\\
PSNR (dB) \(\uparrow\)
& \(29.254\pm0.041\)
& \(29.820\pm0.008\)\\
SSIM \(\uparrow\)
& \(0.89523\pm0.00048\)
& \(0.90387\pm0.00017\)\\
LPIPS \(\downarrow\)
& \(0.11034\pm0.00060\)
& \(0.09356\pm0.00032\)\\
\bottomrule
\end{tabular}
\caption{CelebA-HQ inpainting on the held-out reporting split using the
benchmark mask distribution. Values are mean \(\pm\) SD across three
independently trained seeds and are means over six mask-coverage buckets.
L1 is the mean absolute error over the full image (\(\times100\)); observed
pixels are copied from the input, so only the missing region contributes.
FID was not computed for these runs.}
\label{tab:unet-feeder-supernode}
\end{table}

Compared with the mean performance of the matched U-Net, the graph model reduces L1 by \(6.95\%\), increases PSNR by \(0.566\) dB, increases SSIM by \(0.00864\), and reduces LPIPS by \(15.20\%\). The reduction in L1 is not limited to one range of mask coverage. Each of the three graph runs has lower L1 than each of the three U-Net runs in every one of the six benchmark coverage buckets.

We also evaluate the graph checkpoints using the mask distribution employed during training,

\[
U(0.09,0.49),
\]

as a robustness check. The results over three seeds are

\[
\mathrm{L1}=1.4135\pm0.0031\%,
\qquad
\mathrm{PSNR}=29.347\pm0.005,
\]

\[
\mathrm{SSIM}=0.90539\pm0.00015,
\qquad
\mathrm{LPIPS}=0.09207\pm0.00033.
\]

We do not interpret the lower aggregate as an improvement over the benchmark result. The two mask distributions populate the coverage buckets differently. In particular, the training distribution samples the extreme \(0\)--\(10\%\) and \(50\)--\(60\%\) buckets sparsely. If we average over the four middle buckets, which are well populated under both distributions, L1 agrees within \(0.01\) percentage points and PSNR within \(0.03\) dB for every run. We therefore retain the benchmark-mask evaluation as the primary comparison. The U-Net baseline was not separately evaluated on the training-mask distribution.

We next ask whether the supernode-attention blocks themselves account for the improvement of the graph model. For this purpose, we evaluate the same graph architecture with the attention blocks removed. Since we performed only one no-attention run, this experiment is a mechanism check and not a replicated comparison. On the benchmark masks, the no-attention model gives

\[
\mathrm{L1}=1.4507\%,
\quad
\mathrm{PSNR}=29.843,
\quad
\mathrm{SSIM}=0.90415,
\quad
\mathrm{LPIPS}=0.09332,
\]

while the three-seed means of the attention model are

\[
\mathrm{L1}=1.4533\pm0.0024\%,
\qquad
\mathrm{PSNR}=29.820\pm0.008,
\]

\[
\mathrm{SSIM}=0.90387\pm0.00017,
\qquad
\mathrm{LPIPS}=0.09356\pm0.00032.
\]

We can also compare the no-attention run with the attention run trained using the same seed. The latter gives \(\mathrm{L1}=1.4506\%\), \(\mathrm{PSNR}=29.826\), \(\mathrm{SSIM}=0.90403\), and \(\mathrm{LPIPS}=0.09329\). The differences have mixed directions. The attention model performs better on L1 and LPIPS, while the no-attention model performs better on PSNR and SSIM. These differences are comparable to, or smaller than, the seed-to-seed variation observed in the attention runs.

Compared with the three-seed attention mean, the no-attention run is nominally better on all four metrics, with PSNR and SSIM falling outside the observed three-seed range. However, this compares a single unreplicated run with an ensemble mean. We therefore do not take it as evidence of superiority in either direction, nor as a test of equivalence.

The purpose of the inpainting experiment is not to establish state-of-the-art performance. Large specialized inpainting models remain stronger on this task \cite{Suvorov2022LaMa,Li2022MAT}. Our objective is more specific. We want to show that the final chromatically constrained fine graph can be used directly as a trainable sparse recurrent computational substrate, replacing the central deep pathway of a matched U-Net while supporting stable learned computation.

We should also distinguish this comparison from an ablation of graph topology alone. The graph architecture differs from the U-Net in its recurrent deep pathway, collector and readout interfaces, and graph-specific optimization. The observed improvement is therefore an architectural result. It does not establish that the connectome-like topology by itself causes the improvement.

The single no-attention experiment addresses a more limited question. It shows that the margin over the U-Net cannot be explained by the additional supernode-attention blocks. More generally, the experiment demonstrates the compatibility required by our multiscale interpretation. A fine causal-dependency graph can support recurrent computation, while its fixed supernodes provide a macroscopic connectome-scale description and, at the same time, a locus for same-phase coactivation.

\subsection{Same-Phase Coactivation, the Causal Web, and the Grain of Events}
\label{subsec:coactivation-attention}

The chromatic graph describes causal ordering between phases. But what makes the units ``fire together''?

Many neural assemblies fire together because they share a context, receive a common drive, participate in a local pre-activation event, or become part of a larger wave of cortical activity. Not all of this is coactivation in our sense. Firing that is merely simultaneous, because of a common drive, produces no joint event. We reserve coactivation for firing together in which rapid mutual exchange within a local population settles a joint state, including synchronization the units produce among themselves, and represent that as symmetric coactivation within the population.

We distinguish two mechanisms. Cross-color sparse edges represent causal or phase-separated computation. Same-color coactivation represents non-causal concurrent mixing. The graph-theoretic idealization of this mixing is a complete graph, where every unit in a local coactivation set can interact with every other unit without an edge specifying which must come first.

Self-attention has exactly this graph interpretation. Without a causal mask, attention places a learned complete graph over the tokens, allowing every token to read from every other token \cite{Vaswani2017Attention}. This represents contextual association without specifying temporal precedence. We use the same idea at the supernode level. The relocated parent graph implements phase-constrained recurrent computation, while supernode-level attention or complete same-phase mixing represents local coactivation within the population associated with each supernode.

The extension is biologically motivated at the level of mechanism, not as a literal model of neurons. The brain exhibits both large-scale waves and local packets of coordinated activity \cite{Muller2018TravelingWaves,Luczak2015Packets}. Supernode attention provides a computational analogue of such same-phase coactivation. It allows the same-color nodes within a supernode to fire together without imposing a causal order among them, while nodes of different colors in the same supernode remain causally linked.

The event points in our effective causal description should not be identified with microscopic atomic events. An event may consist of a very small or a very large number of underlying physical interactions. What makes these interactions a single cognitive event is not their number, spatial extent, or microscopic simplicity. It is their effective distinguishability from the rest of the system. The constituent activities are never addressed separately by the cognitive dynamics. Whatever influence they exert is exerted collectively through the state of the packet they form. Thus, even a highly resolved distributed process may constitute one event if all subsequent interactions depend only on its collective state. Conversely, if the organization changes so that an internal distinction becomes separately available to downstream processing, that distinction is no longer invisible, and the effective event acquires a finer decomposition.

The same principle applies to the formation and transmission of messages. A causal message need not pass directly from one already-formed event to another. It may enter a population, participate in a sequence of same-phase coactivations and recurrent internal interactions, be replicated or slightly transformed through several internal routes, and only later emerge as part of another collectively distinguishable packet. These intermediate activities are physically real. However, if their individual differences are never separately available to the rest of the system, they need not appear as events in the effective causal description. They may instead be integrated out, as discussed in Appendix~\ref{app:hierarchical_sde}. Their influence remains in the state of the outgoing packet and, in a reduced dynamical description, in the effective memory, correlations, or noise inherited from the hidden internal dynamics.

We can therefore interpret symmetric coactivation more precisely as an invariance arising from the organization of the system. The effective cognitive dynamics is insensitive to transformations of the underlying activity that preserve every distinguishable collective packet and every subsequent causal influence. The fundamental grain of cognitive events is thus determined by effective distinguishability, not by physical scale or the number of constituents. We adopt this as a postulate of the present interpretation. It offers an explanation for the remarkable stability of cognitive representation despite continual variation in the underlying computational dynamics.

Coactivation is itself a physical process. At the finest level, it need not be instantaneous or symmetric. Nearby units may couple first, with activity spreading outward through local interactions. This propagation may be rapid and diffuse, without addressed signals, and influence may pass back and forth between constituents before the ensemble settles. At the informational level, however, the process is symmetric. Its internal asymmetries are not separately available to the rest of the system. Over the duration of a phase, the ensemble acts as a single joint update whose effect on the rest of the system depends only on its collective state. No transmitter and receiver can be identified among its constituents, and no causal order is resolved. Because the update is joint, an input that reaches any one constituent enters the settling of the whole ensemble and shapes every constituent's outcome. A single cross-phase edge to one constituent therefore orders the whole ensemble relative to the edge's other endpoint, and through it one population can influence many.

Message passing is different. It is the strong collective signal that a formed ensemble sends through cross-phase paths to units outside it. The fine-scale ordering of internal activity is not erased. It is integrated out and contributes to the outgoing packet without appearing as separate events. The converse is also possible. A causal step followed closely by its successor may appear as joint firing if the activity is recorded at insufficient temporal resolution. Apparent simultaneity is therefore not a criterion in either direction. What distinguishes a causal step from coactivation is the informational asymmetry between a transmitter and a receiver.

Although the underlying physical interactions need not be symmetric, they may fail to form distinct clusters that can be separately understood, used, or exchanged by the rest of the network. The internal computation may remain too diffuse and distributed for its substructures to become stable informational objects in their own right. How can we identify such regions? We propose that they are physically connected dynamical clusters whose internal subdivisions are not independently visible or usable by the wider system. These subdivisions do not produce states that can be meaningfully \emph{widely broadcast}, interpreted, or acted upon separately. They therefore do not appear as distinct elements in the effective cognitive description. In fact, the ideas developed here may be viewed as a realization of certain aspects of global-workspace-like brain architecture \cite{Baars1988}.

\subsection{Why the graph constrains routing: weights have addresses}

Why should the structural graph constrain the dynamics? Learned transformations are not independent objects that can be applied anywhere in the network. Learning modifies particular synapses at particular locations, and the acquired transformations remain associated with those synapses. A weight matrix is not a message traveling through the network. It is a local property of the network that determines how arriving activity is transformed. Different populations therefore carry not only different signals, but also different learned transformations. The graph is consequently more than a wiring diagram. It is the address space in which the transformations available to the system are physically distributed.

Routing is therefore a computational necessity. If a task requires several learned transformations whose weights reside in different populations, the relevant activity must reach these populations in an order compatible with the computation. The weights do not move to the representation; the representation moves through the weights. The possible compositions depend on which populations are connected, whether the routes are direct or pass through intermediate populations, and how long transmission along these paths takes. Learning determines the transformations available at the nodes, while anatomy determines which transformations can be composed through message passing.

Routing can remain flexible within these constraints. Attention, inhibition, gain, neuromodulation, and changes in network state may determine which anatomically available pathway becomes effective at a given moment. Thus, the realized functional graph may change continuously while the structural graph remains unchanged. This flexibility, however, selects from an existing repertoire. It does not create a new repertoire. Gating can select which recurrent loop becomes active, but cannot create an absent projection, eliminate the distance along an existing connection, or make a recurrent path shorter than the shortest path provided by the anatomy. The structural graph determines the possible routes, while the dynamics selects among them.

A temporal constraint follows. Every task requiring recurrent passage through a particular collection of learned transformations has a shortest structurally available realization. Its duration cannot be less than the conduction and processing time required to traverse that route. Changes in network state may slow the process, redirect it, or move the computation to another available path. They cannot reduce its duration below that of the fastest route actually present in the substrate. The graph does not determine the instantaneous dynamics, but it imposes a lower bound on the time required for certain distributed computations to close. The dynamics provides the realized clock, while anatomy constrains what can be accomplished within one cycle.

This provides a physical basis for distinguishing local settling from distributed recurrence. Within a population, activity may circulate rapidly while a collective state forms through recurrent excitation, inhibition, competition, synchronization, or other local interactions. Across populations, a settled state must be transmitted, largely through finite axonal paths, and incorporated elsewhere before the larger recurrent sequence can continue. These two processes need not operate on the same timescale. Several local settling processes may occur during the time required for one extended circuit to close.

Concurrency allows activity to reach many distributed clusters of learned transformations rapidly, without requiring these clusters to be visited in a long sequential chain. The multiscale architecture of the brain is unlikely to be incidental. Its nested organization allows specialized local computations to proceed concurrently, while longer-range pathways integrate their results across scales.

So far, we have used a deliberately simplified description to make the underlying ideas clear. In particular, we have often described the chromatic classes as though one entire class becomes active at a time, with activity passing sequentially from one color to another. This should not be understood as a literal description of global neural firing. The coloring represents a bounded causal organization of computation, not a rigid synchronous clock imposed on the entire network.

In a more realistic dynamical description, many causal paths may be active simultaneously, and activity belonging to different color classes may overlap substantially in physical time. A population in one class may already be active when it receives a message from another class, while a different population elsewhere participates in another causal chain. After one recurrent cycle, activity need not return to the same nodes or reproduce the same spatial pattern. What recurs is the constrained computational topology. Local causal transitions continue within a structure of bounded chromatic depth, while the active subgraphs and their trajectories may change from one cycle to another.

We should therefore interpret color classes as causal compatibility classes, not as globally synchronous populations. They constrain which dependencies can be organized concurrently and how messages pass between clusters, but do not require every node of a class to become active together. A realized computation may involve only a subset of a color class, and different subsets may participate at different times. Our essential claim is that rich recurrent computation can proceed along many parallel causal paths while the dependency structure within each sweep remains shallow. We have described a sharp anatomical separation between coactivation and message passing, but this is an idealization. Causal paths are likely more strongly embedded together with coactivations.

\paragraph{Local timing and autonomy.}

The absence of a global clock is not merely a matter of realism. It may have a direct bearing on experience. When transmission takes time, the state of the system cannot be specified entirely by the current activation of its units. It must also include the wavepackets in transit, their departure times, and their expected arrival times. The state of a recurrent system with delays is therefore properly represented by a history over the delay window, not by an instantaneous configuration.

Such a state is difficult to observe. Reading the information traveling through a channel requires intercepting it, and this interception changes what arrives. We call this property \emph{observational irreducibility}. The complete state of the system, including its timing, cannot be read from outside without entering the causal structure being observed.

Within this structure, timing is not imposed entirely from outside. It is partly a degree of freedom of the units themselves. Whether constituents fire together or against one another, and when a packet forms and is released, is determined locally and at every scale. We conjecture that this local temporal autonomy is one ingredient of what is experienced as agency, or will.

A system in which all units are paced by a single external clock, and whose state can be read completely at each tick, lacks this freedom. In our view, it also lacks the temporal organization from which experienced agency could be composed. Like geometry, agency must be present across scales. It cannot arise only at the highest level of a system whose lower levels are entirely clocked. It must be built from the autonomy of the underlying scales.

Two qualifications are necessary. First, unpredictability alone does not constitute autonomy. Noise is unpredictable, but does not organize anything. What is required is timing that is self-generated and nevertheless integrated into a coherent complex. Second, the relevant property is not the physical presence or absence of a hardware clock. A clocked machine can reproduce any asynchronous ordering of events. The distinction is whether timing belongs to the causal structure of the events as a state variable of the units, or is fixed by a schedule external to them.

The loss of consciousness provides a check on this proposal. Its two characteristic failure modes affect temporal freedom in opposite ways. Globally stereotyped activity removes local temporal freedom, while isolated local activity may retain this freedom without integrating it into a coherent whole.

Neural oscillations can nevertheless arise naturally in such a recurrent architecture. Recurrently coupled populations may partially synchronize, produce transient rhythmic modes, or become phase-locked for limited periods. Large-scale brain rhythms may therefore organize or bias the timing and excitability of many local events without acting as a single wave that must traverse an entire color class before computation can proceed. Local recurrent dynamics may settle, cross a threshold, or produce an outgoing message independently of whether a global rhythm has reached other populations of the same class.

The resulting dynamics is more complex than either a purely asynchronous event network or a collection of globally synchronized oscillators. Local bursts, persistent activity, transient synchronization, collective oscillatory modes, and sparse causal message passing may coexist. Oscillations may regulate the probability and timing of event formation, while the chromatic graph constrains the causal organization through which these events influence one another. Collective rhythms may therefore regularize and coordinate activity, while the computational structure remains fundamentally concurrent, recurrent, and of bounded chromatic causal depth.

Thus, the three color classes need not be interpreted as three globally alternating systems in which all of $C_1$ becomes active, followed by all of $C_2$, and then all of $C_3$. The three directed layers may be active concurrently. What is ordered is the local causal flow through them.

Let

\[
\mathcal V
=
\mathcal V_1\oplus\mathcal V_2\oplus\mathcal V_3
\]

denote the composite state space associated with the three phase classes. We can represent the one-sweep evolution operator schematically as

\[
\mathsf H_{\rm cog}
=
\begin{pmatrix}
K_1 & 0 & T_{3\rightarrow1}\\
T_{1\rightarrow2} & K_2 & 0\\
0 & T_{2\rightarrow3} & K_3
\end{pmatrix}.
\]

Here $T_{i\rightarrow j}$ represents directed causal transfer from one color class to the next, while $K_i$ represents the transformation of relational structure within the corresponding same-phase state, including its available coactivation and entanglement structure.

The separation between $K_i$ and $T_{i\rightarrow i+1}$ is a modeling choice. If the same-phase transformation and outgoing causal transfer constitute one indivisible operation, we may combine them into a single effective map,

\[
\widetilde T_{i\rightarrow i+1}
=
T_{i\rightarrow i+1}\circ K_i .
\]

The elementary cognitive event can then be represented as a sheet-to-sheet transformation,

\[
E_\alpha:
\mathcal V_i
\longrightarrow
\mathcal V_{i+1},
\]

rather than as a static state belonging to either sheet. An event packet carries structured information across the causal transition. Changes in coactivation and entanglement contribute to the structure of this transformation without becoming a separate stream of events.

All three channels may operate simultaneously. A particular causal history is therefore not the global activation sequence of the network, but a directed path through the active operator,

\[
\gamma:
e_0\prec e_1\prec e_2\prec\cdots .
\]

If admissible cross-color transitions are restricted to

\[
C_1\rightarrow C_2,\qquad
C_2\rightarrow C_3,\qquad
C_3\rightarrow C_1,
\]

then each causal strand follows the same cyclic color ordering,

\[
C_1\rightarrow C_2\rightarrow C_3\rightarrow C_1\rightarrow\cdots ,
\]

although the actual vertices, branches, convergences, and event packets traversed by different strands may be entirely different. Many causal histories,

\[
\gamma_1,\gamma_2,\ldots,\gamma_N,
\]

can therefore coexist and interleave within the same globally active dynamics. Color specifies the phase structure of causal propagation. It does not impose a single globally synchronized sequence of events.

The distinction is fundamental:

\[
\text{global operator activity}
\neq
\text{individual causal history}.
\]

Causal time is determined by the ordered composition of event packets along a directed history,

\[
E_{\alpha_1}
\prec
E_{\alpha_2}
\prec
E_{\alpha_3}
\prec\cdots ,
\]

while the composite operator determines the entire space of transformations that can occur concurrently. In our proposal, conscious experience is associated with these causally ordered sheet-to-sheet transformations, not with the simultaneous activation of the complete operator.

If the three layers are genuinely active simultaneously, one sweep applies the composite operator at once,

\[
x_{n+1}=\mathsf H_{\rm cog}\,x_n ,
\]

with each class updated from the previous state of the others. This differs from a globally ordered three-step implementation,

\[
U_{\rm cycle}
=
U_{3\rightarrow1}
U_{2\rightarrow3}
U_{1\rightarrow2},
\]

where each phase uses the most recent values of the preceding classes. The two are not generally equivalent unless the phase maps commute. The distinction lies in the causal chains that the ordered product allows within a single sweep. Our interpretation requires only that the operator can be globally active while causal order remains locally recoverable from the directed histories of events.

The brain differs from a generic natural many-body system in an important respect. As a general-purpose computational architecture, its effective causal organization is \emph{partly} encoded in anatomy and expressed functionally, rather than arising entirely as a transient property of the dynamics. The cognitively accessible event structure is embedded in a larger recurrent ontic network. The organization of this network constrains which activities remain internally integrated and which can participate in subsequent causal exchanges. In our construction, we do not begin with a generic causal substrate and arbitrary dynamics, expecting the required inseparable computational organization to emerge afterward. Instead, we build this organization into the admissible causal relations, while the dynamics determines which particular histories are realized.

The clustering of causal events, discussed in the next section, follows from the laws of cognitive geometry. Concurrency minimizes their spread in time. Clusters must be sufficiently large to exert effective control, but sufficiently small for that control to arrive in time to influence processes that continue to unfold in physical time. The requirement of rapid execution favors smaller clusters, while the need to draw the correct overall inferences favors larger ones. Cluster size therefore results from an optimal compromise between these two requirements.

The overlapping clusters reflect differentiation in time, while their size measures integration. Control arising from a maximal cluster can modulate the outputs already emitted by its sub-clusters or overwrite them in working memory. What is eventually stored in long-term memory may differ from what the clusters produced as they formed. The experience, however, consists precisely of the events of the total cluster as they occurred.

We must distinguish computation integrated away within the formation of experience from lower-level sensory and motor processing. The former belongs to the experienced state, although its internal organization is not separately experienced. The latter lies closer to the sensory and motor interfaces of the system, supplying inputs to the experiential organization and receiving its outputs and control actions.

The geometry is defined over the causal web of events. A stray electron may influence an event, and a sudden change in temperature may do the same, without either becoming part of the event structure. Such processes appear as uncertainty in the outcomes of interactions, or as constraints, not as cognitive events. At some point in a chain of causal influence, we reach a boundary beyond which influence does not imply membership in the event assembly.

The geometry belongs to the events and their relations. Processes that merely perturb their formation may remain outside that geometry. This is analogous to points lying near, but not on, a lower-dimensional data manifold. Proximity or influence does not imply membership.

The same distinction applies to the functions through which events are formed. The transformations implementing causal transfer and same-phase coactivation are influenced by factors outside the complex: learned weights, gain and neuromodulatory state, the global context \(\Gamma_t\) (Section~\ref{sec:ads-cft-correspondence}), the physical conditions of the substrate, and activity elsewhere in the network. Some of these influences are events in their own right, as are the excluded alternatives in the recordings discussed in Section~\ref{subsec:cognitive-geometry}.

Relative to a particular carved-out complex, however, everything outside it acts as a parameter. These external factors determine how the events of the complex form without themselves belonging to it. The situation is similar to couplings and external fields that influence a physical system without being among its degrees of freedom. In holographic terms, such influences correspond to sources rather than states (Section~\ref{sec:emergent-geometry}).

The distinction between an event and a parameter is therefore relative to the carving, not intrinsic. The same activity may be an event in one complex and a parameter in another. It becomes part of a complex only when it is incorporated through closure. Furthermore, information can enlarge an experiential cluster only insofar as it can be incorporated without destroying the causal geometry of which clustering is an aspect. Continuous sensory input therefore need not imply continuous incorporation into a single experiential object. Event membership retains a degree of discreteness and dynamical isolation.

Finally, the natural probabilistic object is the joint distribution over vector-valued events and their realized causal relations. Writing a realized configuration as

\[
\mathcal S=(E,R,\{x_e\}),
\]

we are interested in the information geometry of distributions over these causally constrained configuration spaces, including distributions over event clusters.

\subsection{Hierarchical Dynamics and the Status of Reduced Oscillator Models}

Many large-scale models of neural activity begin with an already coarse collective variable. For example, a noisy network of Hopf oscillators may be written as

\[
dz_i
=
\left[
(a_i+i\omega_i-|z_i|^2)z_i
-
\kappa\sum_j L_{ij}z_j
\right]dt
+
\sigma_i\,dW_i .
\]

Here each \(z_i\) represents a collective variable associated with a neural population or anatomical region. Such models can provide useful effective descriptions of large-scale rhythmic activity. However, the lower-level dynamical structure through which the collective variable is formed has already been integrated away. We should therefore regard the oscillator as a reduced dynamical object, not as a fundamental computational unit.

Our framework suggests retaining some of this intermediate structure explicitly. At the simplest level, we may consider two coupled stochastic dynamical states,

\[
\boxed{
dX_t
=
F(X_t,Y_t)\,dt
+
B_X(X_t,Y_t)\,dW_t^{(X)}
}
\]

and

\[
\boxed{
dY_t
=
G(X_t,Y_t)\,dt
+
B_Y(X_t,Y_t)\,dW_t^{(Y)} .
}
\]

This two-block form is standard in the theory of coupled stochastic systems \cite{PavliotisStuart2008}. In generative modeling, it also appears in critically damped Langevin diffusion, where the data coordinate is driven through an auxiliary velocity and \(B_X=0\) \cite{Dockhorn2022CLD}.

Our proposal concerns the interpretation of \(X\) and \(Y\) as levels of description, together with general state-dependent coupling and diffusion and the elimination and recursion discussed below. We take \(X_t\) to represent a more integrated collective state and \(Y_t\) a richer lower-level dynamical state, retained explicitly instead of being completely absorbed into an effective stochastic description. The functions \(F\) and \(G\) specify the coupled dynamics between the levels, while \(B_X\) and \(B_Y\) specify the effective stochastic forcing at each level.

The two levels evolve simultaneously and may influence one another in both directions. The lower-level state \(Y_t\) may supply structured input to the formation of \(X_t\). Conversely, \(X_t\) may modulate, gate, or otherwise constrain the subsequent evolution of \(Y_t\). The higher level need not perform the detailed transformations carried out by the lower level. It may instead integrate their collective consequences and return control signals to the dynamics from which they arise.

The stochastic terms need not have comparable strengths. In particular, the higher collective state may be only weakly stochastic,

\[
\|B_X B_X^{\mathsf T}\|
\ll
\|B_Y B_Y^{\mathsf T}\|,
\]

where \(B_X\rightarrow 0\) removes the noise acting directly on the higher level; \(X_t\) then remains random only through its coupling to \(Y_t\). Thus, the lower-level dynamics may remain strongly influenced by still-finer degrees of freedom while the higher-level state is comparatively stable. More generally, the diffusion amplitudes may depend on the state and need not be constant.

We can introduce additional levels recursively,

\[
X_t^{(0)}
\longrightarrow
X_t^{(1)}
\longrightarrow
\cdots
\longrightarrow
X_t^{(L)},
\]

with each level retaining part of the collective structure of the level below while integrating away other distinctions. The variables at each level are therefore not necessarily latent variables in the usual machine-learning sense. They may be physically realized dynamical states that remain explicit relative to finer levels while being integrated away at coarser levels.

This hierarchy provides a more expressive alternative to assigning a single already-formed oscillator to each macroscopic neural region. A Hopf-type variable may still emerge as an effective description of a particular collective dynamical regime. However, the model does not require oscillation to be fundamental. The recurrent causal architecture, the lower-level computational dynamics, and the formation of collective informational states remain conceptually prior.

In this view, Hopf and related oscillator models remain useful reduced descriptions of particular collective regimes. The more fundamental structure is the multiscale recurrent causal architecture and the hierarchy of dynamical states from which oscillatory regimes, integrated informational states, and their effective stochastic descriptions emerge.

\section{Emergent Geometry from Causal Structure: Correlation, Locality, and Causal Time}
\label{sec:ads-cft-correspondence}

\subsection{What the Construction Gives and Requires}

We must distinguish three things in what follows.

\emph{Derived from the graph.} The graph determines the admissible causal relations, the successive clustering, the graph-theoretic proximity arising from shared causal environment, and the scaling of coarse domains under iteration. Together, these define a potential multiscale organization, the space of geometries that the substrate can support. They do not determine the realized cognitive geometry or the realized coactivation state. These belong to the activity, not to the graph itself (Section~\ref{sec:emergent-geometry}).

\emph{Demanded by the graph, supplied by the theory.} Same-color units have no edges between them. The construction therefore permits coactivation, but does not explain why it occurs or what makes it coherent. This is a gap arising from the construction itself. We propose a global context to account for this coherence. Such an account must be consistent with the empirical evidence for large-scale, frequency-dependent coordination among distributed neural populations.

We denote this global context by $\Gamma_t$. It is a system-wide constraint on same-phase coactivation, but is not the state of any population and is not carried by any edge. Its role is the one identified in Section~\ref{sec:dynamical-realization}: a shared oscillatory context of the kind that large-scale brain rhythms may provide. It is collectively generated, remains stable over the timescales relevant to cognition, and carries no content of its own. What the theory supplies is not the constraint itself, but its place in the account.

\emph{Supplied by interpretation.} The graph, through its evolution operator, determines the admissible causal order: which units may influence one another and in what phase order. We interpret succession between color classes as a primitive temporal order at this level. The causal order of realized experience, however, is a different matter. At any given moment, only some of the admissible paths are active, and many combinations of active paths are possible. The realized causal order is determined by the paths that are actually active. It is therefore a property of the state, rather than of the graph, and belongs to the geometry of that state. This order must respect the admissible causal order, but need not coincide with it. Its determination remains part of the subsequent geometric construction. Cognitive or subjective time is not derived here.

The global context is not itself a cause within the effective causal description. It is not the state of any population and is not carried by a signaling channel. A cause must be locatable among the objects of the description in which it is invoked. What the global context produces therefore appears at this level as correlation, not as consequence. Thus, $\Gamma_t$ acts as a global constraint on the effective dynamics, not as a localized channel of causal signaling. Our claim is limited to this structural distinction.

\subsection{Causal Time and the Global Constraint}

Let \(C_1,C_2,C_3\) denote the three color classes. A proper
three-coloring guarantees only that adjacent vertices have different colors;
it does not by itself assign temporal phases or a causal ordering. In the
recurrent realization used here, the color classes are assigned to a
directed schedule. The concrete recurrent implementation selects the fixed
schedule
\[
C_1\longrightarrow C_2\longrightarrow C_3
\]
and repeats it across recurrent sweeps,
\[
1\to2\to3\;|\;1\to2\to3\;|\;\cdots .
\]
Other directed realizations could instead choose a different schedule,
provided the edge directions and update rule are changed consistently. The
essential structural claim is therefore not supplied by coloring alone:
temporal interpretation comes from the directed dynamics acting on the
coloring. Dependencies may continue across successive sweeps, so the three
scheduled phases do not bound the total causal depth of a computation.

We take a recurrence, or several successive recurrences, as the interval
over which content may become unified. Within a single scheduled phase no
intra-phase causal update is defined, so the primitive temporal resolution
of this implementation is the color-to-color transition. Whether this also
imposes a lower bound on subjective temporal resolution is a further
hypothesis rather than a consequence of coloring alone. Causal events are
color-to-color transitions on the fine graph, and supernodes play no part
in what a causal event is.  Coarse-scale messages are accordingly aggregates rather
than primitives: whether a given transition appears as internal to a
supernode or as a message between two depends on the resolution at which
the coarsening is taken, so coarsening changes how causal steps are
grouped and not what they are.

What supernodes supply is different.  They mark the regions in which
coactivation becomes collective, and they induce proximity.  The proximity
in question cannot be adjacency: same-color units are pairwise nonadjacent
by construction, so no density of relations {\it among} them is available
to define it.  What is available is their relation to a common causal
environment.  We define an effective same-phase proximity by the overlap of the
cross-color populations from which two same-color units receive and to
which they send. Proximity is therefore second-order, a similarity under
shared causal context rather than a density of mutual connection. This
supplies a graph-theoretic precursor of effective space: a second-order
proximity derived from causal structure rather than posited independently.  That a supernode carries both roles is then not a
coincidence: the same concentration of cross-color traffic that makes a
region a cluster is what places many same-color units in a common causal
environment within it.  At the coarse level the trajectory
\[
C_{\alpha_0}
\longrightarrow
C_{\alpha_1}
\longrightarrow
C_{\alpha_2}
\longrightarrow \cdots
\]
records the effective causal direction of the model as a sequence of
bundles.  Same-phase correlation can connect degrees of freedom without
supplying such a trajectory, whereas cross-color transitions change the
state of one population through the state of another and so contribute
to causal time.

Two contributions to coactivation therefore scale differently. Attention
weight weakens as cluster relations rise in the hierarchy, while the
constraint imposed by the global context is intended not to weaken simply
because the constrained class is larger. If $\Gamma_t$ is required to
remain effective independently of cluster size, a natural implementation is
for it to gate or modulate the gain of the coactivation operator rather than
enter as an additive source term. This is a modeling requirement of the
present construction, not a consequence of graph structure alone.  The arrangement
resembles that of a guiding field in the sense of the de Broglie--Bohm
formulation~\cite{bohm1952}, where, in its first-order (guidance) form, the wavefunction fixes the velocity through the gradient of its phase and its influence is
unchanged under rescaling, rather than entering the dynamics as a summed
force.  We take from this the architecture of guidance only, and note
that in that formulation the hidden variables are the particle positions
while the wavefunction is the guiding structure.

\subsection{Symmetry of Tensor Powers and the Origin of the Metric}
\label{subsec:symmetry-metric}

Throughout, $S$ is a finite connected non-bipartite graph on $s$ vertices,
and $S^{\otimes k}$ denotes its $k$-th tensor (Kronecker) power: the vertex
set is $V(S)^k$, and $(u_1,\dots,u_k) \sim (v_1,\dots,v_k)$ if and only if
$u_i \sim v_i$ in $S$ for every $i$.  Non-bipartiteness guarantees that
$S^{\otimes k}$ is connected \cite{Weichsel1962}.  We write $d$ for the
shortest-path metric and call the multiset $P(v) = \{ d(v,w) : w \in V \}$
the \emph{distance profile} of $v$; two vertices are \emph{metrically
distinct} if their profiles differ.

\begin{proposition}[Wreath symmetry of tensor powers]
\label{prop:wreath}
The wreath product $\operatorname{Aut}(S) \wr \operatorname{Sym}(k)
= \operatorname{Aut}(S)^k \rtimes \operatorname{Sym}(k)$ embeds into
$\operatorname{Aut}(S^{\otimes k})$: coordinatewise seed automorphisms
$(u_1,\dots,u_k) \mapsto (\varphi_1 u_1,\dots,\varphi_k u_k)$ and factor
permutations $(u_1,\dots,u_k) \mapsto (u_{\sigma^{-1}(1)},\dots,
u_{\sigma^{-1}(k)})$ are automorphisms, and the map is injective for
$s \ge 2$.
\end{proposition}

\begin{proof}
Adjacency in $S^{\otimes k}$ is the conjunction over coordinates of
adjacency in $S$; coordinatewise automorphisms preserve each conjunct and
factor permutations permute them, so both preserve adjacency, and they
compose according to the semidirect product. For injectivity, if
$(\varphi_1,\dots,\varphi_k;\sigma)$ fixes every tuple, evaluating on tuples
that vary in a single coordinate forces $\sigma = \mathrm{id}$ and each
$\varphi_i = \mathrm{id}$.
\end{proof}

\begin{corollary}[Metric degeneracy of the pure power]
\label{cor:degeneracy}
Let $r$ be the number of $\operatorname{Aut}(S)$-orbits of $V(S)$. Then the
group of Proposition~\ref{prop:wreath} has exactly $\binom{r+k-1}{k}$ orbits
on $V(S)^k$, and consequently $S^{\otimes k}$ has at most
$\binom{r+k-1}{k}$ metrically distinct vertex positions. In particular, the
number of distinguishable positions grows polynomially in $k$ (as
$O(k^{\,r-1})$) while $|V| = s^k$ grows exponentially, so the fraction of
metrically distinct vertices tends to $0$.
\end{corollary}

\begin{proof}
Automorphisms are isometries, so vertices in one orbit have equal distance
profiles; distinct positions are therefore at most as numerous as orbits.
The subgroup $\operatorname{Aut}(S)^k$ acts independently per coordinate,
so its orbits on $V(S)^k$ are the $k$-tuples of seed orbits;
$\operatorname{Sym}(k)$ then identifies tuples with equal multisets of
orbit labels. Two tuples with the same multiset are exchanged by first
permuting coordinates to align labels and then applying coordinatewise
automorphisms, so the orbits are exactly the multisets of size $k$ over $r$
labels, of which there are $\binom{r+k-1}{k}$.
\end{proof}

\begin{proposition}[Symmetry breaking is necessary for vertexwise metric individuation]
\label{prop:breaking}
Every vertex quantity derived from the metric alone (distance profile,
eccentricity, closeness, or any isomorphism-invariant function) is
constant on $\operatorname{Aut}$-orbits. Hence a graph on $n$ vertices in
which a fraction $\alpha$ of vertices are metrically distinct has at least
$\alpha n$ automorphism orbits, and full metric differentiation requires
the orbits to separate almost all vertices. Moreover, any graph that is
again a tensor power \((S')^{\otimes k}\), for some seed \(S'\), retains the
wreath symmetry of Proposition~\ref{prop:wreath} and therefore the
degeneracy of Corollary~\ref{cor:degeneracy}: a nonvanishing fraction of
metrically individuated positions as \(k\) grows requires edits after which
the graph is no longer of this form.
\end{proposition}

\begin{proof}
The first claim is the isometry property of automorphisms; the bound on
orbits follows by counting. For the last claim, $(S')^{\otimes k}$ satisfies
Proposition~\ref{prop:wreath} for any seed $S'$, so by
Corollary~\ref{cor:degeneracy} its distinguishable positions number at
most $\binom{r'+k-1}{k}$, where $r'$ is the number of
$\operatorname{Aut}(S')$-orbits. This is polynomial in $k$, while the
number of vertices grows exponentially, so the fraction of metrically
individuated positions tends to $0$.
\end{proof}

\begin{remark}[The construction's instance]
For the seed of this paper ($s = 7$, $k = 5$, hence $|V| = 16{,}807$),
$\operatorname{Aut}(S)$ has order $4$, generated by the transpositions
$0\leftrightarrow5$ and $1\leftrightarrow3$, with vertex orbits
$\{0,5\},\{1,3\},\{2\},\{4\},\{6\}$; thus $r=5$.  The observed count of
$126 = \binom{9}{5}$ metrically distinct positions in the pure power
equals the orbit bound of Corollary~\ref{cor:degeneracy} exactly, so the
degeneracy is the one forced by symmetry and no further.
Proposition~\ref{prop:diameter-flat} accounts for the diameter of $4$
observed at every scale, including the seed itself: flat diameter is a
property of powering, so agreement with a target network on this
statistic carries no evidence of fit.   After the edits of the local
construction, which break the product form, the parent graph has $16{,}323$
distinct distance profiles ($97.12\%$ of its $16{,}807$ vertices) and its diameter
 opens to $12$: by Proposition~\ref{prop:breaking}, the automorphism group
has been broken to the point of near-rigidity, and the walk-length bound
$\Lambda$ underlying Proposition~\ref{prop:diameter-flat} no longer
applies.  Vertex identity becomes metrically individuated only insofar as the
inherited automorphism symmetry is broken. Symmetry breaking is therefore
necessary for the form of vertexwise differentiation used here, although it
is not by itself sufficient to define the eventual cognitive geometry.
Every added edge joins distinct color classes, and deletions cannot
invalidate the carried proper coloring, so every edited graph satisfies
$\chi\le3$. Each reported parent graph also has nonzero clustering
coefficient and therefore contains a triangle, giving $\chi\ge3$; hence
$\chi=3$ for every parent graph reported here.
\end{remark}

\begin{remark}[On the word ``distance'']
\label{rem:time-emergence}
The distance in the propositions above is the hop metric of the graph and
nothing more. What distance and time properly are in this framework
(which separations are primitive and which derived) is not settled here
and is left to later work. The propositions nonetheless constrain that
later work: any quantity derived solely from the uncolored graph,
including its hop metric, is an isomorphism-invariant function of it,
hence constant on automorphism orbits, so under the symmetry of the pure
power no such quantity can distinguish more than the $126$ orbit
positions. The inherited coloring and the directed phase schedule add
structure that the wreath symmetry need not preserve (coordinate permutations do not, and here neither do the seed automorphisms, which exchange color classes), so the bound
does not automatically extend to phase-aware notions of causal time. The results
above therefore do not presuppose a choice of geometry; they suggest
that the symmetry breaking is a precondition for anything interesting.
\end{remark}

\subsection{Emergent Geometry and the Holographic Analogy}
\label{sec:emergent-geometry}

The placement of same-phase correlation on the clusters is not a choice made
solely for convenience within the construction. Modular community structure
in the human connectome is independently attested, and functional connectivity
is constrained, but not completely determined, by structural connectivity
\cite{Honey2009,BullmoreSporns2009}. Importantly, functional coupling between
regions need not correspond to a direct structural edge, since correlations
may also arise through indirect polysynaptic pathways and common network
inputs. The empirical structure--function relationship therefore supports the
placement of collective coactivation within a modular substrate, while the
causal--correlational distinction is supplied by the construction itself.

The pattern set out above has a limited holographic analogue: in both
settings a deeper relational structure differs from ordinary adjacency,
while coarse-graining reorganizes effective locality and the temporal or
causal evolution is inherited from the underlying dynamics. In the boundary description of a holographic system, entanglement is a property of
the quantum state and is not itself a signaling channel, yet its
organization is deeply related to the geometry and connectivity of the
bulk, relative to which bulk causal propagation takes place. If the boundary
state evolves according to
\[
|\psi_{t+\Delta t}\rangle_{\partial}
=
U_{\partial}(\Delta t)
|\psi_t\rangle_{\partial},
\qquad
U_{\partial}(\Delta t)
=
e^{-iH_{\partial}\Delta t/\hbar},
\]
the corresponding bulk description changes with it, while the distinction
between entanglement structure and causal propagation remains. Boundary time $t_\partial$ supplies the temporal coordinate of the
boundary evolution, while bulk proper times and causal intervals are
determined by the reconstructed bulk metric.

\begin{equation*}
{\setlength{\fboxsep}{6pt}
\fbox{\small$
\begin{array}{r@{\;\longrightarrow\;}l@{\qquad}l}
\text{cross-color edges}
    & \text{causal relations}
    & \text{primitive}\\[2pt]
\text{color transition}
    & \text{causal step}
    & \text{primitive}\\[2pt]
\text{shared causal context}
    & \text{same-phase proximity}
    & \text{derived}\\[2pt]
\Gamma_t+\text{state}
    & \text{coactivation/correlation}
    & \text{state-dependent}
\end{array}
$}}
\end{equation*}

A supernode is therefore not an atomic vertex. It supplies the domain within
which correlation is controlled.  An immediate consequence is that
correlation distance and causal distance need not agree. They are built from
different orders of the same relation (the first from shared causal
environment, the second from traversal of causal edges), and so their
divergence requires no further assumption. Correlations can exist between events
that are far apart in the causal geometry.

A related consequence concerns duration, a consequence of the model's
bookkeeping rather than a claim about physical time. A same-phase relation
has zero color-transition separation within the model's causal bookkeeping.
This does not imply zero physical duration: the relation occurs at a definite
phase and may depend on causal history extending over earlier transitions.

Supernodes accordingly play a role structurally analogous to mass
concentrations. As a supernode grows it contains a larger population of
fine-scale degrees of freedom, supports a richer space of concurrent events
during each recurrent cycle, and admits more same-phase correlations jointly
expressed within it. Increasing concentration therefore gives rise to stronger
effective locality and a greater capacity to organize both causal evolution
and correlation. The analogy nevertheless has a conjectural temporal face. Consider a
video clip presented without sound, and the same clip presented with a
soundtrack shifted in time. We hypothesize that, within a dense causal
environment that binds the images to the sound, the effective duration
assigned to the visual sequence may be dilated so as to maintain that
binding, whereas the same clip playing unattended in the background need
not be. On this conjecture, greater integration within a supernode could
alter the subjective duration assigned to information sequences. If images
are treated as the ticks of a clock, the resulting experience would be
described as a slower clock. This is a structural analogy only and is not a
claim of gravitational time dilation.

The comparison with holography locates a genuine difference. In
semiclassical AdS/CFT, the bulk divides into a causal spacetime and the
quantum fields living on it, while this division has no manifest
counterpart at the boundary. The bulk metric is associated with the boundary
stress-tensor sector, while matter fields correspond to other boundary
operators. Bulk causal structure is encoded nonlocally in the boundary
theory and can, in suitable regimes, be probed through the singularity and
analytic structure of boundary correlators. Horizons have several boundary
signatures, including thermal behavior and characteristic entanglement
structure. Correlation without signaling and causal response are therefore
different observables of a single boundary quantum theory. Our construction,
by contrast, makes their distinction explicit already at the substrate.

This suggests an evolutionary conjecture. We conjecture that an important
evolutionary transition may have occurred when neural architectures
developed a stronger anatomical separation between causal routing and
collective correlational organization. \emph{We conjecture that subjective
experience arises when a nervous system encodes this division anatomically}:
when causal routing and collective correlation become, partially,
architecturally distinct registers, whereby the causal structure can
gate correlational states, select among them, and commit. On the present
hypothesis, a less differentiated substrate would provide less explicit
machinery for governing its own collective states through closure and
commitment. The proposed division is therefore not merely an efficient
architecture but a candidate enabling structure for experience.
\emph{Subjective experience is the duality made explicit.}

The anatomical division need not be primary. The organization displayed by the connectome may have first arisen dynamically, through effective routing and transient coactivation, and then become partly fixed in anatomy through evolution. A fixed route is cheaper and faster to use than one that must be reassembled through activity each time. Once established, the anatomy constrains what can emerge on it. As we argued in the discussion of routing, gating can select among available paths but cannot create absent ones. The distributed patterns formed by coactivation must therefore respect the message-passing constraint. They must produce packets that the existing routes can carry and that downstream complexes can receive. Content is thus formed already shaped for the routes it must follow.

Transformers represent the other extreme. Their attention recomputes every interaction from the current context. Whatever message passing they exhibit is therefore itself dynamic, emerging within the same operator that mixes content, if it emerges at all. Their distributed patterns need not conform to fixed routes. This makes them far more fluid and is one source of their flexibility. But it also means that nothing in the substrate distinguishes the routes carrying causal structure from the mixing that forms content. The brain combines these two regimes. Its organization remains dynamic and flexible within limits, but operates on an anatomy that holds part of this organization fixed. Every pattern formed on it must be compatible with that anatomy.

This compatibility is achieved differently from its counterpart in physics. In a quantum system, the state need not anticipate the dynamics. The Hamiltonian acts on the entire state at once, and the result is consistent with the evolution. In constrained systems, this consistency is secured in advance by the theory itself. Requiring the constraints to be preserved in time leads to further constraints that every admissible state must already satisfy, as in Dirac's treatment of constrained dynamics.

In our framework, formation and transmission are distinct stages that occur successively in physical time. A packet is first formed through coactivation and then transmitted. There is no single operator acting on the entire state to ensure that the two stages agree. The formation process must therefore produce packets, before transmission, that the available routes can carry and the receiving complexes can incorporate. What is secured by physical law in the former case must here be acquired through evolution, development, and plasticity. These processes tune coactivation so that the packets it produces are already compatible with the available routes.

A consequence follows. If the routes change faster than coactivation can be retuned, as after a sudden disconnection, packets should continue to form for routes that can no longer serve them. Recovery should then proceed as the formation process is relearned to fit the remaining routes.

Our conjecture concerns the kind of organization required. Whether a minimum size or complexity is also necessary cannot be determined until the geometry of experience is defined. The holographic parallel suggests that such a requirement may exist. A local, semiclassical bulk description exists only when the boundary system has many degrees of freedom and a clear separation between collective modes and the remaining degrees of freedom \cite{Heemskerk2009}. Once the map from the activity state to cognitive geometry is constructed, an analogous requirement of scale may turn out to be a condition for that geometry.

\paragraph{Local recurrence and global accounts.}
The principal alternative to global accounts holds that recurrence
confined to sensory cortex already suffices for experience.  On the
recurrent-processing view, the feedforward sweep through the visual
hierarchy is unconscious, recurrent exchange within and between visual
areas through horizontal and feedback connections constitutes phenomenal
experience whether or not it is reported, and widespread recurrence
involving frontoparietal areas adds access and report
\cite{LammeRoelfsema2000,Lamme2006}.  A central piece of evidence is
inattentional blindness, in which unreported figures still evoke recurrent
processing that stays within visual areas.  We agree that recurrence is
required, and that what matters is the organization of activity rather
than whether it lies at the back or the front of the head
\cite{Lamme2006}.  We nonetheless side with global accounts, because not
every process that contributes to experience is experienced as a process.
If the dynamics that generate content were themselves experienced,
experience would be as variable as those dynamics; it is instead invariant
(Section~\ref{subsec:coactivation-attention}).  An orange keeps its color
under changing light, and the correction for the light is never
experienced.  Experience accordingly contains two kinds of things.  Some
are never presented apart, such as a color and the extent it covers, which
Husserl called moments as opposed to independent pieces
\cite{Husserl1970LI}.  The binding that produces them stays hidden even
when it fails: an illusory conjunction is seen as a wrongly colored
object, not as a color detached from any shape
\cite{TreismanSchmidt1982}.  Others are presented as relations between
separable things, and judging such relations appears to involve attention moving
from one object to the other \cite{Franconeri2012}.  On our
interpretation, the first kind is produced by same-phase coactivation,
whose internal construction is sealed inside event packets, and the second
by causal relations within a carved-out complex.  Local recurrence in
sensory cortex builds packets.  It may reach closure of its own, but
closure over coactivation alone yields a packet rather than a complex:
there are no experienced relations inside it, and therefore no geometry of
its own.  A packet enters experience as a moment of an object only when
causal routing incorporates it into a complex; until then it acts as a
parameter.  This is the status of the unreported figures, whose
information, in Lamme's words, ``is not inaccessible, it is just not
currently accessed'' \cite{Lamme2006}, and which are reported once the
interactions become widespread.  The global stage is therefore not access
to an experience already completed locally, but the causal closure that
carves the complex and gives its contents a geometry.  Frontoparietal
regions matter not because of where they lie but because they supply the
long-range routing of global accounts \cite{DehaeneChangeux2011}, which on
our reading is the causal routing that wraps same-phase coactivation at
every scale at once.

\paragraph{Where the geometry comes from.}

Another element of the holographic picture has a measurable counterpart in our construction. It concerns symmetry. A maximally symmetric geometry may be highly nontrivial, but its symmetry removes intrinsic differentiation between points. The corresponding concept in graph theory is the automorphism group, which plays the role of the isometry group. Automorphisms preserve distance. Vertices belonging to the same orbit have identical distance profiles, and no quantity derived from the metric can distinguish them.

The pure tensor product has precisely this kind of symmetry. As Proposition~\ref{prop:wreath} and Corollary~\ref{cor:degeneracy} establish, the \(16{,}807\) vertices of $S^{\otimes 5}$ occupy only \(126\) metrically distinct positions. This is exactly the orbit count of the wreath group, and the bound is attained at every power checked. The diameter is \(4\) at every scale, including the seed (Proposition~\ref{prop:diameter-flat}). This property belongs to pure tensor powering, as in Kronecker graphs \cite{LeskovecKronecker2010}. For any connected non-bipartite seed, the diameter of every unedited power is bounded by \(\Lambda(S)\), independently of scale. For our seed, \(\operatorname{diam}(S)=\Lambda(S)\), so the diameter remains constant. In short, position in the pure power carries no metric information beyond its orbit.

This bound does not apply to the pruned and edited graphs compared with the connectome. Their diameters are determined by the edits. The final quotient has diameter \(4\), while Budapest has diameter \(5\) (Table~\ref{tab:nine-criteria-values}). The edits that break the tensor-product form also break its inherited symmetry, which Proposition~\ref{prop:breaking} shows is necessary for metric differentiation. In the final relocation substrate, the \(16{,}807\) vertices occupy \(15{,}505\) automorphism orbits, corresponding to \(92.25\%\) of the vertex count. The final \(1{,}015\)-node quotient is fully asymmetric, with one orbit per vertex. Thus, the construction strongly differentiates positions while preserving the proper three-coloring of the fine substrate.

What symmetry breaking makes possible is metric differentiation. Regions that were previously indistinguishable copies of one another become distinct. The graph already possesses a metric geometry; symmetry breaking makes its positions increasingly distinguishable. Our analogy with holography is limited to this structural correspondence. We have not constructed a dictionary between the two descriptions, an area law, or an analogue of the entanglement--geometry relation.

We should now make explicit where experience stands in this analogy. Experience occupies the position of the bulk. In holography, the bulk is not an additional ingredient introduced into the boundary theory. It is entirely encoded in boundary data, and this encoding has two distinct aspects. The boundary theory, through its operators and Hamiltonian, determines which bulk geometries are possible and the boundary causal structure that every bulk must respect: under the null energy condition, no signal through the bulk outruns a boundary light ray \cite{GaoWald2000}. The boundary state determines which geometry is realized. Different states of the same boundary theory are dual to different bulk geometries, encoded in the entanglement structure of the corresponding states \cite{RyuTakayanagi2006,VanRaamsdonk2010}. The matter fields inhabiting the bulk are encoded in boundary operators.

We propose the same division here. The graph and its evolution operator play the role of the boundary theory. They determine the admissible causal order and the space of geometries that the substrate can support. The supernode hierarchy is part of this structure. It describes what is possible, not what is experienced. Experience corresponds to the bulk of a particular activity state. Its geometry is realized by the causal history of the activity and by the end state of each color phase as it is carried into the next. The contents arranged within this geometry are represented by the values attached to events, playing the role of bulk matter. Experience is therefore entirely determined by the causal--correlational event substrate. It introduces no independent degrees of freedom, but changes as the activity state changes.

Neural firing patterns by themselves do not carry a geometry of experience. That geometry is their collective description. The clustering required for this description is determined by what the system itself can distinguish, not by a partition imposed from outside. The non-causal, same-phase part of the dynamics is also present in experience, but as black-box content. It is carried by the values attached to events, whose internal construction is not separately visible. These values are subsequently modified by the causal interactions that carry them from one phase to another.

Several empirical findings support a particular claim of our framework: causal routing and local coactivation are both necessary, and consciousness requires them to operate together. Under anesthesia, the structural connectome is unchanged, but local dynamics become more homogeneous and more strongly constrained by structure \cite{LopezGonzalez2021}. Conversely, damage to long-range white matter may leave local perceptual detail intact while disrupting the organization of a scene (Section~\ref{subsec:cognitive-geometry}).

Perturbation studies provide evidence for the same two-sided requirement. During non-rapid-eye-movement sleep, cortical stimulation produces activity that remains local and fails to propagate to distant regions \cite{Massimini2005}. The complexity of the evoked response is high only when activity spreads widely while remaining differentiated. This measure tracks the level of consciousness across wakefulness, sleep, anesthesia, and brain injury \cite{Casali2013}.

The response can lose complexity in two different ways. Under propofol, it is weak and local. Under xenon, it is strong but global and stereotyped. In both cases, subjects report no experience upon waking. Under ketamine, however, subjects may be behaviorally unresponsive while the evoked response remains complex, and they report vivid dreams \cite{Sarasso2015}. Consciousness is therefore lost both when activity cannot propagate and when propagation occurs without differentiation. In our construction, the first corresponds to a failure of causal routing, preventing a complex from being carved across regions. The second corresponds to a failure of differentiation, leaving no distinguishable events to carve. The ketamine case further suggests that experience follows the organization of the state, not necessarily responsiveness to the external world.

Experience exists as an organized bulk only in the regime made available by the construction: after symmetry breaking has allowed positions to be individuated (Proposition~\ref{prop:breaking}) and where concentration provides effective locality. We have not constructed the map from activity to realized cognitive geometry, which would correspond to the holographic dictionary. Nor do we claim that the cognitive bulk has a quantum character.

We can now summarize the extent of the analogy by separating what has been constructed from what remains open. In both settings, an admissible causal structure is specified in advance. In holography, this is the boundary light-cone structure; here it is the order determined by the evolution operator. In both settings, the realized causal structure belongs to the state.

Coarse-graining introduces a hierarchy of scale. In AdS/CFT, the boundary spacetime coordinates are rescaled, while the renormalization-group scale becomes an additional radial coordinate in the bulk. The bulk metric, reconstructed from the state, then determines the complete spacetime geometry and its causal structure. In our construction, the evolution operator determines the admissible causal order, while successive supernode formation provides a hierarchy of spatial scales on the same-phase sheets.

We have therefore constructed a multiscale causal--spatial organization of what the substrate permits. What remains open is the map from the activity state to a realized cognitive spacetime geometry, including its realized causal order (Section~\ref{subsec:cognitive-geometry}). Table~\ref{tab:done-and-left} summarizes this distinction. Whether temporal order itself can emerge from a deeper relational structure also remains an open question in our framework (see Remark~\ref{rem:time-emergence}).

Four notions of time occur in this discussion: causal time, physical time, subjective time, and boundary time. They should not be identified with one another. Their respective roles and correspondences are summarized in Table~\ref{tab:done-and-left}.

\paragraph{Clock, drive, and effective description.}

The analogue of holographic boundary time is the brain's own finest temporal resolution, represented by the local color-to-color transitions. These transitions need not be globally synchronized. The large-scale sweep \(1\to2\to3\) is not imposed by an external clock. It emerges when coupled neural populations become partially synchronized.

Once this synchronization develops, the sweep acts as a drive. Within each sweep, the dominant activity moves from one phase class to the next, so that the effective generator changes with phase. The net evolution over a sweep is then dominated by the ordered product \(U_{\rm cycle}\) introduced in Section~\ref{sec:dynamical-realization}. Concurrent activity in the other classes contributes the remaining part. The resulting operator plays a role analogous to the period map of a periodically driven system in Floquet theory, where a time-dependent drive is reduced to a fixed evolution operator over one period.

In our framework, however, this reduction is only approximate, for four reasons.

First, the emergent sweep is not strictly periodic. The corresponding period map therefore changes with time.

Second, external inputs, neuromodulation, and the global context \(\Gamma_t\) act as time-dependent sources. In the holographic analogy, these correspond to couplings of the boundary theory that vary in time.

Third, experience consists of a subset of realized events, carved out by closure after the fact. Its effective dynamics is reduced in two different ways. Everything outside the carved complex acts as its environment. At the same time, the internal dynamics of each coactivation is integrated out, leaving only the collective state of the packet and the effective memory and noise inherited from those dynamics (Section~\ref{subsec:coactivation-attention}). Neither reduction gives the complex a fixed evolution operator of its own.

Fourth, the operator itself changes. Synaptic weights are modified by activity over timescales ranging from short-term plasticity to long-term learning. Only the anatomical graph, which determines the admissible relations, remains approximately fixed over short timescales.

Time-dependent sources can play a similar role in holography, where changing couplings may switch parts of the boundary Hamiltonian on or off. Whether a coupling is an external source or part of the system depends on how the system is divided, just as the distinction between events and parameters depends on the chosen complex. A time-dependent operator for a subsystem may be part of a fixed operator governing a larger system, when viewed from within that subsystem.

Holographic models of black-hole evaporation provide an example. The boundary theory is coupled to an external auxiliary system acting as a heat sink. Once the coupling is established, the combined system evolves under a fixed Hamiltonian, while the holographic subsystem alone evolves as an open system \cite{AlmheiriEngelhardtMarolfMaxfield2019}.

In the brain, the processes that modify the couplings are largely contained within the same network and respond to its activity with memory. We could restore a fixed operator on an enlarged state space by treating synaptic weights as slow dynamical variables. But this would require including the substrate itself as part of the state.

\begin{table}[!htbp]
\centering
\footnotesize
\setlength{\tabcolsep}{4pt}
\renewcommand{\arraystretch}{1.05}
\begin{tabularx}{\textwidth}{>{\raggedright\arraybackslash}p{1.9cm} X X >{\raggedright\arraybackslash}p{2.0cm}}
\hline
 & \textbf{Present construction} & \textbf{Holography} & \textbf{Status here} \\
\hline
Distance &
Hop metric, made informative by symmetry breaking
(Props.~\ref{prop:diameter-flat},~\ref{prop:breaking}); cognitive geometry
to be built from it (Section~\ref{subsec:cognitive-geometry}) &
Emergent: bulk geometry derived from boundary data &
Primitive in hand; construction open \\[1pt]
Mass-like concentration &
Supernode population and degree concentration
(Section~\ref{sec:full-parent-and-budapest}); a property of the graph,
i.e.\ of what is permitted &
Bulk mass and energy, set by the state (a black hole in a thermal state) &
Defined for the graph; state-dependent counterpart open \\[1pt]
Admissible causal order &
Fixed by the evolution operator: who may influence whom, in which phase
order, within a strict light cone &
Boundary light cones and microcausality, fixed by the boundary theory &
Given \\[1pt]
Realized causal order &
Selected by the activity: the order traced by the end states of successive
color phases along the paths actually active &
Bulk causal structure, from the bulk metric of the state; consistent with
boundary causality &
Open; requires the geometry of the state \\[1pt]
Subjective time &
Cognitive construct layered on causal steps
(Section~\ref{subsec:cognitive-geometry}) &
No direct counterpart; bulk proper time is geometric rather than
subjective &
Open \\[1pt]
Physical time &
External embedding; assigns causal steps their duration
(a nonrelativistic idealization) &
No external time beyond the boundary theory's own time coordinate &
Disanalogy, stated \\[1pt]
Dictionary &
Absent: no map from activity state to realized geometry; one constraint
from lesion data (Section~\ref{subsec:cognitive-geometry}) &
Developed but regime-dependent dictionaries; no complete general
reconstruction map &
Left to future work \\
\hline
\end{tabularx}
\caption{Geometry: what the construction has done and what is left, set
against the holographic setting. In both, an admissible causal structure is
given in advance (boundary light cones there, the order fixed by the
evolution operator here), the realized causal structure belongs to the
state, and coarse-graining adds a hierarchy of scale. The map from the
activity state to the realized geometry, and with it the realized causal
order, remains open (Section~\ref{subsec:cognitive-geometry}); whether the
derived/inherited division is final is left open
(Remark~\ref{rem:time-emergence}).}
\label{tab:done-and-left}
\end{table}

The roles of dynamics and coarse-graining should therefore be separated.
The causal direction is already present before coarse-graining: the directed
maps between color classes order events, and their composition supplies the
primitive temporal structure of the substrate. Supernode formation does not
generate that order and does not alter its primitive steps; it changes the
resolution at which those steps are organized.

\medskip
\noindent
What the hierarchy generates is instead a scale structure on the same-phase
sheets. At the fine level a color class contains many simultaneously available
event sites. Clustering collects subsets of those sites, through their shared
causal environments, into effective regions; clustering the resulting
supernodes collects those regions again, and the process may be iterated.
Following one color class through this hierarchy therefore gives a succession
of progressively coarser spatial descriptions,
\[
C_i^{(0)}
\longrightarrow
C_i^{(1)}
\longrightarrow
C_i^{(2)}
\longrightarrow\cdots ,
\]
in which increasingly large portions of the same-phase sheet are represented
as single effective regions. The hierarchy is thus a spatial scale direction
generated by coarse-graining.

\medskip
\noindent
The causal maps remain present at every level. Coarse regions on one
same-phase sheet are related dynamically to coarse regions on the next, so a
level of the hierarchy is not merely a coarse space: it is a coarse
description of the same causal spacetime. The temporal ordering is inherited
from the dynamics, while spatial organization is progressively exposed by
the supernode hierarchy. Their combination yields the multiscale spacetime
structure of the construction.

The construction therefore assigns two different roles to causal structure. Causal order supplies the temporal direction directly: directed transitions between color classes determine which events can succeed which others. Spatial organization, by contrast, is obtained indirectly from the same causal substrate. Same-phase events are grouped according to the density and overlap of their cross-color causal environments, and iterating this clustering produces supernodes, supernodes of supernodes, and progressively coarser descriptions of each color sheet. The resulting hierarchy defines a spatial scale direction without presupposing a spatial metric or adjacency relation among same-phase events. In this sense, causality supplies time directly and space indirectly: temporal order comes from directed propagation, while effective spatial proximity and scale emerge from the multilevel organization of shared causal connectivity.

The relevant comparison is therefore not between the graph and a particular state of the universe, but between the graph and the operator that determines what states and relations can arise. In a quantum theory the Hamiltonian fixes the dynamical architecture: it determines which degrees of freedom can interact, how influence propagates, and which patterns of entanglement can be generated, while the actual state determines which of those possibilities are realized. The same division appears here. The graph is the cognitive operator: its structure encodes the admissible causal relations and the architecture within which entanglement can arise, and it does so in a way that makes the two mutually compatible. The realized neural state supplies the particular content. In the recurrent implementation the corresponding dynamical operator is the composition of the three phase operations, so causal succession, possible entanglement structure, and the states they act upon belong to one composite dynamics rather than to independent constructions.

The analogy with physical dynamics nevertheless has an important asymmetry. The patterns realized by a brain are far more strongly constrained in advance than arbitrary patterns available to a physical system. Neural architecture selects bounded regions in which structured complexity can persist, recur, and become exceptionally information-dense. This boundedness is not incidental: it restricts the degrees of freedom that can participate jointly and thereby helps define the effective operator acting on them. A cognitive object is therefore not merely an unusually complicated state. It is a region of persistent structured complexity whose boundaries constrain the dynamics capable of maintaining it. In this sense the brain does not reproduce the unrestricted relational possibilities of physics; it carves out bounded, recurrent, information-dense sectors of them and subjects those sectors to a much more tightly organized operator structure.

Causal clustering acts on relations that extend across phase classes and
therefore groups structure in both causal and same-phase directions. Its
immediate products are consequently spacetime regions rather than purely
spatial blocks. The spatial hierarchy appears when this multiscale structure
is viewed at a fixed phase: intersecting successive levels of supernodes with
a single color class yields progressively coarser partitions of the
same-phase sheet. Thus the clustering is intrinsically spacetime-like, while
its fixed-phase sections expose the emergent spatial organization. Primitive
temporal order is inherited from the directed dynamics; spatial scale becomes
visible through the repeated coarse-graining of the causal structure.

A single color sheet at one level of the hierarchy is therefore not itself
the radial direction; it is the spatial section at that scale. The radial-like
direction is obtained by following that same-phase sheet through successive
levels of coarse-graining,
\[
C_i^{(0)}
\longrightarrow
C_i^{(1)}
\longrightarrow
C_i^{(2)}
\longrightarrow\cdots .
\]
Each step represents the same sheet at a coarser resolution, so the hierarchy
of sheets, rather than any one sheet, supplies the emergent scale direction.
In this sense the color sheet supplies space at a given scale, while its
trajectory through the supernode hierarchy supplies the radial-like direction
of the construction. This parallels tensor-network models of holography, in which the successive coarse-graining layers of an entanglement-renormalization network play the role of the radial direction \cite{Swingle2012}.

Any completed cognitive geometry should ultimately be uniquely fixed, or at least tightly constrained, by the underlying causal--correlational structure. That requirement applies not only to the final metric or geometric description but also to the coarse-graining procedure itself: the admissible clustering cannot be arbitrary if different clustering methods would produce incompatible geometries from the same substrate. The present construction should therefore be understood as an approximate proposal for the relevant coarse-graining, chosen to expose the causal, correlational, and multiscale structure that a more complete theory would have to recover uniquely.

The construction sits between AdS/CFT and causal-set theory in a useful sense. Like the boundary side of AdS/CFT, it is formulated operator-first: the graph acts as an evolution operator, and iterating it determines how states evolve. As there, the operator fixes an admissible causal order in advance (which units may influence which, in which phase order, within a strict light cone), much as a boundary theory fixes its own light cones. The causal order of experience is not primitive in either setting. It is realized by the activity: at any moment only some admissible paths are active, and the order of experience is the one traced by the end states of successive color phases along those paths. Like bulk causality in AdS/CFT, it belongs to the state, must respect the admissible order, and is not yet reconstructed here. What the construction shares with causal-set theory is therefore not a primitive order but a discrete event structure: once realized, the order of events is part of the physical record, with values attached to the events.

The difference from causal-set theory is, moreover, one of kind.  A causal
set is meant to describe the whole of reality, with nothing outside it and
no inputs or outputs.  The present construction is an information system
embedded in a world it does not contain: it is driven by inputs, must commit
to outputs within the physical time of that world, and its geometry, insofar
as it exists, must represent what the inputs are about and serve what the
outputs must do.  The quantities attached to events differ as well.  In
causal-set theory \cite{BombelliLeeMeyerSorkin1987} the basic quantity beyond order is a count: the number of
elements in a region gives its volume, so that geometry is to follow from
order and counting, and a count is in effect a single intensity.  The brain,
by contrast, communicates through explicit packets of activity whose
amplitude, timing, and route through the substrate carry the information
(Section~\ref{subsec:coactivation-attention})
\cite{Muller2018TravelingWaves,Luczak2015Packets}.  The events of the
present construction are accordingly interaction events (the arrival,
combination, and onward transfer of such packets) rather than points to be
counted.  Geometry therefore cannot be expected to follow from causal order
and counting alone; it must be constructed from the realized interaction
events together with the packets they carry, under the constraints imposed
by the system's function.

Clustering the admissible causal structure generates the coarse spatial organization of the same-phase sheets, so the space of possible geometries is derived from, and remains compatible with, the operator that generates the dynamics; which geometry is realized depends on the state. In this sense the framework combines an operator-first dynamics with an event-based relational foundation in which causal order is realized rather than posited.

At the coarse level, the contents of a same-phase sheet are best regarded as an information field rather than as experience already completed. The sheet specifies which event packets and relations are jointly available at one causal phase, while the directed dynamics carries that field into subsequent sheets, transforming it and incorporating newly generated events. Experience is constituted through this causal stacking: successive same-phase information fields become parts of a temporally extended event complex. The internal fine-scale construction of each packet need not remain accessible at this level; what matters is the coarse field it contributes and how that field is transformed by later causal evolution. Spatial organization is therefore expressed within each sheet, while experiential duration and continuity arise from the causal succession and accumulation of those sheets.

\subsection{Toward Cognitive Geometry: Events, Objects, and State}
\label{subsec:cognitive-geometry}

The position we have been led to should be stated as the sequence of steps
that forced it. Same-phase coactivation need not require anatomical
adjacency, but in the present framework it requires an effective relation of
shared causal context. We therefore derive a graph-theoretic proximity from
causal neighborhoods and clustering. This provides a candidate scaffold for
cognitive geometry rather than the completed geometry itself. Causal
distance therefore plays a primitive role in the construction; under the
bounded-integration assumption used here, events separated by sufficiently
great causal depth need not participate in the same bounded cognitive object
at a given scale, while the same causal organization generates the
supernodes and hence a second, coarser notion of proximity.

There is a further layer of structure not contained in distance alone.
Each realized event $e$ carries a real-valued state
\[
x_e\in\mathbb{R}^{d_e},
\]
but the cognitive state is not merely the flat collection of these values.
Information is also carried by the causal relations among the events: their
ordering, branching, convergence, and distance. The fixed graph therefore
defines a \emph{potential geometry}, while neural activity selects which
events and relations are actually realized within it. A cognitive object may
be written schematically as
\[
O_\alpha=
\left(C_\alpha,R_\alpha,\{x_e:e\in C_\alpha\}\right),
\]
where $C_\alpha$ is a dynamically realized event cluster and $R_\alpha$ its
realized causal structure. Such objects may overlap and occur at multiple
scales. Their values and relational organization can be propagated into
memory, action, report, or further computation.

In the present construction, the realized state is richer than a collection
of values attached to events. It also contains the realized causal relations
among those events, and it is from this combined structure that cognitive
spacetime geometry must be constructed. This differs from the holographic
setting. In AdS/CFT the state carries the correlations and entanglement from
which bulk geometry emerges, but the possible relations among the
underlying degrees of freedom are supplied separately by the boundary theory,
most explicitly through its interaction Hamiltonian and operator structure.
Thus neither construction obtains geometry from a bare vector of values. In
our case, however, the relational structure is itself part of the realized
event state:
\[
\mathcal S=\left(E,R,\{x_e:e\in E\}\right),
\]
so the primitive object already contains both information and its realized
causal organization.

An external readout could therefore use not only the values attached to the
events and relations among those values, but also information about how the
events are causally connected as they occur.

In this respect the construction is closer to placing dynamics on a causal
set than to ordinary quantum mechanics on a fixed background, with one
difference: the admissible relations are fixed in advance by the graph, as
the possible relations of a boundary theory are fixed by its operators,
while the realized causal relations are selected by the activity and become
part of the physical record, with dynamical values attached to the events.

This synthesis arises because the correlational structure is built into our
symmetry-broken causal-event substrate rather than added afterward:
cross-phase paths carry propagation and order, while same-phase patterns
support non-signaling correlation. The resulting system is event-based at
its foundation, like a causal set, but its causal order is realized rather
than primitive, and it is holographic in the way a coarser geometry is meant
to emerge from the organization of those realized causal events.

But the disparity with physics is deeper.
First, our construction is irregular by necessity, not by
accident: a {\it patterned} departure from symmetry (something likely needed
for a physics program) would merely trade one automorphism group for a
smaller one, and whatever regularity survived would reduce the capacity for
vertexwise differentiation for the same symmetry reason as in the pure
product (Corollary~\ref{cor:degeneracy} and Proposition~\ref{prop:breaking}). The capacity of the graph
lives precisely in the irregularity of its differentiation. Second, graph
entanglement is robust, and this robustness should be read as expressive
freedom: because the symmetric background is maximally sensitive (a
handful of edits suffices to collapse the orbit structure almost entirely),
the edit process can encode a broad class of differentiated correlational
structures upon it. Third, the native operations of this substrate include
gating, pruning, and commitment, which are generally non-invertible at the
effective level. This contrasts with closed-system unitary quantum
evolution, which is reversible and norm-preserving. The contrast should not
be taken to imply that reversible or unitary systems cannot implement
decision procedures; rather, the present architecture treats effective
irreversibility and selective commitment as primitive computational
operations.

The correlational layer may be approximately information preserving: same-phase
coactivation plausibly redistributes structure rather than destroying
it, in the manner of an invertible mixing representation, though
whether it does so without loss is not settled from the architecture
alone.

Physics is a ``coherent'' theory in a sense in which the event networks
introduced here are not. This was apparent from the start: the
operations that break the product form and its symmetry, unlike vector addition in a
tensor-product Hilbert space whose algebraic scope is sharply defined,
include a broad class of pruning operations and chromatically admissible edge additions.
Likewise, both the correlational and causal dynamics are supplied by
learnable functions rather than derived from a common physical law. The
construction therefore has no reason to reproduce specifically quantum
phenomena such as the double-slit experiment or violations of Bell
inequalities.

\begin{figure}[t]
\centering
\includegraphics[width=0.95\textwidth]
{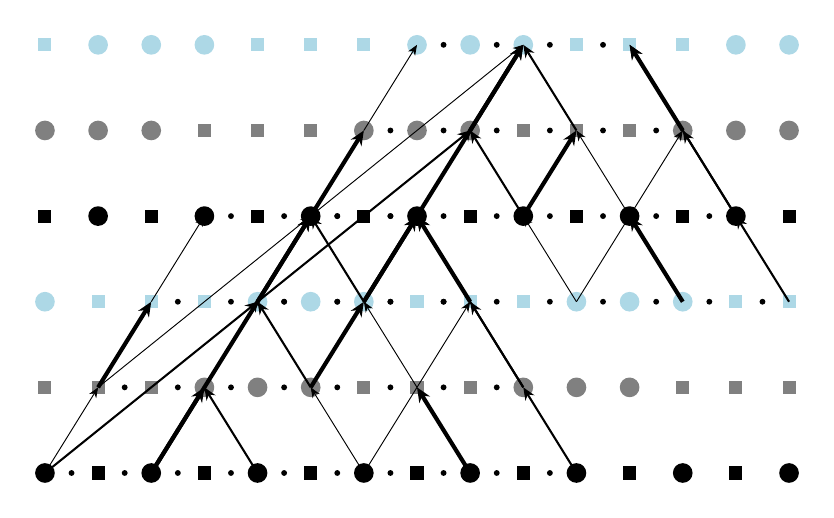}
\caption{Schematic multiscale cognitive event geometry. The fixed causal
substrate defines a potential causal structure, while neural activity realizes
particular causal and coactivation patterns within it. Directed arrows denote
causal relations, with line width indicating interaction strength, and the
small same-phase nodes denote coactivation. The overlapping structures
illustrate that cognitive objects may share constituent events.}
\label{fig:event-geometry}
\end{figure}

Figure~\ref{fig:event-geometry} illustrates this distinction between the fixed
causal substrate and the event geometry dynamically realized within it. The
large nodes represent possible event sites, the arrows represent realized
causal relations, and the small same-phase nodes, arranged horizontally, represent coactivation.
The two overlaid structures share events, emphasizing that cognitive objects
need not form a disjoint partition but may overlap within a larger
multiscale state. Clusters are spacetime {\it columns} nested together, not flat states.

Causal clustering was originally used to define the domains of same-phase
interaction, but what emerged afterward is that correlation itself acquires
a geometric organization. Same-phase coactivation provides a fine-scale
notion of realized proximity, while shared causal environments, supernodes,
and their quotient provide its coarse organization. The two scales are
dynamically related and arise from the same causal substrate.

This forces a question the present work cannot fully answer. When we look at
a scene there is a cognitive geometry: things are arranged near and far,
beside and behind, and they have meaning. If the framework is right, that geometry must arise from this event structure as a cognitive field defined over it. Here the
holographic analogy ceases to be constructive. AdS/CFT provides several
controlled boundary--bulk dictionaries and reconstruction procedures,
whereas the present model lacks an explicit map from its event structure to
experienced geometry. What is missing is therefore not the relational ground
but the map from this ground to the geometry of a scene.

One property of that map can nevertheless be read off the clinical finding
\cite{Chechlacz2012,Neitzel2016}. Anomalies of white matter (the brain's
long-range, inter-cluster wiring and the physical carrier of coarse
organization) can distort high-level scene organization while leaving
detail perception intact: the elements are seen, their arrangement is not.
These findings are consistent with a partial dissociation between coarse
scene organization and local perceptual detail. They support, but do not by
themselves establish, the identification of coarse organization with
long-range structure and fine organization with local coactivation. Moreover, subjective experience supports
several simultaneous notions of distance, including physical, object-level,
and semantic proximity, so cognitive distance need not be represented by a
single scalar metric.

The complexity of the causal event lattice already permits such distinct
notions of distance to emerge from one structure. Causal distance is measured
along ordered chains and records how many causal transitions separate two
events. These transitions may differ in kind and strength while still functioning
as repeated primitive causal operations in the model.

Among concurrent events, a different proximity may be derived from
their higher-order position in the lattice: for example, overlap of causal
neighborhoods, convergence of histories, or participation in the same
recurrent or coactivation structure. Two events may therefore be far apart
under one effective metric yet close under an object-level or semantic one.
The correlation structure of coactivations can be propagated along causal paths and survive them, but
it might not be perceived as proximity in the metrics tied to those paths.

Whatever the map from this ground to experienced geometry turns out to be, it
must respect this multiscale organization. Cognitive space and cognitive time
also cannot be independent in this framework. Both arise from the same causal
event structure: causal transitions provide primitive temporal order, while
distance, clustering, and effective locality are derived from relations among
those same events. A solution for cognitive space must therefore constrain
cognitive time, and conversely.

Because causal trajectories are dynamically realized and may terminate, the
existence of an extended cognitive object cannot be inferred from its early
causal fragments. The object exists only when the causal information required
for its constitution has reached closure; conditional on such closure, its
earliest possible existence is determined by the \emph{longest} required
causal path. Thus, if its temporal structure contains sequences of the form
\[
E_0\to E_1\to\cdots\to E_T,
\]
under the unit-step timing used here, and for branches initiated from the
same reference phase, its earliest possible completion is determined by the
required branch of greatest causal depth. If transition latencies differ, the
corresponding quantity is the longest weighted causal path.

Decision formation therefore has a natural critical-path structure. If
physical latency is to be minimized, the relevant optimization is a min--max
problem: minimize the completion time of the longest required causal branch.
Concurrency serves precisely this role by moving mutually independent
computations into the same execution phases, flattening the completion
frontier and reducing the causal depth required for a decision.

Information about relations in the sensory input does not become part of cognitive geometry merely by being represented somewhere in the nervous system. When a coactivation process comes into contact with processed sensory signals, the correlations already present in those signals constrain the joint dynamics of the participating events: they alter how the coactivation operator acts and therefore become part of the relational structure realized by that activity. In this sense, an input correlation becomes cognitively meaningful only when it is converted into a dynamical property of an event complex. Alternatively, the relation may be established causally, when information from one event structure propagates into the formation of another. If neither coactivation nor causal propagation links two processed features, then even if those features are strongly correlated in the external scene, that correlation never becomes part of the realized cognitive object. Cognitive geometry is therefore constituted not by correlations merely encoded in neural values, but by correlations that enter the dynamics and thereby help define how events cohere into objects. And importantly, the realized causal and correlational history by which cognitive events became related is itself part of the cognitive structure. What appears phenomenologically as a single spatially extensive moment may be understood as a multiscale event structure: dense coactivation supplies local experiential detail, while temporally extended causal relations supply the continuity and organization of the scene.

An ambiguity remains in the formation of cognitive objects: where, in physical time, does subjective experience begin and end? If the highest-level clusters determine experienced objects, then during their physical formation there will generally be partial structures that already exist but have not yet reached the organization of the completed cluster. Why are these partial objects not themselves experienced? One possible answer is that membership in a cognitive object need not be
determined locally at the instant each event occurs. It may instead be
assigned once the relevant causal structure has reached closure. Earlier
events are not altered by later ones, and no future-to-present causation is
required; rather, the completed structure determines retrospectively which
earlier events belong to the object. This picture also offers a way to
reconcile discrete and continuous experience. Cognitive objects can form
distinct yet overlapping clusters with definite completed boundaries, while
the physical realization of successive clusters overlaps in time. One object
may therefore be reaching closure while events belonging to the next are
already forming. Experience can consequently appear continuous through the
overlap of temporally extended but individually bounded event complexes.

Recent recordings sharpen this picture \cite{Hesse2026}. Hesse et al.\ recorded with Neuropixels probes across face patches in macaque inferotemporal and prefrontal cortex during binocular flash suppression. The stimulus was constant, and so was the percept. Yet the same neural populations alternated, roughly fifteen times per second, between micro-epochs in which they carried the consciously perceived image and micro-epochs in which they carried the suppressed one.

Suppose conscious content were simply whatever these populations carry. Then the percept would alternate with them. It does not. In the earlier essay we described the mind as a theater with spotlights, whose audience somehow never perceives the darkness \cite{Ramezani2014CausalEvent}. Here, the same actors step onto the stage and back into the wings again and again, and the audience sees one unbroken scene.

Whatever its final interpretation, the result bears on our framework in three ways. First, participation in conscious content is not a fixed property of a region or of a cell: the same areas, and the same neurons, are included in and excluded from the emergent whole within tens of milliseconds, while activity is present throughout. Inclusion must then be governed by timing, by loops in space and time that determine which events cohere, and not by which units are active. Second, subjective experience then appears as a coordinated geometry of realized events, the event-factuals, embedded among subconscious event-counterfactual alternatives and ongoing inputs, which remain represented and continue to evolve without belonging to the experienced object. Third, the causal structure does more than support this geometry; it times it, carves it out of ongoing activity, activates it and controls it.

On this view, causation and correlation together define a timed global
geometry. We do not yet know its clustering laws, but it is a particular
geometry: it selects a specific subset of the events taking place inside
the brain. The inclusion of time does not mean that transmission itself enters
experience. The events from which the experienced geometry is built are
milestone events: the formation of a packet through coactivation, and
the interaction of packets when information carried from different
sources meets. A packet traveling along a neural pathway, or replicated
on its way, is not itself an event in this sense. This also shows how coactivation and causal relation separate at this level. Coactivation is the joint creation of a packet: a single event, generated together and available to the rest of the network only as a whole. Simultaneity alone does not produce it. Two distant populations in different color classes may each form a packet at the same moment, independently, and each packet may later reach the other. That is not a symmetric exchange but two separate formations, each followed by an interaction when its packet arrives, and the arriving packet is a causal input to a later formation on the receiving side. The two were simply at different points in their local phase sequences: color classes are causal compatibility classes, not globally synchronous populations.
Timing matters because it determines which formations and interactions occur and which events
they join, so which events belong together cannot be read off from a
snapshot of activity, only from how that activity unfolds. This is
distinct from the observational irreducibility introduced earlier, which
concerns the complete state of the system: to write that state down so
that the system could be replicated exactly from a given moment, one
would also need the timing of every packet in transit, even though
transit itself contributes no events to experience. It follows that the highest-level cluster
corresponding to subjective experience is a dynamic subset of events even
within the substrate that generates consciousness, since, as these
recordings suggest, parts that contribute to a conscious event at one
moment can be excluded from it at the next. Even with the limited neural
primitives available in artificial systems, these principles (timed
causal selection, membership by closure, and active but excluded
alternatives) may suffice to produce nontrivial behavior, and
understanding the clustering laws of this geometry would take us a
substantial part of the way.

Finally, coactivations are weighted in both brain and artificial neural networks. 
At some level of weight asymmetry, we might transition to a causal regime. Causal relations
may therefore be embedded in coactivations in a way that our abstract framework does not entirely capture.

\paragraph{
Is AI self-aware?} None of the existing AI-LLM architectures is remotely close to observational irreducibility. We could stop the computation at any moment of inference or generation, read off the registers, save the weights together with every intermediate value, reload them a decade later, run the experiment for a few more seconds, and repeat until we reach the result. Spread over a century, the run would give  the same outputs as an uninterrupted one. Even if these machines passed the Turing test, they would, according to the ideas developed here, still be as conscious as a rock. In fact, as a thought experiment, we could transcribe the data into cuneiform carvings on the walls of a cave and continue the process by reading from them; it would be as self-aware as the wall.

This does not mean that existing computer architectures cannot emulate our minds closely. Event-driven multiscale multi-agent systems working asynchronously could emulate the kind of temporally distributed self-organized autonomy that we have conjectured is essential to experience. Yet these are still register-based, highly controllable processes, and at the fundamental level they are observationally reducible. But because they could emulate our minds, perhaps they could also simulate our will. We might not have to worry about whether they experience discomfort or pain, but we may have to worry about how such multi-agent systems decide to act once they are embodied and capable in the physical world. They might act as if they had wills entirely of their own.

\medskip
This paper suggests what kind of architecture might generate subjective experience. Any step in that direction should be taken with great caution.

\section{Conclusion and Future Work}
\label{sec:conclusion}

We began from a simple computational question: How can a computational
system continue to increase its architectural complexity while preserving
bounded chromatic causal depth?  The motivation came from two observations.
During mammalian evolution, cortical complexity increased primarily through
expansion of the cortical sheet rather than through a proportional increase
in cortical thickness, while characteristic perceptual timescales remained
comparatively stable.  Together these observations motivate the possibility
that increasing computational capacity in higher-level brain processes may
have been achieved primarily through greater concurrency rather than through
a continually increasing number of sequential processing stages.

This led to a graph-theoretic construction based on repeated tensor-product
growth together with chromatically constrained edits that break the product
form and its symmetry.  Tensor-product growth provides a mechanism for
compositional expansion, while the subsequent constrained edge modifications
increase structural complexity without increasing the chromatic causal depth:
the minimum number of causal phase classes required per recurrent sweep
remains three.  Dependencies may nevertheless extend across repeated sweeps,
so this bound is not a bound on the total causal depth of a computation.  The
resulting graph is interpreted as a sparse recurrent computational substrate,
not as an anatomical reconstruction.

The construction also supplies the basis for a causal--correlational
description.  The graph acts as an evolution operator that fixes the
admissible causal order: which units may influence which, and in which phase
order.  Activity realizes a web of events within it, ordered through the
cross-phase paths actually active, together with same-phase correlations that
introduce no additional causal ordering.  We conjecture that subjective
experience stands to this realized web in a relation structurally analogous
to that between bulk and boundary descriptions in holography, with the graph
in the role of the boundary theory and the activity in the role of the
boundary state.  On this account, experience is not an additional ingredient
placed beside the causal web, but an alternative description of organization
already encoded within it, and it changes as the state changes.  The analogy
becomes nontrivial only after the inherited tensor-product symmetry has been
broken: as Proposition~\ref{prop:breaking} shows, an extensively individuated
metric structure requires breaking the symmetry of the pure power, while the
resulting concentrations and coarse organization supply the basis for
effective locality.  What the analogy still lacks is precisely what
holography possesses: a dictionary mapping the realized state of the
underlying relational structure to the emergent geometry of experience, and
with it the causal order of experience itself.

\subsection{Future Work}
\label{subsec:future-work}

A first direction is a systematic search over the parameters of the
construction, including the chromatic constraint itself.  The present work
uses
\[
D_{\chi}(G)=\chi(G)=3
\]
as a bounded chromatic causal-depth prototype.  Preliminary experiments with
four and five colors did not produce comparably strong results, although these
trials were exploratory rather than systematic.  Larger chromatic numbers
remain of interest.  For example, graphs with six, nine, or more causal phase
classes might admit useful higher-level groupings into a smaller number of
functional macro-phases.  More generally, different chromatic causal depths,
seed graphs, and hierarchical organizations may produce closer agreement with
the statistical properties of empirical connectomes.

The relevant parameter space is large, combinatorial, and multi-objective.
One could vary the seed, chromatic causal depth, tensor-growth schedule,
pruning and repair rules, bridge formation, multiscale-targeted
sparsification, the community partition used to define the mesoscale
scaffold, the inverse coarse-graining map, the edge-relocation procedure, and
the dynamical realization.  These choices interact nonlinearly: changing one
stage generally alters the conditions presented to the stages that follow.
The present construction should therefore be read as a constructive existence
demonstration and proof of concept rather than as an optimized solution.
Systematic ablation, automated parameter search, and multi-objective
optimization are important next steps.

Graphs for which tensor products strictly decrease chromatic number may also
be especially interesting.  Hedetniemi counterexamples
\cite{shitov2019counterexamples,Wrochna2022} show that there exist factors
whose tensor product requires fewer colors than either factor individually.
In the present language, such a product can reduce chromatic causal depth
under composition rather than merely preserve a bound on it.  This suggests
a distinct computational regime in which composition itself buys additional
concurrency.  There is a loose analogy with settings in quantum computation
where composition and entanglement enlarge the accessible computational
structure and can alter attainable evolution times
\cite{GiovannettiLloydMaccone2003Entanglement,
JozsaLinden2003EntanglementSpeedup}; the analogy is only structural, however,
and no quantum mechanism is assumed here.

The two-level coupled stochastic dynamics introduced in Section~\ref{sec:dynamical-realization} can be extended with jump terms at both levels, with the sparse graph constraining where such
event-like transitions may occur:
\[
dX_t
=
F(X_t,Y_t)\,dt
+
B_X(X_t,Y_t)\,dW_t^{(X)}
+
J_X(X_{t^-},Y_{t^-})\,dN_t^{(X)},
\]
\[
dY_t
=
G(X_t,Y_t)\,dt
+
B_Y(X_t,Y_t)\,dW_t^{(Y)}
+
J_Y(X_{t^-},Y_{t^-})\,dN_t^{(Y)},
\]
where \(W_t^{(X)}\) and \(W_t^{(Y)}\) are independent Wiener processes
and \(N_t^{(X)}\) and \(N_t^{(Y)}\) are counting processes with
state-dependent intensities
\(\lambda_X(X_t,Y_t)\) and \(\lambda_Y(X_t,Y_t)\). Drift and diffusion
describe continuous local dynamics, while the jump terms provide a natural
representation of discrete event-like messages capable of gating,
amplifying, or switching the state at either level.  Their support can be
restricted by the sparse causal graph so that learned dynamical transitions
remain compatible with the structural substrate.

Generator matching
\cite{HolderriethGeneratorMatching2025} provides a natural mathematical
framework for learning such processes.  Writing \(z=(x,y)\) for the joint
state, a flow--diffusion--jump generator takes the form
\[
\begin{aligned}
\mathcal L_t f(z)
&=
\underbrace{
\nabla f(z)^{\!\top}
\begin{pmatrix}
F(z)\\[2pt]
G(z)
\end{pmatrix}
}_{\text{flow}}
+
\underbrace{
\tfrac12\,\nabla^2 f(z)\cdot\sigma_t^2(z)
}_{\text{diffusion}}
\\
&\qquad+
\underbrace{
\int
\bigl[f(z')-f(z)\bigr]\,
Q_t(\mathrm dz',z)
}_{\text{jump}},
\end{aligned}
\]
with
\[
\sigma_t^2
=
\operatorname{diag}
\!\left(
B_XB_X^{\top},
B_YB_Y^{\top}
\right),
\]
and with the jump measure written schematically as
\[
Q_t(\mathrm dz',z)
=
\lambda_t(z)\,
J_t(\mathrm dz',z).
\]

Within the Markov setting addressed by generator matching, this decomposition
provides a general framework encompassing deterministic flow, diffusion, and
jump dynamics.  The coefficients
\[
(F,G,B_X,B_Y,\lambda,J)
\]
could therefore be learned jointly while the graph acts as a structural prior:
its sparse edge set constrains where event-like transitions may occur, whereas
their intensities and state updates remain learned.  This would turn the
present construction from a fixed recurrent substrate into a structured
generative dynamical model and provide a direct setting in which to study how
continuous local dynamics, discrete causal events, and multiscale
coarse-graining interact.

A structural direction is to introduce an explicit spatial embedding.
The present inverse construction solves for a valid fine substrate under a
fixed coarse-graining map, so that the strict quotient lies in the desired
connectome regime, but it does not assign anatomical coordinates to the resulting
supernodes.  Spatial information could therefore be introduced as an
additional constraint on edge relocation, for example through wiring-cost
penalties, distance-dependent connection priors, or explicit geometric
organization of the supernodes.

This may be particularly important for bilateral structure.  The Budapest
Reference Connectome contains a pronounced paired large-scale organization,
whereas the present construction contains no left--right anatomical axis and
does not recover that organization.  This residual difference is visible in
the spectral structure as well as in the community organization.  A natural
extension would therefore build a bilateral coordinate, hemispheric
partition, or corresponding bottleneck directly into the parent substrate
before solving the inverse coarse-graining problem.  Bilateral organization
is unlikely to emerge reliably from relocation alone when no variable in the
construction distinguishes the two sides.

\begin{figure}[htbp]
\centering
\includegraphics[width=\textwidth]{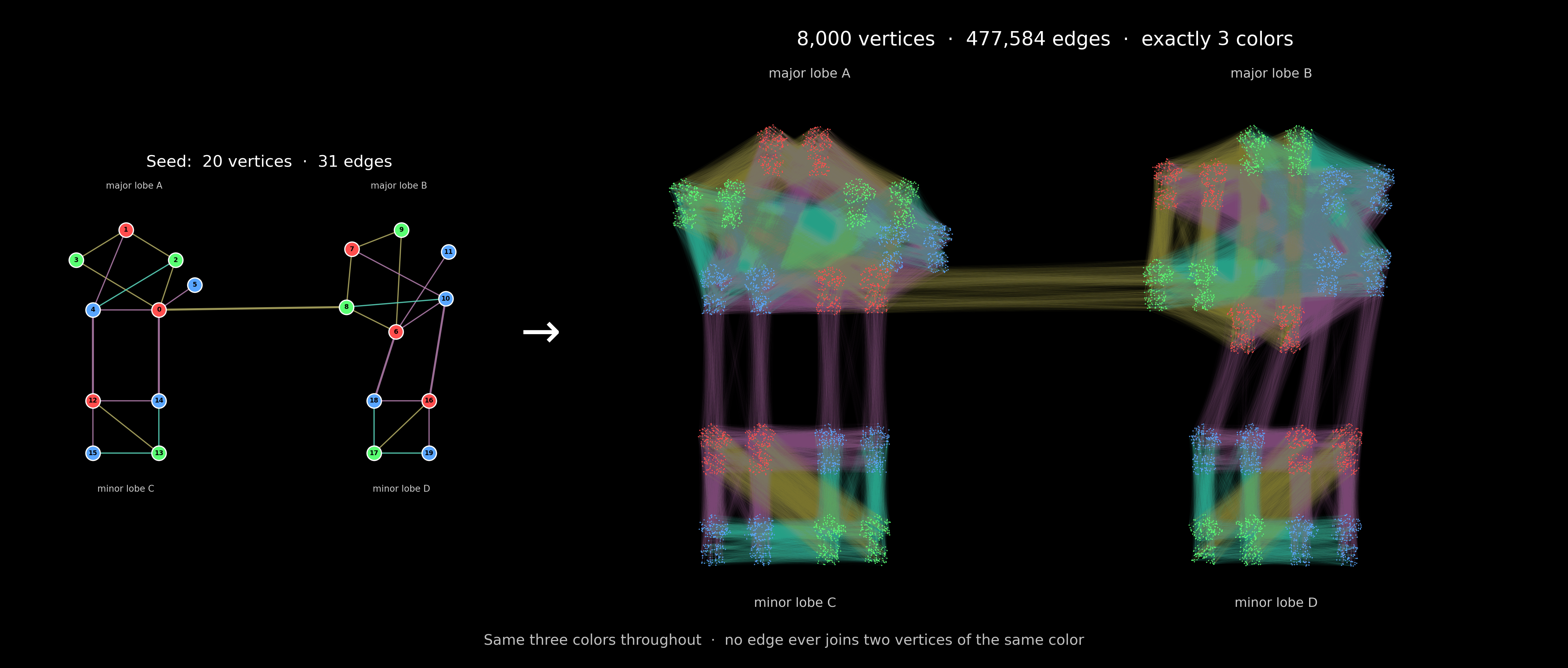}
\caption{A bilaterally designed twenty-vertex seed (left) and the completed
\(8{,}000\)-vertex parent grown from it (right), drawn in the same layout.
Vertex color is the carried three-coloring (\(C_1\) red, \(C_2\) green,
\(C_3\) blue); edge color blends the colors of its two endpoints, and no
edge joins two vertices of the same color. Major lobes \(A\) and \(B\) are
on top, minor lobes \(C\) and \(D\) below, with one hemisphere per column.
Because color and lobe are read off the leading tensor coordinate, each
seed vertex becomes a cloud of \(400\) vertices of its own color, placed at
the seed vertex's position. The single cross-hemisphere seed edge becomes
the \(3{,}844\)-edge bridge between \(A\) and \(B\). Vertex positions are
chosen for display and carry no metric meaning.}
\label{fig:seed-to-brain}
\end{figure}

Such symmetry can be imposed by hand: a twenty-vertex seed carrying a
four-block bilateral pattern, with one hemisphere per side and a single
cross-hemisphere edge, grows into a parent graph whose organization
inherits that pattern (Figure~\ref{fig:seed-to-brain}). There, however, the
bilateral organization is placed in the seed rather than predicted; how it
could emerge automatically from the construction is not yet known.

The dynamics placed on the graph are likewise only toy models of real neural
dynamics.  The recurrent updates, same-phase attention, and
backpropagation-based learning procedure should not be interpreted as literal
biological mechanisms.  Their purpose is more limited: to demonstrate that
the constructed topology can serve as a trainable recurrent substrate rather
than merely reproducing static graph statistics.  More realistic dynamics
could incorporate heterogeneous conduction delays, excitation and
inhibition, synaptic adaptation, oscillatory modulation, and local
population-level dynamics while retaining the same underlying structural
constraints.

A further direction concerns how control over the system's own collective
states could arise.  In biological evolution the lower neural structure came
first, and control over its collective activity is likely to have begun as an
embedded and emergent property of that activity itself: a global context of
the kind introduced in Section~\ref{sec:ads-cft-correspondence}, carried by no
dedicated population.  Only later would such control have become partly
differentiated anatomically, and even then much of it would have remained embedded in
the dynamics it regulates.  The recordings discussed in
Section~\ref{subsec:cognitive-geometry} point in the same direction
\cite{Hesse2026}.  The populations that alternate between carrying the
consciously perceived image and the suppressed alternative belong to the
inferotemporal and prefrontal circuitry associated with conscious content
itself, and what alternates in them is not an input being attached and
detached but higher-level content: the factual alternative and its
counterfactual, both represented at the same level.  The alternation
therefore concerns the formation of the total content of experience, that
is, which of the available contents joins the coherent whole.  That formation
is partly open to control: in binocular rivalry, a closely related case in
which the competition between alternatives becomes perceptually visible,
attention and intention can bias which alternative dominates, although they
cannot fix it at will \cite{MengTong2004,ChongTadinBlake2005}.  Control over the formation of experienced content is
thus at least partly embedded in the same distributed system that
experiences, and from there it can reach back into lower levels, including
perceptual areas.  Architectural evolution may therefore have favored an
organization that remains largely distributed but has become efficient at
creating control messages and carrying them where they are needed, rather
than one that concentrates control in a separate center.  Taken together,
these observations suggest that the emergent geometry that clusters events
into experience is realized by a dynamic subset of the activity of a recurrent substrate
of the kind developed here, and that the mechanisms which construct this
geometry are likely to have evolved as a mixture: partly embedded in the
collective dynamics, as a global context carried by no dedicated population,
and partly as message passing along causal routes that carry control.  The
developmental order matters for a further reason: coupling a meta-level
controller to a lower-level one is not innocuous, and if the coupling is
introduced abruptly, the collective states that the meta-level is meant to
govern can be destabilized, or absorbed into regimes from which they can no
longer be read out.  The sequence can be studied directly on the present
substrate.  Beginning with the simplest repeated tasks, one can sweep the
coupling strengths of the recurrent loops, locate the bifurcations at which
the dynamics changes character, such as the onset of a rotation through the
three phase classes, and examine near them both the sensitivity of the
dynamics and the emergence of persistent, localized collective modes that are
carried from phase to phase in a particle-like way.  Meta-control can then be
introduced in stages, first embedded, as a context computed from the
population state itself, and then with units of its own; each stage can be
compared for where the bifurcations fall, whether the collective modes
survive, whether embedded control can attach such modes to the coherent whole
and detach them again, and whether the coupling drives the system into
regimes in which they are lost.  The natural setting for these experiments is
the one the construction already provides: multiscale, spatially wide,
recurrent and quasi-periodic concurrent loops in which same-phase coactivation
is wrapped by tensored causal paths, with this wrapping present at
every scale at once (Section~\ref{subsec:full-parent-operators}).  We regard
this sequence (embedded emergence followed by partial anatomical
differentiation) as a concrete route for examining, in artificial systems,
the organizational conditions that the conjecture of
Section~\ref{sec:emergent-geometry} associates with experience.

A related empirical direction concerns structure--function coupling.  In an
artificial neural network much of the computation is encoded directly in
learned weights.  In the brain, computation is also strongly constrained by
which pathways physically exist and by how activity is routed through them.
The structural connectome specifies a repertoire of possible interactions,
whereas conduction delays, excitation and inhibition, oscillatory phase,
neuromodulation, and changing network state determine which part of that
repertoire is functionally expressed at a particular time.  The framework
developed here suggests studying functional connectivity as a
state-dependent selection from a structurally constrained set of possible
causal routes.  A direct comparison between the functional dynamics generated
by the constructed parent and empirical structure--function coupling would
therefore provide a substantially stronger test than static topology alone.

The dynamical freedom permitted by the chromatic organization also remains
largely unexplored.  The concrete inpainting implementation uses a fixed
three-phase sweep,
\[
1\longrightarrow2\longrightarrow3,
\]
repeated across recurrent iterations.  This is one admissible realization of
the three causal phase classes, not a requirement imposed by the coloring
itself.  Future architectures could resolve the active dependencies using
different phase orders, different subsets of each color class, or trajectories
that continue across successive recurrent sweeps before returning to the same
region of the graph.  In such systems the coloring would become not only a
bound on chromatic causal depth but also a routing resource: the same
structural substrate could support different computations through different
realized causal trajectories while retaining the same three-class bound per
sweep.

A deeper dynamical question is whether the combination of nested causal routing wrapping 
same-phase coactivation, multiscale organization, and recurrent synchronization
produces distinctive structure in the evolving joint distributions.  The
relevant effects need not require Bell-inequality violation or microscopic
quantum dynamics.  Instead, the architecture may support many differentiated
variables simultaneously while selectively preserving, coupling, and routing
them across recurrent cycles.  Future work should determine whether such
dynamics exhibit unusual information-theoretic properties (for example,
persistent mutual information, high-dimensional separability, higher-order
synergy, or approximately information-preserving transformations) relative
to more conventional static or feed-forward learning architectures. 

If such a regime exists, it may also provide a more precise basis for the
epistemic interpretation developed here.  The causal graph and its dynamics
would constitute the underlying ontic process, while the structured joint
state made available by that process would constitute the system's internal
epistemic reality.  Subjective experience could then be investigated as an
emergent property of a recurrent system whose multiscale coactivation
structure remains compatible with, and helps control, the particular multiscale causal
organization developed here. 

The aim is not to imitate the structure of physical reality, but to identify
properties and dynamical laws that may recur across the emergence of different
kinds of reality.  Such a program may not answer the \emph{hard problem} \cite{Chalmers1996} of consciousness in its traditional form.  It may instead replace it with a
different and potentially harder problem: why do certain recurrent
causal--relational structures generate coherent epistemic realities at all,
whether in the emergence of physical description or in the emergence of
cognitive experience?  On this view, the important object of study is not a
literal correspondence between physics and cognition, but the possibility
that both instantiate more general principles governing how an underlying
ontic process gives rise to an internally accessible world.

\section*{Use of AI Tools}
The author used Claude (Anthropic), a large language model, to help write
the code for the constructions and analyses, to help edit the text, and to
check reported values against the analysis code. Each graph metric in the
analysis code was tested individually, against graphs whose values are known
and against small examples from the construction whose values can be worked
out by hand. The author reviewed all AI-assisted content and takes full
responsibility for the manuscript.

\section*{Code Availability}
The code for the graph constructions, topology tests, and inpainting experiments is
available at \url{https://github.com/poetlifeart/tensormind}. Scripts for
the remaining specialized analyses are available from the author on request.

\appendix

\section{Parent Graphs}
\label{app:global_parent_graph}

This appendix records the fine-scale graphs used at the main stages of the
construction and distinguishes the original integrated parent from the final
relocation substrate.  The local and integrated constructions each produce a
\(16{,}807\)-node graph.  The inverse procedure then changes the placement of
fine edges under a fixed \(1{,}015\)-supernode map.  The final relocation
substrate is the fine graph used directly by the recurrent neural model.  No
additional reconstruction is applied after the final quotient is derived.

\subsection{Local parent graph}

The local construction produces a parent graph with
\[
n=16{,}807,
\qquad
m=277{,}319.
\]
Against \(100\) tripartite degree-preserving  nulls, this graph passes seven of the
nine scored criteria.  It passes clustering enrichment, small-world sigma,
modularity enrichment, local-efficiency enrichment, connector-hub,
rich-club, and multi-resolution modular-stability criteria, while failing
the two near-random integration criteria: path length and global efficiency.
\begin{table}[H]
\centering
\small
\setlength{\tabcolsep}{4pt}
\renewcommand{\arraystretch}{1.12}
\begin{tabularx}{\textwidth}{p{0.38\textwidth}Xc}
\toprule
Criterion & Interpretation for the local parent & Verdict \\
\midrule
Clustering enriched &
Local clustering is enriched relative to tripartite degree-preserving nulls. &
\pass{} \\
Path length near random &
Paths remain too long relative to the corresponding null ensemble. &
\fail{} \\
Small-world sigma &
The graph has an enriched clustering/path-length ratio. &
\pass{} \\
Modularity enriched &
Community structure is enriched relative to nulls. &
\pass{} \\
Global efficiency near random &
Global efficiency is too low relative to the corresponding null ensemble. &
\fail{} \\
Local efficiency enriched &
Local efficiency is enriched relative to nulls. &
\pass{} \\
Connector hubs &
Participation-based connector hubs are present. &
\pass{} \\
Rich-club organization &
A significant rich-club regime is present. &
\pass{} \\
Multi-resolution VI stability &
A stable modular scale is present. &
\pass{} \\
\midrule
\multicolumn{2}{p{0.78\textwidth}}{
\textbf{Score:} \(7/9\).  The failures are precisely the two
near-random integration criteria.} &
\(7/9\) \\
\bottomrule
\end{tabularx}
\caption{
Topology test for the local parent graph.  The fine graph already exhibits
strong local and mesoscale organization, whereas short-path integration
emerges more strongly after coarsening.
}
\label{tab:local-repair-parent-test}
\end{table}

\subsection{Original integrated parent graph}

The integrated parent introduced in
Section~\ref{sec:full-parent-and-budapest} has
\[
n=16{,}807,
\qquad
m=337{,}983,
\qquad
\rho=0.002393,
\]
with color-class sizes
\[
4{,}802/4{,}802/7{,}203.
\]
It retains the proper three-coloring and contains triangles, and therefore has chromatic causal depth
three.

Using \(1{,}000\) tripartite degree-preserving nulls, the integrated parent
passes all nine scored criteria.  Its clustering coefficient is \(0.0366\),
\(2.29\times\) the null expectation; its transitivity is \(0.0317\),
\(2.04\times\) null; and its modularity is \(0.4605\), \(4.23\times\) null.
Its local efficiency is \(0.1208\), \(2.68\times\) null.  The small-world
statistics are
\[
\sigma_{\rm dp}=1.68,
\qquad
\sigma_{\rm ER}=10.24.
\]

The two weakest passes are the near-random integration criteria.  Its average
path length is
\[
L=3.7871,
\]
or \(1.217\times\) the degree-preserving null expectation, while its global
efficiency is
\[
E_{\rm glob}=0.2844,
\]
or \(0.845\times\) the corresponding null value.  Both remain inside the
predefined acceptance range.
\begin{table}[H]
\centering
\small
\setlength{\tabcolsep}{4pt}
\begin{tabular}{p{0.48\linewidth}p{0.39\linewidth}}
\toprule
Criterion & Original integrated parent \\
\midrule
Clustering enriched & \pass{} (\(C=0.0366\), \(2.29\times\) null) \\
Path length near random & \pass{} (\(L=3.7871\), ratio \(1.217\)) \\
Small-world organization & \pass{} (\(\sigma_{\rm dp}=1.68\), \(\sigma_{\rm ER}=10.24\)) \\
Modularity enriched & \pass{} (\(Q=0.4605\), \(4.23\times\) null) \\
Global efficiency near random & \pass{} (\(E_{\rm glob}=0.2844\), ratio \(0.845\)) \\
Local efficiency enriched & \pass{} (\(E_{\rm loc}=0.1208\), \(2.68\times\) null) \\
Connector hubs & \pass{} (\(185\) hubs) \\
Rich-club organization & \pass{} (longest significant run \(290\)) \\
Multi-resolution VI stability & \pass{} (\(1\) stable scale) \\
\midrule
Overall score & \(\mathbf{9/9}\) \\
\bottomrule
\end{tabular}
\caption{
Topology test of the original integrated parent graph against
\(1{,}000\) tripartite degree-preserving nulls.  The graph passes all nine
scored criteria.  Box-covering scaling is descriptive and is not included in
the \(9/9\) score.
}
\label{tab:global_parent_test}
\end{table}

\paragraph{Descriptive multiscale scaling.}
The original integrated parent also exhibits descriptive box-covering
scaling.  Under the test convention the fitted slope is approximately
\[
d_B=4.03,
\qquad
R^2=0.94.
\]
This quantity is not one of the nine scored criteria and is not treated as a
direct Budapest target.  Its role is to characterize the multiscale
organization present at the fine scale.

\subsection{Final relocation substrate}

The final relocation substrate has
\[
n=16{,}807,
\qquad
m=337{,}980.
\]
It therefore differs in total edge count from the original integrated parent
by only three edges, although the edge placement changes substantially.  The
two graphs share \(50.3\%\) of their edges, corresponding to a Jaccard
similarity of \(0.336\).

The relocation substantially strengthens local and mesoscale organization.
Clustering rises from \(0.0366\) to \(0.2281\), transitivity from \(0.0317\)
to \(0.3089\), local efficiency from \(0.1208\) to \(0.3497\), and
modularity from \(0.4605\) to \(0.8104\).  Degree assortativity increases
from \(+0.2542\) to \(+0.5802\).

\begin{table}[H]
\centering
\small
\setlength{\tabcolsep}{5pt}
\begin{tabular}{lcc}
\toprule
Quantity & Original integrated parent & Relocation substrate \\
\midrule
Nodes & \(16{,}807\) & \(16{,}807\) \\
Edges & \(337{,}983\) & \(337{,}980\) \\
Clustering & \(0.0366\) & \(0.2281\) \\
Transitivity & \(0.0317\) & \(0.3089\) \\
Local efficiency & \(0.1208\) & \(0.3497\) \\
Modularity & \(0.4605\) & \(0.8104\) \\
\(\sigma_{\rm dp}\) & \(1.68\) & \(15.37\) \\
\(\sigma_{\rm ER}\) & \(10.24\) & \(83.80\) \\
Assortativity & \(+0.2542\) & \(+0.5802\) \\
Average path length & \(3.7871\) & \(4.5114\) \\
Global efficiency & \(0.2844\) & \(0.2439\) \\
Rich-club longest run & \(290\) & \(215\) \\
Test score & \(9/9\) & \(7/9\) \\
\bottomrule
\end{tabular}
\caption{
Fine-scale structural comparison before and after inverse relocation.
Most segregation and small-world enrichment measures become substantially
stronger after relocation.  The two criteria lost are specifically the
near-random path-length and near-random global-efficiency criteria.
}
\label{tab:parent-relocation-comparison}
\end{table}

Using \(100\) tripartite degree-preserving nulls, and \(1{,}000\) for the
rich-club test, the relocation substrate passes seven of the nine scored
criteria. It fails exactly
\emph{path length near random} and \emph{global efficiency near random}.
This should not be interpreted as a loss of small-world organization:
\[
\sigma_{\rm dp}=15.37
\]
is much larger than for the original parent.  Rather, the stricter
near-random-path criterion asks whether the observed path length remains
within \(1.25\times\) that of a degree-preserving random rewiring, and the
relocation substrate lies outside that tolerance.

The distinction is clearer in absolute terms:
\[
\begin{array}{lc}
\toprule
\text{graph} & \text{average path length}\\
\midrule
\text{degree-preserving null of the substrate} & 3.13\\
\text{original integrated parent} & 3.79\\
\text{\(1.25\times\) threshold for the substrate} & 3.91\\
\text{relocation substrate} & 4.51\\
\text{same-size, same-degree ring lattice} & 210.6\\
\bottomrule
\end{array}
\]
Approximately \(93\%\) of node pairs are separated by six hops or fewer,
and the maximum hop distance is \(19\), compared with \(421\) in
the corresponding lattice.  The two test failures therefore identify a
departure from \emph{near-random} integration rather than intrinsically long
paths.

The relocation substrate also retains descriptive multiscale scaling,
with
\[
d_B=3.625,
\qquad
R^2=0.952.
\]
As above, this measure is descriptive rather than part of the nine scored
criteria.

The recurrent neural model uses this final relocation substrate directly.
The final quotient is subsequently recomputed from it by complete
fine-to-coarse counting under the fixed supernode assignment.

\section{Coarsening Obstruction and Inverse Relocation}
\label{app:inverse-relocation}

\subsection{Forward community quotient and the obstruction}

The forward construction fixes the integrated parent, derives a community
partition, and then collapses it strictly.  Under strict collapse, every
parent edge whose endpoints lie in different supernodes contributes to the
corresponding quotient relation; only intra-supernode edges disappear.  For a
map \(\pi\), the weighted quotient is therefore
\[
W_{ab}
=
\left|
\left\{
\{u,v\}\in E:
\pi(u)=a,
\pi(v)=b
\right\}
\right|,
\qquad a\neq b.
\]

For the original community partition, complete representation produces far
more cross-supernode relations than the selected mesoscale support pattern.
If the unsupported relations are instead removed physically from the fine
substrate, \(101{,}816\) parent edges supporting \(73{,}141\) coarse
relations must be deleted.  The resulting graph has \(236{,}167\) edges,
splits into three connected components, and has minimum degree one.  Thus the
selected mesoscale pattern cannot be treated as a quotient of the unchanged
parent while simultaneously preserving the fine-substrate constraints.

The obstruction is not specific to one partition.  Twenty-five strict
coarsenings were tested, all forced to exactly \(1{,}015\) groups and all
using complete representation.  The families included Leiden RB, Leiden CPM,
balanced METIS, Louvain-plus-METIS size-capped maps, nested two-level maps,
box-covering ball maps, and a random balanced control.  The strongest tested
modularity was
\[
Q=0.288
\]
for Leiden RB at resolution \(40\), still well below Budapest's
\[
Q=0.5574.
\]
The random balanced control gives \(Q=0.040\).

\begin{table}[H]
\centering
\small
\setlength{\tabcolsep}{5pt}
\begin{tabular}{lrrrr}
\toprule
Map & Quotient edges & \(Q\) & \(C\) & APL \\
\midrule
Budapest & \(70{,}654\) & \(0.5574\) & \(0.6696\) & \(2.219\) \\
Leiden RB, res.~40 & \(60{,}513\) & \(0.288\) & \(0.275\) & \(1.882\) \\
Leiden CPM, \(0.05\) & \(13{,}080\) & \(0.217\) & \(0.541\) & \(2.014\) \\
Leiden RB, res.~30 & \(90{,}277\) & \(0.204\) & \(0.330\) & \(1.825\) \\
METIS balanced & \(84{,}729\) & \(0.196\) & \(0.389\) & \(1.863\) \\
Leiden RB, res.~20 & \(141{,}633\) & \(0.124\) & \(0.373\) & \(1.725\) \\
Random balanced & \(234{,}240\) & \(0.040\) & \(0.516\) & \(1.545\) \\
\bottomrule
\end{tabular}
\caption{
Representative members of the \(25\)-map strict-coarsening sweep.  No tested
strict coarsening of the fixed integrated parent reaches the modular regime of
the Budapest Reference Connectome.
}
\label{tab:strict-coarsening-sweep}
\end{table}

Community clustering nevertheless remains essential to the inverse solution.
For the reported inverse construction, the community-derived
\(1{,}015\)-supernode map is held fixed and the fine edge placement is solved
under that map.  The obstruction therefore changes the object being adjusted:
rather than suppressing coarse relations after quotienting, the fine substrate
is modified until strict quotienting produces the desired coarse regime.

\subsection{Reported inverse construction under the fixed map}

Conceptually, an inverse coarse-graining problem could allow both a fine
substrate and a coarse map to vary.  The reported implementation uses the more
restricted construction
\[
(G_{\rm int},\pi)
\longrightarrow
(\widetilde G,\pi)
\xrightarrow{\text{strict collapse}}
H,
\]
where \(G_{\rm int}\) is the original integrated parent and \(\pi\) is fixed.
The quotient is always computed from the parent, never edited directly.

The map \(\pi\) is recomputed from the original integrated parent by Louvain
community detection at resolution
\[
\gamma_{\rm init}=20.0
\]
with seed \(42\), followed by deterministic merging to exactly
\[
N=1{,}015
\]
supernodes.  At each merge, a smallest community \(C_i\) is merged into the
adjacent community \(C_j\) maximizing
\[
\frac{\operatorname{edges}(C_i,C_j)}{|C_i|\,|C_j|}.
\]
If \(C_i\) has no external neighbor, it is merged into the smallest remaining
community.  The resulting member lists are kept fixed through the inverse
stages.

For any fine graph used below, define the complete cross-supernode bundle
\[
E_{ab}
=
\left\{
\{u,v\}\in E:
\{\pi(u),\pi(v)\}=\{a,b\}
\right\},
\qquad a\neq b,
\]
and its multiplicity
\[
W_{ab}=|E_{ab}|.
\]
Every final quotient relation and every final quotient weight are obtained
from these sets by direct counting.

The reported inverse pipeline is
\[
\begin{aligned}
G_{\rm int}
&\xrightarrow{\text{bundle redistribution}} G_B,\\
G_B
&\xrightarrow{\text{mesoscale support refinement}} G_R,\\
G_R
&\xrightarrow{\text{degree-profile adjustment}} G_{\rm final},\\
G_{\rm final}
&\xrightarrow{\text{strict collapse}} H_{\rm final}.
\end{aligned}
\]
The following subsections give the implementation-level rules.
\subsection{Mesoscale target-support scaffold}

The bundle-redistribution stage uses a target set \(\mathcal S\) of
supernode pairs.  This set is constructed from the weighted witness graph of
the original integrated parent under \(\pi\); it is a design scaffold, not an
independently valid quotient of the unchanged parent.

At the coarsest level, unweighted Louvain community detection is applied to
the witness graph at resolution
\[
\gamma_0=0.95
\]
with seed \(42\).  Our run gives four large blocks of sizes
\[
389,
\qquad
265,
\qquad
239,
\qquad
122.
\]
Within these blocks, two finer partitions are computed using
\[
\begin{array}{c|cc}
\text{block} & \gamma_{L1} & \gamma_{L2}\\
\hline
0 & 0.9 & 1.0\\
1 & 0.7 & 1.0\\
2 & 1.0 & 2.0\\
3 & 1.0 & 2.0
\end{array}
\]
with seed \(42\).  Relations lying within the resulting nested blocks are
retained.  For relations crossing the four largest blocks, pairs of
\(L1\)-groups are ranked by total witness weight; the ten strongest such
cross-block patches are retained, and individual relations within those
patches are taken in decreasing witness-weight order toward the coarse edge
budget
\[
M_{\rm target}
=
\left\lfloor\frac{139N}{2}\right\rfloor.
\]
The resulting stored support set contains \(70{,}538\) relations and is used
only as the target support \(\mathcal S\) for the fine-edge redistribution
below.

\subsection{Bundle-weight redistribution}
\label{app:bundle-redistribution}

This stage starts from the original integrated parent and uses NumPy random
seed \(42\).  Let \(E_0\) be the original parent edge count and let
\(\mathcal S\) denote the target support set above.

\paragraph{Removing unsupported cross-supernode relations.}
For every cross-supernode pair \(p=(a,b)\) present in the original parent but
absent from \(\mathcal S\), all fine edges in \(E_{ab}\) are removed from the
working graph.  At this point the operation is a deletion, not a one-for-one
rewiring.  The total parent edge budget is restored only after the target
bundle adjustment and repair operations below.

\paragraph{Rank assignment of Budapest bundle targets.}
Let the target pairs in \(\mathcal S\) be sorted in ascending order of their
original witness multiplicity:
\[
p_1,\ldots,p_M.
\]
From the Budapest \texttt{fiber\_count\_mean} values, the implementation draws
\(M\) values without replacement using seed \(42\), sorts them in ascending
order,
\[
b_{(1)}\leq\cdots\leq b_{(M)},
\]
and assigns them rank-for-rank.  The integer target multiplicity of
\(p_i\) is
\[
t_{p_i}
=
\max\!\left\{1,\operatorname{round}(b_{(i)})\right\}.
\]
This is a distributional match only.  No pair in the generated graph is
identified with the anatomical Budapest edge that supplied a particular
fiber-count value.

\paragraph{Trimming over-weight pairs.}
If the current bundle multiplicity \(w_p\) exceeds its target \(t_p\), each
supporting fine edge \(\{u,v\}\in E_p\) is scored by common-neighbor support
\[
s(u,v)=|\Gamma(u)\cap\Gamma(v)|.
\]
Exactly
\[
w_p-t_p
\]
supporting edges are removed, beginning with the smallest values of
\(s(u,v)\).  Thus the weakest triangle-supported fine edges are trimmed first.

\paragraph{Densifying under-weight pairs.}
If \(w_p<t_p\), the candidate additions are all absent fine edges
\[
\{u,v\},
\qquad
u\in\pi^{-1}(a),
\quad
v\in\pi^{-1}(b),
\]
satisfying the chromatic constraint
\[
c(u)\neq c(v).
\]
Candidates are ranked by decreasing
\[
|\Gamma(u)\cap\Gamma(v)|
\]
and added until the bundle reaches \(t_p\).  Thus new bundle support is placed
where it maximizes local triangle support subject to the coloring.

\paragraph{Fine-vertex degree repair.}
After the bundle adjustments, each fine vertex with degree below three is
repaired.  For a deficient vertex \(u\), the algorithm first forms its
second-neighborhood candidate set, removes existing neighbors and \(u\)
itself, keeps only vertices \(v\) satisfying \(c(v)\neq c(u)\), and sorts
those candidates by decreasing common-neighbor support.  Edges are added in
that order until
\[
\deg(u)\geq3.
\]
If this local candidate pool is exhausted, random admissible non-neighbors are
drawn with seed \(42\) until the same degree floor is reached.

\paragraph{Connectivity repair.}
Connected components are then computed.  If more than one component remains,
smaller components are joined to the current largest component by a
chromatically admissible missing edge.  For each smaller component, vertices are
scanned until one can be connected to one of the first \(400\) stored
vertices of the current largest component.

\paragraph{Restoring the parent edge budget.}
With parent-size preservation enabled, let
\[
\Delta=E_0-|E|
\]
be the remaining edge deficit after bundle adjustment and repairs.  The
algorithm processes supernodes in decreasing size order.  Within each
supernode it enumerates absent cross-color pairs, scores them by
common-neighbor support, and adds them in decreasing score order until
\(\Delta\) edges have been inserted.  These edges remain explicit in the fine
substrate but have both endpoints in the same supernode and therefore
disappear under strict coarse-graining.

After this stage, the strict quotient is recomputed from the full modified
parent.  Because the degree and connectivity repairs are allowed to introduce
cross-supernode edges when needed, this complete quotient need not contain
exactly the target set \(\mathcal S\).  In our run it contains
\(71{,}696\) coarse relations and already passes the full nine-criterion
test.

\subsection{Mesoscale support refinement}
\label{app:mesoscale-support-refinement}

The next stage acts on the complete strict quotient derived from the
bundle-redistributed parent while simultaneously making the corresponding
changes in the fine substrate.  It uses seed \(42\).

First, an unweighted Louvain partition of the current quotient is computed at
resolution
\[
\gamma_0=0.95.
\]
Within the four largest blocks, unweighted Louvain partitions are then
computed with resolutions
\[
0.9,
\qquad
0.7,
\qquad
1.0,
\qquad
1.0,
\]
respectively.

Among existing quotient relations whose endpoints lie in the same large
block but in different within-block groups, relations are sorted by ascending
bundle multiplicity.  The first
\[
6{,}500
\]
relations are selected for replacement.  Removing one selected coarse
relation means removing \emph{all} fine parent edges supporting that
supernode pair.

A replacement pool is formed from quotient non-edges whose endpoints lie in
the same within-block group.  Each candidate pair \((a,b)\) is scored by its
number of common neighbors in the current quotient,
\[
s_Q(a,b)=|\Gamma_Q(a)\cap\Gamma_Q(b)|,
\]
and the \(6{,}500\) highest-scoring candidates are selected.

For each selected new coarse pair \((a,b)\), the implementation provisionally
adds the pair to the coarse working graph and searches the first \(30\) stored
members of supernode \(a\) and the first \(30\) stored members of supernode
\(b\).  Among absent cross-color fine edges in this \(30\times30\) candidate
set, it chooses the edge maximizing parent-level common-neighbor support and
inserts that edge when such a candidate exists; in our run one existed for all \(6{,}500\) pairs, so each new coarse relation is backed by exactly one parent edge.  The parent, rather than this
provisional coarse working graph, is the object passed to the final stage.
The final degree-profile adjustment reconstructs its quotient from the parent,
so only coarse relations with explicit fine support survive.

\subsection{Final degree-profile adjustment}
\label{app:degree-profile-adjustment}

The final adjustment starts from the parent produced above and uses NumPy
random seed \(11\).  The same fixed community-derived map \(\pi\) is
recomputed from the original integrated parent, and the complete strict
quotient of the current parent is formed.

Let
\[
d_a
\]
be the current quotient degree of supernode \(a\), and let
\[
d^{\rm Bud}_{(1)}
\leq\cdots\leq
d^{\rm Bud}_{(1015)}
\]
be the sorted Budapest degree sequence.  The generated supernodes are sorted
by their current degree and assigned Budapest targets rank-for-rank.  If
\(r(a)\) denotes the rank of supernode \(a\), its target is
\[
d_a^*=d^{\rm Bud}_{(r(a))}.
\]
No anatomical node correspondence is implied by this assignment.

\paragraph{Reducing degrees above target.}
Supernodes satisfying
\[
d_a>d_a^*
\]
are processed beginning with the largest degree surplus.  The current
quotient relations incident to \(a\) are ordered by increasing bundle
multiplicity.  The weakest relations are considered first.  A relation
\((a,b)\) is removed only if its partner is also above target,
\[
d_b>d_b^*,
\]
so that removing the pair does not deliberately push the partner below its
assigned target.  When a relation is removed, all fine parent edges supporting
that pair are removed.  These removed fine edges define the budget available
for creating new coarse relations.

\paragraph{Increasing degrees below target.}
Supernodes satisfying
\[
d_a<d_a^*
\]
are processed beginning with the largest deficit.  While a supernode remains
below target and the released fine-edge budget remains positive, the
implementation draws \(60\) candidate supernode indices with replacement
using seed \(11\).  A candidate partner \(b\) is admissible when
\[
b\neq a,
\qquad
(a,b)\notin E_Q,
\qquad
d_b<d_b^*.
\]
The first admissible sampled partner is selected.

The algorithm then scans the stored member lists of the two supernodes in
their existing order and inserts the first available chromatically admissible
fine edge.  The nested scan permits up to two fine edges to be inserted for
the selected new supernode pair before the next partner is considered.  Once
at least one fine edge is inserted, the new pair becomes a quotient relation
and both quotient degrees are incremented by one.

\paragraph{Restoring the fine-edge count.}
After the degree adjustment, if the parent contains fewer edges than it did at
the start of this stage, the remaining deficit is restored inside
supernodes.  Supernodes are processed in decreasing size order and absent
cross-color intra-supernode pairs are added in descending common-neighbor
order.  The strict quotient is then recomputed from the completed parent.
The bundle-rank-matching procedure of
Section~\ref{app:bundle-redistribution} is not rerun at this stage; the final
bundle distribution is instead measured from the resulting parent and is
found to remain close to Budapest.

Approximately \(2{,}146\) fine-edge changes are involved in this final
degree-profile adjustment.  The resulting final parent has
\[
16{,}807\ \text{vertices},
\qquad
337{,}980\ \text{edges},
\]
and its strict quotient has
\[
1{,}015\ \text{vertices},
\qquad
71{,}098\ \text{edges}.
\]
This is the single final quotient reported in the main results.

\subsection{Strict verification and bundle accounting}

For every final parent edge \(\{u,v\}\), let
\[
a=\pi(u),
\qquad
b=\pi(v).
\]
If \(a\neq b\), the edge contributes one unit to \(W_{ab}\); if \(a=b\), it
remains in the parent but disappears under coarse-graining.  In the final
substrate, the \(337{,}980\) parent edges divide exactly into
\[
147{,}217
\quad\text{cross-supernode edges}
\]
and
\[
190{,}763
\quad\text{intra-supernode edges},
\]
so that
\[
147{,}217+190{,}763=337{,}980.
\]
Recomputing the quotient from the final parent reproduces the saved quotient
edge set and weights exactly.  Thus every coarse relation has explicit fine
support and no cross-supernode parent edge is omitted from the quotient.

The final bundle-weight distribution closely tracks the Budapest
\texttt{fiber\_count\_mean} distribution:

\begin{table}[H]
\centering
\small
\begin{tabular}{lcc}
\toprule
Bundle statistic & Final quotient & Budapest \\
\midrule
Mean & \(2.071\) & \(2.033\) \\
Median & \(2.000\) & \(1.667\) \\
Maximum & \(155\) & \(154.9\) \\
Gini coefficient & \(0.306\) & \(0.305\) \\
Edges with weight \(>18\) & \(41\) & \(40\) \\
\bottomrule
\end{tabular}
\caption{
Distribution of fine-edge multiplicities underlying the final quotient
relations.  The agreement is distributional rather than anatomical: no
individual quotient edge is identified with a particular Budapest tract.
}
\label{tab:inverse-bundle-weights}
\end{table}

\section{Graph Construction and Parameter Accounting}
\label{app:construction-parameters}

This appendix section collects the construction parameters in one place.  The
integrated parent is built by tensor-product growth followed by chromatically
admissible pruning, repair, triangle closure, cross-community bridge insertion,
and multiscale-targeted sparsification.  The forward community quotient then
exposes the fixed-parent coarsening obstruction.  The inverse stages retain
the resulting \(1{,}015\)-supernode map and modify the fine edge placement so
that the complete strict quotient enters the target connectome regime.

A key tool used in the construction is Louvain community detection, a
standard modularity-optimization heuristic for partitioning a graph into
communities \cite{Blondel2008}.  Here it is used as an algorithmic partitioning
tool and as part of the mesoscale scaffold; no anatomical interpretation is
assigned to the detected communities.

\paragraph{Seed graph.}
The construction begins from a fixed tripartite seed graph
\[
S=(V_S,E_S),
\qquad
|V_S|=7,
\qquad
|E_S|=9,
\]
with vertex colors
\[
(0,0,1,1,2,2,2).
\]
The seed edges used in the reported tensor expansion are
\[
(0,2),(0,5),(1,3),(1,4),(1,6),
(2,5),(2,6),(3,4),(3,6).
\]
Every seed edge joins vertices of different colors.

\paragraph{Stage A: chromatic tensor-product expansion and pruning.}
Starting from the seed graph, the construction applies four tensor-product
expansion steps.  If \(M_{k-1}\) denotes the graph at step \(k-1\), then the
unpruned graph at the next step is
\[
M_k=M_{k-1}\otimes S.
\]
Since the initial seed already contributes one tensor factor, four additional
expansions produce
\[
n=7^{4+1}=16{,}807
\]
vertices.

The coloring is tracked as explicit vertex metadata.  A product vertex is
stored in flattened form as
\[
(u,a)\longmapsto u|V_S|+a,
\]
where \(u\in V(M_{k-1})\) is the parent-coordinate vertex and
\(a\in V(S)\) is the seed-coordinate vertex.  The product color is inherited
from the parent coordinate:
\[
c_k(u,a)=c_{k-1}(u).
\]
Thus each parent vertex produces a block of \(|V_S|\) children with the same
color as the parent.  Every tensor-product edge projects to an edge of the
previous graph, so the carried coloring remains proper.

Pruning is applied after each tensor-product expansion.  At current vertex
count \(n\), our run uses edge budget
\[
B(n)=\operatorname{round}\!\big(n\log_2(n)m_{\rm mult}\big),
\qquad
m_{\rm mult}=1.0.
\]
A Louvain partition of the current graph is computed with seed \(42\) and
resolution \(1.0\).  Each edge \((u,v)\) receives score
\[
s_A(u,v)
=
T(u,v)+10^{-7}\big(\deg(u)+\deg(v)\big),
\]
where
\[
T(u,v)=|\Gamma(u)\cap\Gamma(v)|.
\]
Triangle support is therefore the primary score, with the degree term acting
only as a small tie-breaker.

The edge budget is split into between-community and within-community portions.
The reported between-community fraction is
\[
\eta=0.35.
\]
Thus
\[
B_{\rm inter}
=
\min\{\eta B(n),|E_{\rm inter}|\},
\]
and
\[
B_{\rm intra}
=
\min\{B(n)-B_{\rm inter},|E_{\rm intra}|\}.
\]
If the intra-community pool is too small, unused budget is redirected to the
inter-community pool.  Within each pool, the highest-scoring edges are
retained.

After pruning, a minimum-degree repair adds cross-color edges until every
vertex has degree at least
\[
d_{\min}^{\rm repair}=3.
\]
If randomized selection fails, an exhaustive-search fallback is used.  If the
graph is disconnected, connectivity repair joins smaller components to the
largest component using cross-color edges.

After the four expansion steps and their associated pruning and repair
operations, an additional
\[
1.2n=20{,}168
\]
triangle-closing edges are added.  Candidate edges must be absent,
cross-color, and close at least one triangle.  The candidate list is shuffled
with seed \(42\), and the first \(20{,}168\) valid candidates are inserted.
This produces a \(16{,}807\)-node graph with
\[
273{,}695
\]
edges.

\paragraph{Stage B: multiscale-local cross-community bridges.}
The box-covering step is used as a multiscale locality test, inspired by
box-covering methods for self-similar complex networks
\cite{Song2005,Song2007memb}.  For each scale
\[
l_B\in\{3,4,5,6,7\},
\]
a randomized greedy covering is constructed using the strict membership rule
\[
\operatorname{dist}(c,x)<l_B.
\]
Multiple random vertex orderings are tried, and the covering with the fewest
boxes is retained.  If \(b_{l_B}(u)\) is the resulting box label of vertex
\(u\), define
\[
\kappa(u)=\big(b_3(u),b_4(u),b_5(u),b_6(u),b_7(u)\big).
\]
A candidate bridge \((u,v)\) is allowed only when
\[
\kappa(u)=\kappa(v),
\qquad
\operatorname{comm}(u)\neq\operatorname{comm}(v),
\qquad
c(u)\neq c(v),
\qquad
(u,v)\notin E.
\]
The valid candidates are shuffled with seed \(42\), and up to
\[
84{,}000\approx5n
\]
are added.  Our run thereby reaches an intermediate edge count of
\[
357{,}695.
\]

\paragraph{Stage C: multiscale-targeted sparsification.}
The bridge-enriched graph is sparsified while monitoring the descriptive
box-covering slope.  The stopping target is
\[
d_B\leq3.8.
\]
This is an internal multiscale target and has no Budapest counterpart.

For this stage the box-membership rule is
\[
\operatorname{dist}(c,x)\leq l_B,
\]
and the tested scales are
\[
l_B\in\{2,3,4,5,6,8,10,12\},
\]
restricted to scales not exceeding the measured graph diameter.  At each
scale, randomized greedy coverings are computed over multiple trials and the
smallest box count is retained.

Edges are ranked for removal by
\[
s_C(u,v)
=
100\,T(u,v)
+
\frac{1}{\deg(u)+\deg(v)+1}.
\]
Edges with the lowest score are considered first.  Graph bridges are excluded
from the candidate-removal list at the beginning of each sparsification
round, and a removal is skipped if it would reduce either endpoint below
\[
d_{\min}^{\rm sparse}=2.
\]
Up to \(20{,}000\) candidate removals are attempted per round.  The procedure
stops when the multiscale target is reached, subject to edge-count floor
\[
m_{\min}=100{,}000.
\]
The resulting integrated parent has
\[
16{,}807\text{ vertices},
\qquad
337{,}983\text{ edges}.
\]

\paragraph{Stage D: fixed community-derived map.}
Louvain community detection is applied to the integrated parent at resolution
\[
\gamma_{\rm init}=20.0
\]
with seed \(42\), producing approximately \(1{,}054\) communities.  The
number is reduced deterministically to \(N=1{,}015\) by repeatedly merging a
smallest community into the neighboring community maximizing intercommunity
edge density.  This map is used first for the forward obstruction test and is
then held fixed throughout the reported inverse relocation.

\paragraph{Stage E: strict-coarsening sweep.}
To test whether the obstruction is specific to this map, \(25\) additional
strict coarsenings are forced to \(1{,}015\) groups, spanning Leiden RB and
CPM, balanced METIS, size-capped hybrid maps, nested maps, box-covering maps,
and a random balanced control.  No tested map reaches the Budapest modular
regime; the best observed modularity is \(0.288\).

\paragraph{Stages F--H: inverse fine-edge redistribution.}
The implementation-level rules are given in
Sections~\ref{app:bundle-redistribution}--\ref{app:degree-profile-adjustment}.
The key point is that the target bundle distribution and degree distribution
are imposed only through changes to the fine parent.  The reported quotient
is always the complete quotient subsequently counted from that parent.

\paragraph{Reported-run parameters.}
The following table collects the parameters and deterministic search limits
needed to reproduce our run.

\begin{table}[H]
\centering
\footnotesize
\setlength{\tabcolsep}{4pt}
\renewcommand{\arraystretch}{0.9}
\begin{tabularx}{\textwidth}{p{0.34\textwidth}X}
\toprule
Parameter & Role \\
\midrule
\(K=4\) & Number of tensor-product expansion steps. \\
\(m_{\rm mult}=1.0\) & Multiplier in \(B(n)=n\log_2(n)m_{\rm mult}\). \\
\(\eta=0.35\) & Fraction of pruning budget reserved for inter-community edges. \\
\(\gamma_{\rm prune}=1.0\) & Louvain resolution during tensor-step pruning. \\
\(d_{\min}^{\rm repair}=3\) & Fine-vertex degree floor after tensor-step pruning and during bundle redistribution. \\
\(\alpha_{\triangle}=1.2\) & Post-expansion triangle-closing budget \(\alpha_{\triangle}n\). \\
\(\mathcal L_{\rm bridge}=\{3,4,5,6,7\}\) & Box-covering scales used for bridge candidates. \\
\(\alpha_{\rm bridge}=5\) & Maximum bridge budget, set to \(84{,}000\approx\alpha_{\rm bridge}n\). \\
\(d_B^*=3.8\) & Target descriptive box-covering slope during sparsification. \\
\(\mathcal L_{\rm sparse}=\{2,3,4,5,6,8,10,12\}\) & Box-covering scales during sparsification. \\
\(\Delta m=20{,}000\) & Maximum attempted removals per sparsification round. \\
\(m_{\min}=100{,}000\) & Edge-count stopping threshold during sparsification. \\
\(d_{\min}^{\rm sparse}=2\) & Fine-vertex degree floor protected during sparsification. \\
\(N=1{,}015\) & Number of supernodes. \\
\(\gamma_{\rm init}=20.0\), seed \(42\) & Community map used for both the forward obstruction and reported inverse run. \\
\(\gamma_0=0.95\) & Coarsest witness/quotient hierarchy resolution. \\
\((\gamma_{L1},\gamma_{L2})\) & \((0.9,1.0),(0.7,1.0),(1.0,2.0),(1.0,2.0)\) for the four large blocks in the target-support scaffold. \\
10 & Number of strongest cross-block \(L1\)-pair patches retained in the target-support scaffold. \\
\(\lfloor139N/2\rfloor\) & Nominal mesoscale support budget used to form the target pair set. \\
seed \(42\) & Random seed for Budapest bundle-value sampling and Stage-F repairs. \\
without replacement & Sampling rule for Budapest \texttt{fiber\_count\_mean} values before rank assignment. \\
400 & Maximum number of stored vertices of the current largest component scanned during connectivity repair. \\
\(6{,}500\) & Number of coarse relations replaced in the mesoscale support-refinement stage. \\
30 & Number of stored fine vertices examined from each supernode when realizing a new refinement-stage coarse relation. \\
seed \(11\) & Random seed for the final degree-profile adjustment. \\
60 & Number of candidate partner supernodes sampled per degree-deficit search attempt. \\
\(\Delta m_{\rm degree}\approx2{,}146\) & Approximate fine-edge changes in the final degree-profile adjustment. \\
\bottomrule
\end{tabularx}
\caption{Parameters and deterministic search limits used in the reported construction.}
\label{tab:reported-run-parameters}
\end{table}

The coarse node count, approximate edge scale, degree distribution, and
bundle-weight distribution are explicit constraints on the inverse solution.
No individual Budapest node, edge, or tract is fitted anatomically.

\section{Additional Diagnostics for the Final Quotient}
\label{app:inverse-quotient-diagnostics}

The single final quotient reported in the paper contains
\[
n=1{,}015,
\qquad
m=71{,}098,
\]
compared with
\[
n=1{,}015,
\qquad
m=70{,}654
\]
for the Budapest Reference Connectome.  The corresponding densities are
\(0.1382\) and \(0.1373\).  The empirical degree distributions are closely
aligned, with
\[
KS=0.0187.
\]
The final quotient passes all nine scored topology criteria against
\(1{,}000\) degree-preserving nulls.

Its principal scalar comparisons with Budapest are
\[
\begin{array}{lcc}
\toprule
\text{quantity} & \text{final quotient} & \text{Budapest}\\
\midrule
Q & 0.4908 & 0.5574\\
C & 0.4724 & 0.6696\\
\text{transitivity} & 0.4234 & 0.5578\\
\text{APL} & 1.9811 & 2.2190\\
\text{diameter} & 4 & 5\\
\text{effective diameter} & 2.162 & 2.731\\
\text{mean degree} & 140.1 & 139.2\\
\text{median degree} & 130 & 127\\
\text{minimum degree} & 13 & 7\\
\text{maximum degree} & 420 & 466\\
\text{assortativity} & +0.0826 & +0.0215\\
\bottomrule
\end{array}
\]
These comparisons are distributional rather than anatomical.  No quotient
node is identified with a particular Budapest region.

\paragraph{Six static graph tests.}
We additionally compare degree distribution, hop plot, singular-value scree,
network value, triangle participation, and effective diameter.  The final
quotient has degree-distribution \(KS\approx0.019\), network-value
\(KS\approx0.089\), and triangle-participation \(KS\approx0.154\).  Its
six-test mean normalized error is \(0.3033\).  The degree and cumulative
network-value profiles are close, whereas the scree plot retains the clearest
residual difference.  This is consistent with the bilateral organization
present in Budapest but absent from the present construction.

\begin{figure}[p]
\centering
\includegraphics[width=\textwidth,height=0.78\textheight,keepaspectratio]
{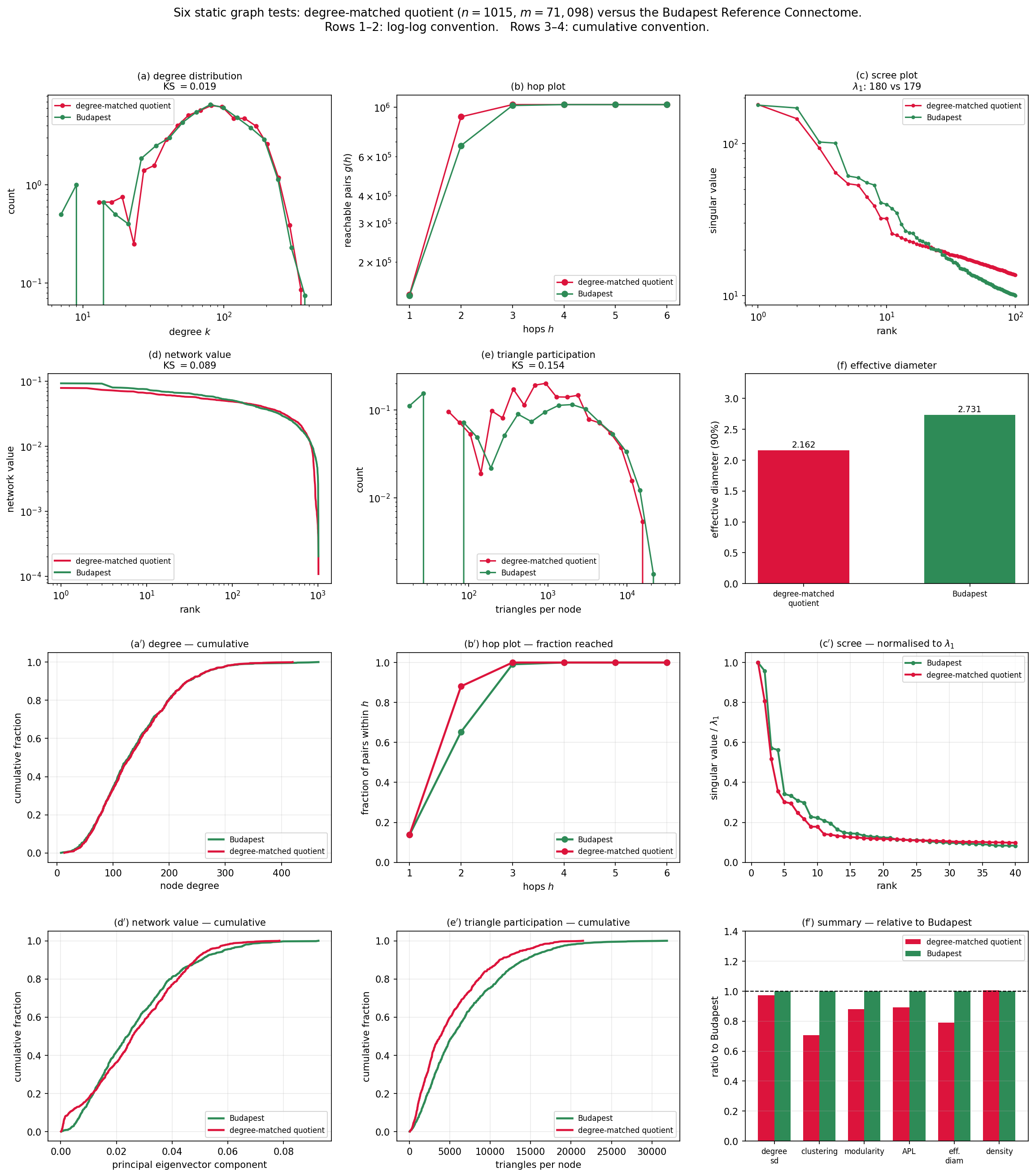}
\captionsetup{font=small}
\caption{Six static graph tests comparing the final quotient
(\(n=1{,}015\), \(m=71{,}098\)) with the Budapest Reference Connectome
(\(n=1{,}015\), \(m=70{,}654\)).  Rows 1--2 use the log--log convention;
rows 3--4 show cumulative versions.  Panels show degree distribution, hop
plot, singular-value scree, network value, triangle participation, and
effective diameter, followed by normalized cumulative summaries.  The degree
match is especially close (\(KS=0.0187\)); the principal residual discrepancy
is spectral and is consistent with the bilateral organization absent from the
present construction.}
\label{fig:graph-tests}
\end{figure}

\begin{figure}[t]
\centering
\includegraphics[width=\textwidth]{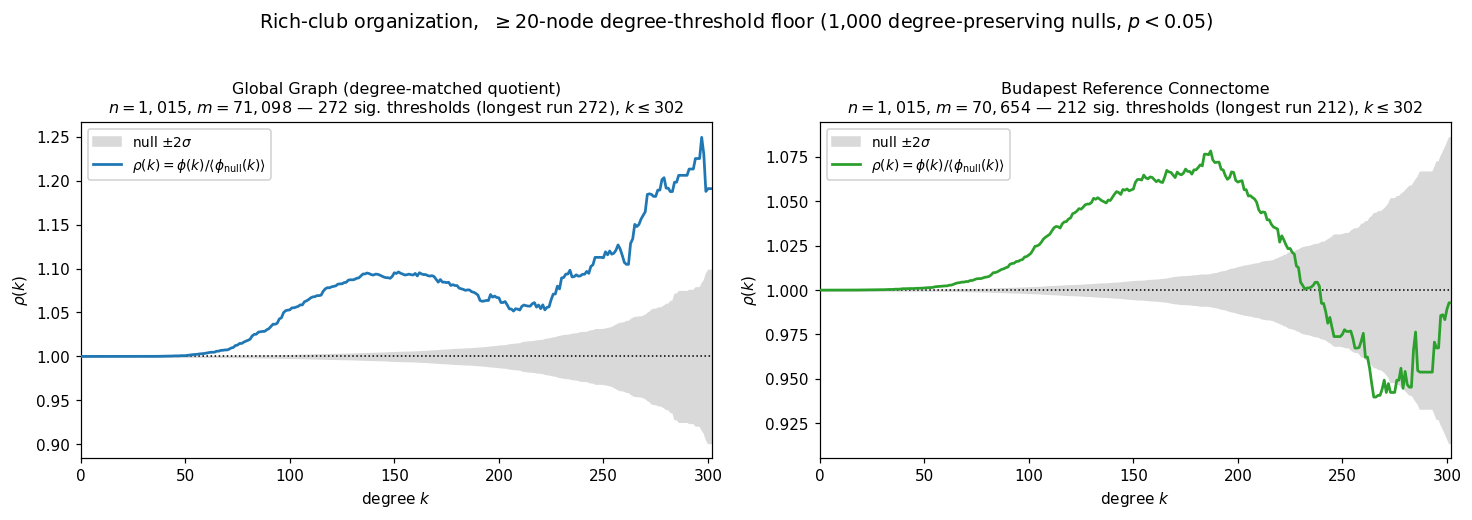}
\caption{Normalized rich-club organization of the final quotient (left) and
the Budapest Reference Connectome (right), each evaluated against
\(1{,}000\) degree-preserving null graphs.  The shaded region denotes the null
mean \(\pm2\sigma\), and the dotted line marks \(\rho(k)=1\).  The plot is
restricted to thresholds retaining at least \(20\) nodes.  Under this
convention the quotient has \(272\) significant thresholds, all in one
consecutive run, and Budapest has \(212\), likewise in one consecutive run.
Both graphs show an extended rich-club enrichment regime, although their
high-\(k\) profiles differ.}
\label{fig:richclub-rho}
\end{figure}

\paragraph{Rich-club organization.}
Figure~\ref{fig:richclub-rho} shows the normalized rich-club coefficient
\[
\rho(k)
=
\frac{R(k)}{\langle R_{\mathrm{null}}(k)\rangle}
\]
against \(1{,}000\) degree-preserving nulls, using the
\(\geq20\)-node degree-threshold floor and \(p<0.05\).  Under this plotted
convention, the final quotient has \(272\) significant thresholds with a
longest consecutive run of \(272\), while Budapest has \(212\) significant
thresholds with a longest run of \(212\).

Together these diagnostics support a restricted structural claim.  The final
quotient is not an anatomical reconstruction of the Budapest connectome and
does not reproduce its bilateral organization, but it occupies a closely
related graph-theoretic regime while being derived strictly from an explicit
sparse fine-scale substrate.

\FloatBarrier
\section{Image Inpainting Model and Training Details}
\label{app:inpainting_model}

\paragraph{Task and data.}
We evaluate the graph module in a \(256\times256\) CelebA-HQ image-inpainting
task. Each training example consists of an image \(x\), a binary mask \(m\),
and a masked observation
\[
x_{\mathrm{obs}}=m\odot x.
\]
The model receives the four-channel tensor \([x_{\mathrm{obs}},m]\) and
predicts a raw reconstruction \(\hat x\). The composited output preserves the
observed pixels and fills the missing region:
\[
\hat x_{\mathrm{final}}
=
m\odot x_{\mathrm{obs}}+(1-m)\odot\hat x.
\]

Training uses \(28{,}000\) images and a \(2{,}000\)-image held-out set.
Validation during training uses a fixed half of the held-out set; the remaining
\(1{,}000\) images form the reporting half used for the final benchmark.
Checkpoint selection and final reporting therefore use disjoint halves of the
held-out data.

The recurrent module operates directly on the final relocation substrate:
\[
16{,}807\text{ nodes},
\qquad
337{,}980\text{ edges},
\qquad
1{,}015\text{ supernodes},
\]
with the inherited proper three-coloring. No additional lifting or
coarse-to-fine reconstruction is performed before the graph is used by the
neural model.

\paragraph{Architecture.}
The graph model uses the same overall encoder--decoder setting as the matched
U-Net baseline, but replaces the ordinary deepest pathway by a sparse recurrent
graph module. In the graph model the convolutional bottleneck is removed, the
two deepest encoder skips are not passed directly to the decoder, and graph
readouts supply the corresponding deep decoder inputs. The graph is therefore
the deep computational pathway rather than an auxiliary branch.

The encoder maps the masked image to six multi-scale feature maps
\[
s_0,s_1,s_2,s_3,s_4,s_5,
\]
with spatial resolutions
\[
256,\ 128,\ 64,\ 32,\ 16,\ 8
\]
and channel dimensions
\[
64,\ 128,\ 256,\ 512,\ 512,\ 512.
\]

The graph has three node populations,
\[
n_1=4{,}802,
\qquad
n_2=4{,}802,
\qquad
n_3=7{,}203,
\]
for a total of \(16{,}807\) nodes, with graph channel width \(C=32\).
The allowed directed sparse operators are
\[
1\to2,
\qquad
1\to3,
\qquad
2\to3,
\]
with recurrent feedback
\[
3\to2,
\qquad
3\to1.
\]
Trainable sparse weights exist only on allowed graph connections; absent
connections carry no sparse edge parameter. A sparse layer mixes information
across graph nodes while leaving the feature channels separate:
\[
Y_{b,u,c}
=
\sum_{v:\,v\to u\ \mathrm{allowed}}
\theta_{uv}X_{b,v,c},
\]
where \(\theta_{uv}\) is the learned scalar associated with an allowed sparse
connection. There is no node bias.

\paragraph{Collector and recurrent update.}
The collector maps encoder feature maps into graph-node features by
\(1\times1\) projection followed by spatial flattening or pooling. The
allocation over the \(n_1=4{,}802\) nodes is
\[
\begin{array}{llr}
s_3: & 32\times32,\ \text{no collapse}         & 1024\\
s_4: & 16\times16,\ \text{no collapse}         & 256\\
s_5: & 8\times8,\ \text{no collapse}           & 64\\
s_2: & 64\times64\to32\times64,\ 2\!:\!1       & 2048\\
s_1: & 128\times128\to30\times32               & 960\\
s_0: & 256\times256\to18\times25               & 450\\
\end{array}
\]
in the order
\[
[s_3,s_4,s_5,s_2,s_1,s_0],
\]
summing to \(4{,}802\) exactly.

Let \(a_1,a_2,a_3\) denote the states of the three populations. The collector
initializes
\[
a_1^0
=
\operatorname{LN}_1\!\left(
\operatorname{Collector}(s_0,\ldots,s_5)
\right),
\qquad
a_3\leftarrow0.
\]
Each of five recurrent passes then uses
\[
a_2
=
\operatorname{GELU}\!\left(
\operatorname{LN}_2\!\left(
W_{12}a_1+
\operatorname{LN}_{fb}(W_{32}a_3)
\right)
\right),
\]
\[
a_1
=
a_1^0+
\operatorname{LN}_{fb1}(W_{31}a_3),
\]
\[
a_3
=
\operatorname{LN}_3\!\left(
W_{23}a_2+W_{13}a_1
\right).
\]

The update order is part of the recurrent dynamics. In particular, \(a_2\) is
computed before the \(3\to1\) feedback updates \(a_1\), so \(a_2\) receives
the effect of that feedback on the following recurrent pass, whereas \(a_3\)
uses the freshly updated \(a_1\).

The implementation therefore selects one concrete causal schedule for the
three chromatic phase classes. The coloring fixes the minimum number of phase
classes required by the structural dependency graph, but not a unique temporal
ordering. Dependencies may continue across recurrent passes, so chromatic
causal depth does not bound total computational depth.

\paragraph{Same-phase attention.}
The reported graph model includes supernode-level linear attention applied
within same-color groups inside each supernode. When enabled, the corresponding
attention module is applied immediately after the update of that population.
The attention blocks provide same-phase mixing without adding same-color edges
to the structural graph.

The attention mechanism contributes \(12{,}864\) trainable parameters. The
complete attention-equipped graph model contains
\[
75{,}965{,}171
\]
trainable parameters, of which
\[
579{,}248
\]
are associated with the sparse graph and \(12{,}864\) with the supernode
attention blocks.

A no-attention variant removes only the three attention modules and their
calls in the recurrent loop. The graph, collector, recurrent update, twin
path, encoder, decoder, data pipeline, and training procedure are otherwise
unchanged. Only one no-attention run was performed on the final relocation
substrate, so this condition is used as a mechanism check rather than as a
separately replicated experimental arm.

\paragraph{Twin path and decoder coupling.}
The twin path maps graph state back into decoder features at six scales. The
two deepest outputs are learned projections over all \(n_3=7{,}203\) nodes
and are read from intermediate recurrent passes:
\[
t_5:\ 8\times8=64\text{ slots, from pass }3,
\qquad
t_4:\ 16\times16=256\text{ slots, from pass }4.
\]
The remaining four outputs are formed from the final recurrent pass:
\[
\begin{array}{llr}
t_3: & \to32\times32                         & 1024\\
t_2: & \to32\times64\to64\times64           & 2048\\
t_1: & \to32\times64\to128\times128         & 2048\\
t_0: & \to32\times64\to256\times256         & 2048.
\end{array}
\]

At the deep end, graph readouts supply the bottleneck and the two deepest
decoder inputs. At the three shallow levels, ordinary encoder skips are
retained and modulated by graph outputs:
\[
s_0\mapsto s_0+t_0,
\qquad
s_1\mapsto s_1+t_1,
\qquad
s_2\mapsto s_2+t_2.
\]
The graph therefore carries the deep semantic pathway while the shallow
localization pathway is preserved.

\paragraph{Parameter counts.}
The matched U-Net baseline contains
\[
87{,}103{,}619
\]
trainable parameters. The graph model is therefore smaller overall despite
the addition of the recurrent sparse substrate.

\begin{table}[H]
\centering
\small
\setlength{\tabcolsep}{5pt}
\renewcommand{\arraystretch}{1.08}
\begin{tabular}{@{}lrr@{}}
\toprule
Component
& Graph + attention
& Graph without attention\\
\midrule
Encoder
& \(26{,}108{,}928\)
& \(26{,}108{,}928\)\\
Decoder
& \(46{,}829{,}699\)
& \(46{,}829{,}699\)\\
Twin path
& \(2{,}370{,}752\)
& \(2{,}370{,}752\)\\
Sparse graph
& \(579{,}248\)
& \(579{,}248\)\\
Collector
& \(63{,}680\)
& \(63{,}680\)\\
Supernode attention
& \(12{,}864\)
& \(0\)\\
\midrule
Total
& \(75{,}965{,}171\)
& \(75{,}952{,}307\)\\
\bottomrule
\end{tabular}
\caption{Parameter accounting for the graph-model variants. The final
relocation substrate fixes the sparse routing pattern; training learns the
parameters carried by the permitted sparse operators. The matched U-Net
baseline contains \(87{,}103{,}619\) trainable parameters.}
\label{tab:inpainting-parameters}
\end{table}

\paragraph{Training and checkpoint selection.}
Models are trained from scratch for a fixed \(50\) epochs with AdamW, batch
size \(2\), gradient clipping at norm \(1.0\), and a cosine learning-rate
schedule decaying to zero. Runs are not terminated early.

The encoder, decoder, and twin-path parameter groups use learning rate
\[
3\times10^{-4}
\]
with weight decay \(10^{-4}\). The group holding the sparse graph, its layer
normalizations, the supernode attention, and the collector uses a
\(10\times\) learning-rate multiplier,
\[
3\times10^{-3},
\]
with weight decay \(10^{-5}\).

The reported checkpoint is selected by a rule fixed before the final runs:
\begin{quote}
train for the full \(50\) epochs; round validation loss to four decimal
places; after epoch \(40\), identify the first sequence of three consecutive
epochs with the same rounded validation loss; report the earliest checkpoint
in that sequence.
\end{quote}
Checkpoint selection uses the selection half of the held-out set, while the
final metrics below use the separate reporting half.

For the three attention-equipped graph runs the rule selects
\[
\text{seed 0: epoch }45,
\qquad
\text{seed 1: epoch }44,
\qquad
\text{seed 2: epoch }45.
\]
For the three U-Net runs it selects
\[
\text{seed 0: epoch }47,
\qquad
\text{seed 1: epoch }46,
\qquad
\text{seed 2: epoch }46.
\]
The single no-attention mechanism-check run selects epoch \(47\).

The pixel loss weights the missing region more heavily than the observed
region,
\[
\mathcal L_{\mathrm{pix}}
=
5\,\frac{\|(1-m)\odot(\hat x-x)\|_1}{3\,\|1-m\|_1}
+
\frac{\|m\odot(\hat x-x)\|_1}{3\,\|m\|_1},
\]
with the observed-region term applied to the raw network output. A VGG16
perceptual loss on the hidden region is added with weight \(0.1\).

\paragraph{Mask distributions.}
Training and evaluation use the same irregular-mask routine with different
coverage inputs. During training, at the nominal \(0.25\) mask setting, the
target coverage is drawn from
\[
U(0.09,0.49).
\]
Across \(30{,}000\) training draws, the empirically achieved missing fraction
spans
\[
[0.0904,0.6040].
\]

The primary reported evaluation uses the broader benchmark setting
\[
U(0.02,0.70),
\]
because this is the protocol used for the matched U-Net comparison. Among the
\(4{,}893\) scored benchmark masks, \(579\), or \(11.8\%\), fall outside the
achieved training range. All lie below the training floor and within the
\(0\)--\(20\%\) coverage band. The \(20\)--\(40\%\) and \(40\)--\(60\%\)
bands lie entirely within the achieved training range.

The graph checkpoints are additionally evaluated on the training-mask
distribution as a robustness check. The benchmark and training-band aggregate
values should not be compared directly because the two distributions fill the
0--10\% and 50--60\% buckets with masks of different coverage. Averaged over
the four middle buckets, which are well populated under both distributions,
L1 differs by less than \(0.01\) percentage points and PSNR by less than
\(0.03\) dB in every measured graph run.

The U-Net baseline was evaluated on the benchmark-mask distribution only; no
U-Net training-mask result is inferred.

\paragraph{Evaluation protocol.}
L1, PSNR, SSIM, and LPIPS are computed separately within six achieved
mask-coverage buckets and then averaged across the six bucket values. The
reported aggregate is therefore a mean of bucket-level metrics rather than a
direct per-image mean. This convention is used identically for the U-Net and
graph benchmark results.

\paragraph{Matched U-Net comparison.}
The primary inpainting comparison consists of three independently trained
U-Net runs and three independently trained graph-plus-attention runs evaluated
under the same benchmark-mask protocol.

\begin{table}[H]
\centering
\small
\setlength{\tabcolsep}{7pt}
\renewcommand{\arraystretch}{1.12}
\begin{tabular}{@{}lcc@{}}
\toprule
Metric
& U-Net
& Graph + attention\\
\midrule
L1 (\%) \(\downarrow\)
& \(1.5618\pm0.0105\)
& \(1.4533\pm0.0024\)\\

PSNR (dB) \(\uparrow\)
& \(29.254\pm0.041\)
& \(29.820\pm0.008\)\\

SSIM \(\uparrow\)
& \(0.89523\pm0.00048\)
& \(0.90387\pm0.00017\)\\

LPIPS \(\downarrow\)
& \(0.11034\pm0.00060\)
& \(0.09356\pm0.00032\)\\
\bottomrule
\end{tabular}
\caption{Benchmark-mask image-inpainting performance of the matched U-Net
baseline and the recurrent graph model instantiated on the final relocation
substrate. Values are mean \(\pm\) SD across three independently trained
seeds and are means over the six coverage buckets. L1 is the mean absolute
error over the full image (\(\times100\)); observed pixels are copied from the
input, so only the missing region contributes. FID was not computed for these
runs.}
\label{tab:inpainting-unet-graph}
\end{table}

Relative to the U-Net mean, the graph model reduces L1 by \(6.95\%\),
increases PSNR by \(0.566\) dB, increases SSIM by \(0.00864\), and reduces
LPIPS by \(15.20\%\). The L1 advantage is present throughout the coverage
range: each graph seed has lower L1 than each U-Net seed in every one of the
six benchmark buckets.

The purpose of this comparison is not to claim state-of-the-art image
inpainting. Rather, it tests whether the constructed fine graph can serve
directly as a trainable sparse recurrent computational substrate. Because the
graph model also changes the deepest computational pathway and its recurrent
organization, the comparison is architectural rather than a topology-only
ablation.

\paragraph{Three-seed graph results.}
The individual benchmark-mask results for the attention-equipped graph model
are
\[
\begin{array}{lrrrr}
\toprule
\text{Run}
& \text{L1 (\%)}
& \text{PSNR}
& \text{SSIM}
& \text{LPIPS}\\
\midrule
\text{Seed 0 (ep45)}
&1.4506&29.826&0.90403&0.09329\\
\text{Seed 1 (ep44)}
&1.4539&29.823&0.90388&0.09349\\
\text{Seed 2 (ep45)}
&1.4554&29.810&0.90369&0.09391\\
\midrule
\text{Mean}
&1.4533&29.820&0.90387&0.09356\\
\text{SD}
&0.0024&0.008&0.00017&0.00032\\
\bottomrule
\end{array}
\]

The same three checkpoints evaluated on the training-mask distribution give
\[
\begin{array}{lrrrr}
\toprule
\text{Run}
& \text{L1 (\%)}
& \text{PSNR}
& \text{SSIM}
& \text{LPIPS}\\
\midrule
\text{Seed 0 (ep45)}
&1.4099&29.352&0.90553&0.09179\\
\text{Seed 1 (ep44)}
&1.4152&29.346&0.90540&0.09200\\
\text{Seed 2 (ep45)}
&1.4154&29.343&0.90524&0.09243\\
\midrule
\text{Mean}
&1.4135&29.347&0.90539&0.09207\\
\text{SD}
&0.0031&0.005&0.00015&0.00033\\
\bottomrule
\end{array}
\]
The lower training-band aggregate should not be interpreted as an improvement
over the benchmark aggregate: the change is driven by the different mask
coverage within the 0--10\% and 50--60\% buckets rather than by a change in
the model.
\paragraph{Attention mechanism check.}
The no-attention variant was evaluated in one run only and is therefore not
included as a separately replicated arm in
Table~\ref{tab:inpainting-unet-graph}. On the benchmark distribution the
single no-attention run gives
\[
\mathrm{L1}=1.4507\%,
\quad
\mathrm{PSNR}=29.843,
\quad
\mathrm{SSIM}=0.90415,
\quad
\mathrm{LPIPS}=0.09332.
\]
On the training-mask distribution it gives
\[
\mathrm{L1}=1.4088\%,
\quad
\mathrm{PSNR}=29.389,
\quad
\mathrm{SSIM}=0.90565,
\quad
\mathrm{LPIPS}=0.09174.
\]

For comparison, the three-seed attention-equipped benchmark result is
\[
\mathrm{L1}=1.4533\pm0.0024\%,
\qquad
\mathrm{PSNR}=29.820\pm0.008,
\]
\[
\mathrm{SSIM}=0.90387\pm0.00017,
\qquad
\mathrm{LPIPS}=0.09356\pm0.00032.
\]
Compared with the attention run trained under the same seed, the
direction of the difference is mixed (attention better on L1 and LPIPS,
no attention better on PSNR and SSIM), by margins comparable to or
smaller than the attention arm's seed-to-seed range. Against the
three-seed attention mean, the single no-attention run is nominally
better on all four metrics. We therefore do not detect a substantial
attention-dependent performance effect in this mechanism check.

Because only one no-attention run was performed, this comparison is not
interpreted as a replicated equivalence test and no superiority is claimed in
either direction. Its role is narrower: the close no-attention result indicates
that the large separation between the three-seed graph model and the matched
U-Net baseline is not explained by the addition of the supernode-attention
blocks.

The reported experimental comparison is therefore the three-seed
attention-equipped graph model against the three-seed matched U-Net baseline.
The no-attention condition is retained only as the one-run mechanism check
described above.

\section{Network Topology}
\label{app:network-topology}

We evaluate the generated graphs using a topology test designed to check
whether a graph reproduces the main graph-theoretic signatures commonly reported
for structural brain networks.  The test consists of nine scored criteria,
together with unscored descriptive measures: box-covering/fractal scaling and broad-degree checks. Unless
stated otherwise, each scored comparison is evaluated against
degree-preserving null graphs, generated by edge-swap randomization, so that the result is not merely a consequence of the degree sequence
\cite{MaslovSneppen2002,VasaMisic2022}.  Enrichment criteria (clustering, modularity, local efficiency, and the significance clauses of the rich-club test) require the empirical value to exceed the null ensemble at \(p<0.05\), i.e.\ above the 95th percentile of the null distribution.

The descriptive box-covering and broad-degree checks are reported to
characterize the graph family, but they are not counted among the nine scored
quotient-level criteria.  This is especially important at the dense
\(1{,}015\)-node quotient scale, where graph distances are too compressed for
box-covering fits to be interpreted in the same way as on the large sparse
parent graph.

\paragraph{1. Clustering enrichment.}
Brain networks are locally clustered: neighbors of a node tend to be
connected to one another more often than in a comparable random graph.
We therefore require clustering, or transitivity, to be significantly
larger than the degree-preserving null ensemble. This captures the
segregated local structure emphasized in small-world network theory
\cite{WattsStrogatz1998}.  Clustering enrichment is not by itself a small-world criterion. Rather,
it measures the local-segregation component that distinguishes
small-world networks from ordinary random graphs. A graph is small-world
only when this local clustering is combined with near-random path length. 

\paragraph{2. Path length near random.}
Small-world organization also requires short global paths. We compare
the average shortest-path length of the graph to that of matched random
graphs and require it to remain close to random. Operationally, the test
uses a tolerance of approximately 25 percent above the random-graph path
length. This prevents graphs from passing merely because they are highly
clustered but globally lattice-like.  Shortest-path quantities were computed exactly on the largest connected
component using sparse all-pairs shortest paths.  We used SciPy's
\texttt{csgraph.shortest\_path} routine with Dijkstra mode on the
unweighted, undirected adjacency matrix.  Average path length was
computed as the mean of all finite nonzero shortest-path distances.
Global efficiency was computed from the same distance matrix by averaging
reciprocal distances, with unreachable pairs contributing zero.

\paragraph{3. Small-world index.}
The small-world index combines the preceding two effects: clustering
should be much larger than random, while path length should remain near
random. We use the Humphries--Gurney statistic
\[
\sigma =
\frac{C/C_{\mathrm{null}}}{L/L_{\mathrm{null}}},
\]
where $C$ is transitivity and $L$ is average path length.  Transitivity
is used in both null conventions; neither reported index uses the mean
local clustering coefficient.
The graph is small-world when $\sigma>1$ \cite{HumphriesGurney2008}. We compute the statistic against both null conventions: the
ER-normalized coefficient \(\sigma_{\mathrm{ER}}\) and the stricter
degree-preserving analogue \(\sigma_{\mathrm{dp}}\), computed against the
degree-preserving ensemble.  The criterion requires \(\sigma>1\), and every
graph tested in this paper satisfies it under both conventions; tables
report \(\sigma_{\mathrm{dp}}\) as the more conservative value unless
labeled otherwise.

\paragraph{4. Modularity enrichment.}
Modularity enrichment measures whether the graph separates into
communities more strongly than expected from its degree sequence alone.
After community detection, each node $i$ is assigned a community label
$c_i$. For an undirected unweighted graph with adjacency matrix $A$,
degrees $k_i$, and total edge count $m$, the modularity of a partition is
\[
Q =
\frac{1}{2m}
\sum_{i,j}
\left(
A_{ij} - \frac{k_i\,k_j}{2m}
\right)
\mathbf{1}\{c_i=c_j\}.
\]
Here $A_{ij}$ records whether nodes $i$ and $j$ are connected, while
$k_i k_j/(2m)$ is the expected adjacency contribution between the same
two nodes under a degree-preserving configuration-model null. The
indicator $\mathbf{1}\{c_i=c_j\}$ restricts the sum to node pairs assigned
to the same community. Thus $Q$ is large when communities contain more
internal connectivity than would be expected from node degrees alone
\cite{NewmanGirvan2004}. In our test, modularity enrichment passes only
when the detected partition has nontrivial modularity, $Q>0.3$, and the
observed value is significant against degree-preserving null graphs.

The community labels \(c_i\) are obtained by first applying a community
detection algorithm to the graph.  In our test, we use Louvain community
detection, a modularity-optimization heuristic in the same community-detection
tradition as earlier large-network methods
\cite{ClausetNewmanMoore2004,Blondel2008}.  Thus \(c_i\) denotes the
Louvain community assigned to node \(i\).

\paragraph{5. Global efficiency near random.}
Global efficiency measures how efficiently information can travel across
the whole graph. If $d_{ij}$ denotes the shortest-path distance between
nodes $i$ and $j$, then
\[
E_{\mathrm{glob}}
=
\frac{1}{n(n-1)}
\sum_{i\neq j}
\frac{1}{d_{ij}} .
\]
Thus short paths contribute strongly, long paths contribute weakly, and
disconnected pairs contribute zero. We compare global efficiency with
matched degree-preserving null graphs. The criterion is not that global
efficiency should be maximized, but that it should remain close to random:
brain-like networks combine efficient long-range communication with
non-random local and modular structure. In our test, this criterion
passes when $E_{\mathrm{glob}}/E_{\mathrm{glob,null}}$ lies in the range $[0.8,1.0)$ (near but below one) \cite{LatoraMarchiori2001,IturriaMedina2008}.

\paragraph{6. Local efficiency enrichment.}
Local efficiency measures how efficiently the neighbors of a node can
communicate with one another after that node is removed. For each node
$i$, let $N(i)$ be its neighborhood and let $G_i$ be the subgraph induced
by $N(i)$. The local efficiency of the whole graph is
\[
E_{\mathrm{loc}}(G)
=
\frac{1}{n}
\sum_{i=1}^{n}
E_{\mathrm{glob}}(G_i),
\]
where
\[
E_{\mathrm{glob}}(G_i)
=
\frac{1}{k_i(k_i-1)}
\sum_{\substack{j,h \in N(i)\\ j\neq h}}
\frac{1}{d^{(i)}_{jh}} .
\]
Here $k_i$ is the degree of node $i$, and $d^{(i)}_{jh}$ is the
shortest-path distance between neighbors $j$ and $h$ inside the
neighbor-induced subgraph $G_i$. If $k_i<2$, the contribution of node
$i$ is taken to be zero. Thus local efficiency is high when the neighbors
of a node remain well connected even without that node. We require local
efficiency to be enriched relative to degree-preserving null networks,
indicating more local fault-tolerant neighborhood structure than expected
from the degree sequence alone \cite{LatoraMarchiori2001}.

\paragraph{7. Connector hubs.}
Connectomes contain hubs that link multiple communities. After community
detection, let $c_i$ be the community of node $i$, let $k_i$ be its total
degree, and let $k_{is}$ be the number of edges from node $i$ to
community $s$. The participation coefficient of node $i$ is
\[
P_i
=
1 -
\sum_s
\left(
\frac{k_{is}}{k_i}
\right)^2 .
\]
This quantity is small when most edges of node $i$ remain inside one
community, and large when its edges are distributed across several
communities. We also compute the within-module degree score
\[
z_i
=
\frac{k_i^{\mathrm{in}}-\mu_{c_i}}{\sigma_{c_i}},
\]
where $k_i^{\mathrm{in}}$ is the number of edges from node $i$ to nodes
inside its own community, and $\mu_{c_i}$ and $\sigma_{c_i}$ are the mean
and standard deviation of within-community degree among nodes in the same
community as $i$. Following the Guimera--Amaral cartographic
classification, we call node $i$ a connector hub when
\[
z_i \geq 2.5
\qquad\text{and}\qquad
0.30 < P_i \leq 0.75 .
\]
Thus connector hubs are nodes that are hub-like inside their own module
while also linking substantially to other modules. The presence of at
least one such connector hub is used as evidence of cross-module
integration \cite{Guimera2005}.

\paragraph{8. Rich-club organization.}
Rich-club organization tests whether high-degree nodes are more densely
interconnected with one another than expected under a null model. For a
degree threshold $k$, let $N_{>k}$ be the number of nodes with degree
greater than $k$, and let $E_{>k}$ be the number of edges among those
nodes. The rich-club coefficient is
\[
R(k)
=
\frac{2E_{>k}}{N_{>k}(N_{>k}-1)} .
\]

This is the edge density of the subgraph induced by nodes whose degree
exceeds $k$. To test whether this density is larger than expected from
the degree sequence alone, we normalize by a degree-preserving null
ensemble:
\[
\rho(k)=\frac{R(k)}
{\langle R_{\mathrm{null}}(k) \rangle}.
\]
A value $\rho(k)>1$ means that high-degree nodes are more
interconnected than in matched null graphs. In our test, rich-club
organization passes when $\rho(k)>1$ with statistical significance over a run of at least three consecutive degree thresholds, indicating a supported high-degree backbone \cite{Colizza2006,vandenHeuvel2011}.

\paragraph{9. Multi-resolution VI stability.}

This criterion follows the multi-scale brain-network view that community
structure should be examined across topological scales rather than at a
single fixed resolution \cite{BetzelBassett2017}; variation of information
provides the partition-distance measure used to identify reproducible
scales \cite{Meila2007}.

Hierarchical modular organization is tested by applying Louvain community
detection at several resolution parameters $\gamma$. The resolution
parameter controls the scale of the detected communities: smaller values
favor larger communities, while larger values favor smaller communities.
At each fixed value of $\gamma$, we repeat Louvain several times with
different random seeds. These repeats test whether the community structure
at that particular resolution is reproducible.

We compare the repeated community partitions using variation of
information,

\[
\mathrm{VI}(\mathcal{P},\mathcal{Q})
=
H(\mathcal{P})
+
H(\mathcal{Q})
-
2I(\mathcal{P};\mathcal{Q}),
\]

where $\mathcal{P}$ and $\mathcal{Q}$ are two detected community
partitions, $H$ denotes partition entropy, and $I$ denotes mutual
information. Low VI means that repeated runs found similar community
assignments. We then look across the tested values of $\gamma$ for local
minima of the mean VI. A local minimum identifies a resolution at which
the community structure is especially reproducible \cite{Betzel2017modular}. We count such a
scale only when the corresponding modularity is also nontrivial, with
$Q>0.3$ \cite{NewmanGirvan2004}. The criterion passes when at
least one VI-stable modular scale is present.

\paragraph{10. Box-covering scaling (descriptive).}
Box covering asks how many boxes of a given size are needed to cover a graph.
For a box size \(l_B\), let \(N_B(l_B)\) be the number of boxes in a covering;
where the count follows a power law over a range of sizes, the box-covering
dimension is the negative slope
\[
d_B
=
-\,\frac{d\log N_B(l_B)}{d\log l_B},
\]
fitted by least squares on \(\log N_B\) against \(\log l_B\)
\cite{Song2005,Song2007memb}.  Exact minimum box covering is NP-hard, so the
count is always obtained by an approximation.  This work uses two different
approximations in two different roles, and they are not interchangeable; both
are given below.

\emph{(a) In the test, as a measurement.}  Box counts are obtained by the
maximum-excluded-mass burning (MEMB) algorithm of Song et al.\
\cite{Song2007memb}, run on the exact all-pairs shortest-path matrix \(d\) of
the largest connected component.  For a burning radius \(r_B\):

\begin{enumerate}
\item Initialize every vertex as uncovered and as a non-center.  For each
vertex \(u\), set its excluded mass
\(\mathrm{mass}(u)=|\{v: d(u,v)\le r_B\}|\).
\item While some vertex is neither covered nor a center:
  \begin{enumerate}
  \item Among all non-center vertices (including those already covered,
        following Song's step (ii)), select the vertex \(c\) of maximum
        excluded mass.
  \item If that maximum is zero, give each remaining uncovered non-center
        vertex its own box and stop.
  \item Mark \(c\) a center.  Assign to its box every uncovered vertex \(v\)
        with \(d(c,v)\le r_B\), mark those vertices covered, and increment the
        box count.
  \item For every vertex \(u\), reduce \(\mathrm{mass}(u)\) by the number of
        newly covered vertices within \(r_B\) of \(u\); set the mass of every
        center to zero.
  \end{enumerate}
\item Report \(N_B(l_B)\) as the resulting box count, with box size
\(l_B=2r_B\).
\end{enumerate}

The procedure is deterministic: no random choices enter it.  Burning within
radius \(r_B\) bounds pairwise distances inside a box by \(2r_B\), which is
why \(l_B\) is defined as \(2r_B\); Song's text writes \(l_B=2r_B+1\) under a
strict distance bound, and we take the offset that reproduces the published
benchmark values more closely.  The procedure is repeated for
\(r_B=1,\dots,\lfloor \mathrm{diam}/2\rfloor\), so the tested scales are
even. Scales at which the graph is covered by a single box are discarded,
the remaining points are fitted in log--log to give \(d_B\), and an
exponential \(N_B \propto e^{-\beta l_B}\) is fitted to the same points for
comparison.  This is the only place \(d_B\) is measured.  It is descriptive
and is not one of the nine scored criteria: it is reported to characterize the
graph rather than to assert fractality, and only where the graph affords
enough distance range for a fit.

\emph{(b) In the construction, as a mechanism.}  The multiscale-targeted
sparsification stage covers the graph in boxes in order to impose a property
rather than to measure one: edges of low triangle support are removed until
the stage's own estimate of \(d_B\) falls to its threshold.  Its covering is a
randomized greedy approximation, not MEMB.  For a scale \(l_B\):

\begin{enumerate}
\item Repeat for \(30\) independent trials:
  \begin{enumerate}
  \item Draw a uniformly random permutation of the vertices; mark all
        vertices uncovered and set the box count to zero.
  \item Scan the permutation in order.  Skip a vertex if it is already
        covered.  Otherwise take it as a center \(c\), mark covered every
        vertex \(v\) with \(d(c,v)\le l_B\), and increment the box count.
  \item Stop the scan once every vertex is covered.
  \end{enumerate}
\item Take \(N_B(l_B)\) to be the smallest box count over the \(30\) trials.
\end{enumerate}

This is repeated over the scales \(\{2,3,4,5,6,8,10,12\}\), restricted to
those not exceeding the graph diameter; single-box scales are discarded and
the remaining points fitted in log--log as above.  Two differences from (a)
are worth making explicit.  Centers are taken in random order rather than by
excluded mass, so the covering is a best-of-\(30\) heuristic rather than a
deterministic construction.  And the ball is taken at radius \(l_B\) itself,
not at \(l_B/2\): for the same numeric label, a box here has twice the radius
of a box in (a).  The bridge stage uses a covering of the same greedy kind
over the scales \(\{3,4,5,6,7\}\), differing in taking the strict ball
\(d(c,v)<l_B\) and in using \(50\) trials rather than \(30\); its box labels
are used solely as a multiscale locality filter on candidate edges; no
dimension is fitted there.

Because the center rule, the meaning of \(l_B\), and the scale set all differ,
the threshold at which sparsification halts and the dimension the test
subsequently reports are not the same quantity, and the two should not be read
as a target and a miss.

\section{Hierarchical Retention of Dynamical Variables}
\label{app:hierarchical_sde}

A useful precedent from statistical mechanics for the hierarchical description introduced in the main text is given by Plyukhin~\cite{Plyukhin2019}. The model considers a system coupled to a thermal environment with two groups of degrees of freedom having well-separated characteristic timescales. The slow part of the bath is retained explicitly as a collection of harmonic modes, while the faster degrees of freedom enter through dissipative and stochastic Langevin forces acting on these modes. Thus, the model provides a simple example in which one level of environmental dynamics is described explicitly, while a finer level is already represented by an effective stochastic description.

We can express this structure using notation closer to our framework. Let

\[
X_t
=
\begin{pmatrix}
Q_t\\
P_t
\end{pmatrix}
\]

denote the state of the higher-level system, and let

\[
Y_t
=
\begin{pmatrix}
q_t\\
p_t
\end{pmatrix},
\qquad
q_t=
\begin{pmatrix}
q_1(t)\\
\vdots\\
q_N(t)
\end{pmatrix},
\qquad
p_t=
\begin{pmatrix}
p_1(t)\\
\vdots\\
p_N(t)
\end{pmatrix},
\]

represent the explicitly retained slow-bath variables. We also define

\[
\Omega^2
=
\operatorname{diag}
\left(
\omega_1^2,\ldots,\omega_N^2
\right),
\qquad
\gamma
=
\begin{pmatrix}
\gamma_1\\
\vdots\\
\gamma_N
\end{pmatrix},
\]

and

\[
\kappa_Q
=
\gamma^{\mathsf T}\Omega^{-2}\gamma
=
\sum_{i=1}^{N}
\frac{\gamma_i^2}{\omega_i^2}.
\]

Using this notation, we can write Plyukhin's coupled equations as

\[
dQ_t
=
\frac{P_t}{M}\,dt ,
\]

\[
dP_t
=
\left(
\gamma^{\mathsf T}q_t
-
\kappa_Q Q_t
\right)dt ,
\]

together with

\[
dq_t
=
p_t\,dt ,
\]

and

\[
dp_t
=
\left(
-\Omega^2 q_t
+
\gamma Q_t
-
2\lambda p_t
\right)dt
+
\sqrt{4\lambda k_{\mathrm B}T}\,dW_t .
\]

Here \(W_t\) is an \(N\)-dimensional Wiener process. The last equation is the It\^o representation of the white-noise forces used in the original model,

\[
\left\langle
\xi_i(t)\xi_j(t')
\right\rangle
=
4\lambda k_{\mathrm B}T\,
\delta_{ij}\delta(t-t').
\]

We introduce the Wiener-process notation only for convenience. Plyukhin's original formulation uses the white-noise variables \(\xi_i(t)\).

We can now collect the equations into a two-level system,

\[
dX_t
=
F_{\mathrm P}(X_t,Y_t)\,dt ,
\]

\[
dY_t
=
G_{\mathrm P}(X_t,Y_t)\,dt
+
B_{\mathrm P}\,dW_t ,
\]

where the subscript \({\mathrm P}\) indicates that the functions belong specifically to Plyukhin's model. Explicitly,

\[
F_{\mathrm P}(X,Y)
=
\begin{pmatrix}
P/M\\[4pt]
\gamma^{\mathsf T}q-\kappa_Q Q
\end{pmatrix},
\]

and

\[
G_{\mathrm P}(X,Y)
=
\begin{pmatrix}
p\\[4pt]
-\Omega^2 q+\gamma Q-2\lambda p
\end{pmatrix},
\qquad
B_{\mathrm P}
=
\begin{pmatrix}
0\\[4pt]
\sqrt{4\lambda k_{\mathrm B}T}\,I_N
\end{pmatrix}.
\]

This form makes the hierarchy clear. Both \(X_t\) and the intermediate state \(Y_t\) are retained as explicit dynamical variables. The faster bath, however, does not appear as another set of resolved variables. Its influence is represented by the damping and stochastic forcing acting on \(Y_t\). Thus, in the notation of a general two-level stochastic system, we have

\[
dX_t
=
F(X_t,Y_t)\,dt ,
\qquad
dY_t
=
G(X_t,Y_t)\,dt
+
B_Y\,dW_t .
\]

In particular, there is no diffusion term acting directly on the upper state,

\[
B_X=0.
\]

What happens if we also eliminate the intermediate variables \(Y_t\)? Their influence on \(X_t\) does not, in general, disappear. In Plyukhin's construction, eliminating the slow oscillators produces a generalized Langevin equation for the particle momentum,

\[
\dot P(t)
=
-
\int_0^t
K(t-s)P(s)\,ds
+
\mathcal{F}(t),
\]

where \(K\) is a memory kernel determined by the spectrum and coupling of the retained slow modes and by the damping \(\lambda\) of the fast bath. The term \(\mathcal{F}(t)\) represents the corresponding effective stochastic force.

We can therefore describe the same hierarchical system in two ways. We may retain the intermediate variables explicitly, or eliminate them and represent their influence through memory and correlated stochastic forcing at the higher level. The two descriptions correspond to different levels of coarse-graining of the same construction.

The neural setting considered in our work motivates a more general form of this hierarchy. We do not assume the specific harmonic dynamics or additive noise of the bath model. Instead, we allow the two retained levels to follow general coupled stochastic dynamics,

\[
dX_t
=
F(X_t,Y_t)\,dt
+
B_X(X_t,Y_t)\,dW_t^{(X)},
\]

\[
dY_t
=
G(X_t,Y_t)\,dt
+
B_Y(X_t,Y_t)\,dW_t^{(Y)} .
\]

Here \(X_t\) may represent a more integrated collective state, while \(Y_t\) represents a richer lower-level state whose internal dynamics remains explicit. The stochastic terms describe the influence of degrees of freedom below the resolution of the retained variables. These influences need not have equal strengths. In particular, we may have

\[
\|B_X B_X^{\mathsf T}\|
\ll
\|B_Y B_Y^{\mathsf T}\|,
\]

so that the higher collective state is only weakly stochastic, even though the lower-level dynamics remains strongly influenced by unresolved finer-scale activity.

We emphasize that this general pair of equations is our proposed extension, not a result derived in Ref.~\cite{Plyukhin2019}. The correspondence is structural. Plyukhin's model provides a concrete physical example of the limiting case

\[
B_X=0,
\]

in which an intermediate dynamical level remains explicitly resolved while a faster level enters through stochastic forces. We extend this organization by allowing general coupled dynamics at both retained levels and, when appropriate, stochastic forcing acting on each of them. The same procedure can be repeated recursively to introduce additional levels.

\end{document}